\documentclass[sigconf, nonacm]{acmart}

\usepackage{verbatim}
\usepackage{graphicx}
\usepackage{multirow}
\usepackage[ruled,vlined,linesnumbered,noend]{algorithm2e}
\PassOptionsToPackage{noend}{algpseudocode}
\usepackage{algpseudocode}
\usepackage[labelfont=bf,textfont=bf]{caption}
\usepackage[labelfont=bf,textfont=bf,singlelinecheck=off,justification=raggedright]{subcaption}

\usepackage[normalem]{ulem}
\usepackage{longtable}
\usepackage{array}
\usepackage{url}
\usepackage{balance}
\usepackage{color}
\usepackage{hhline}
\usepackage{paralist}
\usepackage{amsmath}

\usepackage{amsthm}
\usepackage{bm}
\usepackage[normalem]{ulem} 
\newtheorem{defn}{Definition}

\graphicspath{{img/}}
\definecolor{myred}{RGB}{189, 52, 67}
\definecolor{mygreen}{RGB}{19, 136, 8}
\definecolor{myblue}{RGB}{16, 52, 166}
\usepackage{float}
\usepackage{hyperref}
\newcommand{\here}[1]{{\bf [[#1]]}}
\newcommand{\ke}[1]{{\color{black} #1}\normalcolor}

\newcommand{\hidevldb}[1]{}

\newcommand\vldbdoi{10.14778/3547305.3547308}
\newcommand\vldbpages{2005 - 2018}
\newcommand\vldbvolume{15}
\newcommand\vldbissue{10}
\newcommand\vldbyear{2022}
\newcommand\vldbauthors{\authors}
\newcommand\vldbtitle{\shorttitle} 
\newcommand\vldbavailabilityurl{http://www.mi.parisdescartes.fr/~themisp/hercules}
\newcommand\vldbpagestyle{empty}

\hypersetup{pdfcreator={Karima Echihabi}}
\hypersetup{pdfproducer={}}

\begin{document}
\title{Hercules Against Data Series Similarity Search}

\author{Karima Echihabi}
\affiliation{%
	\institution{Mohammed VI Polytech. Univ.}
}
\email{karima.echihabi@um6p.ma}

\author{Panagiota Fatourou}
\affiliation{%
	\institution{Universit{\'e} Paris Cit{\'e} \& FORTH}
}
\email{faturu@csd.uoc.gr}

\author{Kostas Zoumpatianos}
\affiliation{%
	\institution{Snowflake Inc.\thanks{Work done while at the University of Paris.}}
}
\email{kostas.zoumpatianos@snowflake.com}

\author{Themis Palpanas}
\affiliation{%
	\institution{Universit{\'e} Paris Cit{\'e} \& IUF}
}
\email{themis@mi.parisdescartes.fr}

\author{Houda Benbrahim}
\affiliation{%
	\institution{IRDA, Rabat IT Center \& ENSIAS}}
\email{houda.benbrahim@um5.ac.ma}

\begin{abstract}
We propose Hercules, a parallel tree-based technique for exact similarity search on massive disk-based data series collections. We present novel index construction and query answering algorithms that leverage different summarization techniques, carefully schedule costly operations, optimize memory and disk accesses, and exploit the multi-threading and SIMD capabilities of modern hardware to perform CPU-intensive calculations. 
We demonstrate the superiority and robustness of Hercules with an extensive experimental evaluation against state-of-the-art techniques, using many synthetic and real datasets, and query workloads of varying difficulty. 
The results show that Hercules performs 
up to one order of magnitude faster than the best competitor (which is not always the same). 
Moreover, Hercules is the only index that outperforms the optimized \ke{scan} on all scenarios, including the hard query workloads on disk-based datasets. This paper was published in the Proceedings of the VLDB Endowment, Volume 15, Number 10, June 2022.
\end{abstract}

\maketitle

\pagestyle{\vldbpagestyle}
\begingroup\small\noindent\raggedright\textbf{PVLDB Reference Format:}\\
\vldbauthors. \vldbtitle. PVLDB, \vldbvolume(\vldbissue): \vldbpages, \vldbyear.\\
\href{https://doi.org/\vldbdoi}{doi:\vldbdoi}
\endgroup

\ifdefempty{\vldbavailabilityurl}{}{
\vspace{.3cm}
\begingroup\small\noindent\raggedright\textbf{PVLDB Artifact Availability:}\\
The source code, data, and/or other artifacts have been made available at \url{\vldbavailabilityurl}.
\endgroup
}

\section{Introduction}
\label{sec:introduction}

Data series are one of the most common data types, 
covering virtually every scientific and social domain\hidevldb{, 
such as astrophysics, neuroscience, seismology, environmental monitoring, 
biology, health care, energy, finance, criminology, social studies, video 
and audio recordings, and many others}~\cite{KashinoSM99,Shasha99,humanbehaviorpatterns,volker,DBLP:conf/edbt/MirylenkaCPPM16,percomJournal,windturbines,spikesorting,VALMOD,DBLP:conf/eenergy/LavironDHP21,DBLP:journals/datamine/LinardiZPK20,dcam,benchref}.  
A data series is a sequence of ordered real values. 
When the sequence is ordered on time, it is called a \emph{time series}.
However, the order can be defined by other measures, such as angle or mass~\cite{conf/sofsem/Palpanas2016}.
\hidevldb{ {\bf ??? remove previous sentence ???} 
However, the order can be defined by angle (e.g., in radial profiles), mass (e.g., in mass spectroscopy), position (e.g., in genome sequences), and others~\cite{conf/sofsem/Palpanas2016}. 
A data series is typically represented as a high-dimensional vector of floating point values, where each value represents an observation and neighboring values are correlated.
In this paper, we use the terms \emph{data series}, \emph{time series} and \emph{sequence}, interchangeably. }

\noindent{\bf Motivation.} 
Similarity search 
lies at the core of many critical data science applications related to data series~\cite{DBLP:journals/sigmod/Palpanas15,fulfillingtheneed,Palpanas2019,dagstuhl-report}, 
but also in other domains, where data are represented as (or transformed to) high-dimensional vectors~\cite{DBLP:conf/wims/EchihabiZP20,icde21tutorial,vldb21tutorial}.
The problem of efficient similarity search over large \hidevldb{data }series collections has been studied heavily in the past 30 
years~\cite{DBLP:conf/sigmod/FaloutsosRM94,shieh2009isax,conf/vldb/Wang2013,ulisse,DBLP:journals/vldb/KondylakisDZP19,evolutionofanindex,conf/sigmod/echihabi2020,conf/sigmod/gogolou20,DBLP:conf/icdm/YagoubiAMP17,peng2021sing,DBLP:journals/kais/LevchenkoKYAMPS21,DBLP:conf/edbt/Chatzigeorgakidis21,messijournal,seanet,tsindex},  
and will continue to attract attention as massive sequence collections are becoming omnipresent in various domains~\cite{Palpanas2019,dagstuhl-report}.

\noindent{\bf Challenges.} 
Designing highly-efficient disk-based data series indexes is thus crucial.
ParIS+~\cite{parisplus} is a disk-based data series parallel index, 
which exploited the parallelism capabilities provided by 
multi-socket and multi-core architectures. 
ParIS+ builds an index based on the iSAX summaries of the data series in the collection,
whose leaves are populated only at 
search time. 
This results in a highly reduced cost for constructing the index, but as we will demonstrate, it often incurs \ke{a} high query answering cost, despite the \ke{superior} pruning ratio offered by the iSAX summaries~\cite{journal/pvldb/echihabi2018}. 

For this reason, other data series indexes, such as DSTree~\cite{conf/vldb/Wang2013}, have recently attracted attention. 
Extensive experimental evaluations~\cite{journal/pvldb/echihabi2018,journal/pvldb/echihabi2019} have shown that sometimes DSTree exhibits better performance in query answering than iSAX-based indexes (like ParIS+). 
This is because DSTree spends a significant amount of time during indexing to adapt its splitting policy and node summarizations to the data distribution, leading to better data clustering (in the index leaf nodes), and thus more efficient \ke{exact} query answering for workloads~\cite{journal/pvldb/echihabi2018}. 
In contrast, ParIS+ uses pre-defined split points and a fixed maximum resolution for the iSAX summaries. 
This helps ParIS+ build the index faster, but hurts the quality of data clustering. 
Therefore, similar data series may end up in different leaves, making query answering slower, especially on hard workloads that are the most challenging. 
As a result, no single data series similarity search indexing approach among the existing techniques
wins across all \ke{popular} query workloads~\cite{journal/pvldb/echihabi2018}. 
\hidevldb{This is the problem we tackle in this paper: we propose Hercules, a new data series index that enjoys the advantages of both approaches (discussed above), by employing new ideas to overcome their limitations. }

\noindent{\bf The Hercules Approach.} 
In this paper, we present Hercules, 
the first data series index that answers queries faster than all \ke{recent}  state-of-the-art techniques
across all \ke{popular} workloads. 
Hercules achieves this by exploiting the following key ideas. 
First, it leverages two different data series summarization techniques, EAPCA and iSAX, 
utilized by DSTree and ParIS+, respectively, to get a well-clustered tree structure during index building, 
and at the same time a high leaf pruning ratio during query answering. Hercules 
enjoys the advantages of both approaches and employs new ideas to overcome their limitations.
Second, it exploits a two-level buffer management technique to optimize memory accesses. 
This design utilizes (i) a large 
memory buffer (HBuffer) which contains the raw data of all leaves and is flushed to disk whenever it becomes full,
and (ii) a small buffer (\ke{SBuffer}) for each leaf containing pointers to the raw data of the leaf that are stored in HBuffer.  
This way, it reduces the number of system calls, guards against the occurrence of out-of-memory management issues, 
and improves the performance of index construction.
It also performs scheduling of external storage requests more efficiently.
While state-of-the-art techniques~\cite{conf/vldb/Wang2013,parisplus} typically store the data of each leaf in a separate file, 
Hercules stores the raw data series in a single file\hidevldb{ (called LRDFile)}, following the order of an inorder traversal of the tree leaves. 
This helps Hercules reduce the cost of random I/O operations, and \hidevldb{as we will elaborate later, }efficiently support both easy and hard queries. 
Moreover, Hercules schedules costly operations (e.g., statistics updates, calculating iSAX summaries, etc.)
more efficiently than in previous works~\cite{conf/vldb/Wang2013,parisplus}. 

Last but not least, Hercules uses parallelization\hidevldb{ whenever needed,} to accelerate the execution of CPU-intensive calculations. 
It does so judiciously, adapting its access path selection decisions to each \hidevldb{single }query in a workload during query answering (e.g., Hercules decides on when to parallelize a query based on the pruning ratio of the data series summaries, EAPCA and iSAX), and carefully scheduling index insertions and disk flushes while guaranteeing data integrity\hidevldb{ during index construction}.
This required novel designs for the following reasons. 
The parallelization ideas utilized for state-of-the-art data series indexes~\cite{conf/bigdata/peng18,conf/icde/peng20,parisplus} 
are 
relevant to tree-based indexes with a very large root fanout (such as the iSAX-based indexes~\cite{evolutionofanindex}). 
In such indexes, parallelization can be easily achieved by having different threads 
work in different root subtrees of the index tree\hidevldb{, and interfering with each other only in order to decide to 
which root subtrees they will work on}. 
The Hercules index tree is an unbalanced \emph{binary} tree, and thus it has a small fanout and uneven subtrees. 
In such index trees, heavier synchronization is necessary, as different threads may traverse similar paths 
and may need to process the same nodes. 
\hidevldb{Such threads have to synchronize repeatedly as they traverse the tree \hidevldb{to ensure data consistency}.}Moreover, working on trees of small fan-out may result in more severe load balancing problems\hidevldb{ (especially
when the tree is dynamically processed and its subtrees have different sizes)}. 

Our experimental evaluation with synthetic and real datasets from diverse domains shows that\ke{, in terms of query efficiency, Hercules} consistently outperforms 
both Paris+ (by 5.5x-63x) 
and our optimized implementation of DSTree (by 1.5x-10x). 
This is true for all synthetic and real query workloads of different hardness that we tried (note that all algorithms return the same, exact results). 

\noindent{\bf Contributions}. 
Our contributions are as follows: 
	
\noindent$\bullet$ We propose a new parallel data series index that exhibits better query answering performance than all \ke{recent} state-of-the-art approaches across all \ke{popular} query workloads.
The new index achieves better pruning \hidevldb{degree}than previous approaches by exploiting the benefits of two different summarization techniques, and a new query answering mechanism. 

%

\noindent$\bullet$ We propose a new parallel data series index construction algorithm, 
which leverages a novel protocol for constructing the tree and flushing data to disk. 
This protocol achieves load balancing (among workers) on the binary index tree, due to a careful scheduling mechanism of costly operations. 


\noindent$\bullet$ We realize an ablation study \hidevldb{on Hercules to better understand the contribution of the different key design choices to its overall performance. 
This study provides a performance breakdown }that explains the contribution of individual design choices to the final performance improvements for queries of different hardness, and could support further future advancements in the field. 

\noindent$\bullet$ We demonstrate the superiority and robustness of Hercules with an extensive experimental evaluation against the state-of-the-art techniques, using a variety of synthetic and real datasets\hidevldb{, including three of the largest publicly available real datasets from the domains of neuroscience, seismology and computer vision}.
The experiments, with query workloads of varying degrees of difficulty, 
show that Hercules is between 1.3x-9.4x faster than the best competitor, which is different for the different datasets and workloads.
Hercules is also the only index that outperforms the optimized \ke{parallel} scan on \emph{all} scenarios, including the case of hard query workloads\hidevldb{ on large, disk-based sequence collections}.

\vspace{-0.3cm}

\section{Preliminaries \& Related Work}
\label{sec:preliminaries}

%
%
A \textit{\textbf{data series}} $S(p_1,p_2,...,p_n)$ is an ordered sequence of points, $p_i$, $1 \leq i \leq n$.
The number of points, $|S|=n$, is the length of the series. We use $\mathbb{S}$ to represent all the series in a collection (dataset). A data series of length $n$ is typically represented as a single point in an $n$-dimensional space. 
Then the values and length of $S$ are referred to as \emph{dimensions} and \emph{dimensionality}, respectively.

Data series similarity search is typically reduced to a k-Nearest Neighbor  (k-NN) query\hidevldb{ over all data series in a collection}.
\hidevldb{
\begin{defn} \label{def:knnquery}}
Given an integer $k$, a dataset $\mathbb{S}$ and a query series $S_Q$, a \textit{\textbf{k-NN query}} 
retrieves the set of $k$ series $\mathbb{A} = \{ \{S_{1},...,S_{k}\} \in \mathbb{S}~|~ \forall \ S \in \mathbb{A} \ and \ \forall \ S' \notin \mathbb{A}, \ d(S_Q,S) \leq d(S_Q,S')\}$, where $d(S_Q,S_C)$ is the distance between $S_Q$ and any data series $S_C \in \mathbb{S}$.
\hidevldb{\end{defn}}

This work focuses on whole matching queries (i.e., compute similarity between an entire query series and an entire candidate series) using the Euclidean distance. \ke{Hercules can support any distance measure equipped with a lower-bounding distance, e.g. Dynamic Time Warping (DTW)~\cite{journal/kis/Keogh2005} (similarly to how other indices support DTW~\cite{messijournal}).}
\hidevldb{{\bf ??? remove previous parenthesis ???} A whole matching query consists of computing similarity between an entire query series and an entire candidate series. 
All series involved in a whole-matching query have the same length.}


\hidevldb{Similarity search methods over massive data collections achieve query efficiency by exploiting summarization techniques and efficient data structures and search algorithms.}

\begin{figure}[tb]
	\captionsetup{justification=centering}
	\captionsetup[subfigure]{justification=centering}
	\centering
	\begin{subfigure}{0.5\columnwidth}
		\centering
		\includegraphics[width=\columnwidth]{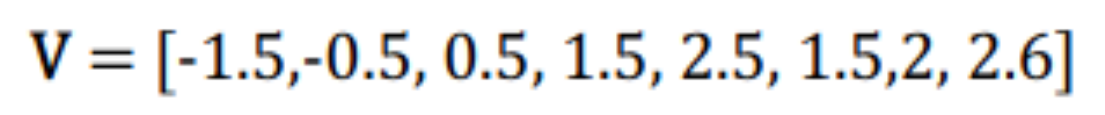}
	\end{subfigure}\\
	\begin{subfigure}{0.24\columnwidth}
		\centering
		\includegraphics[width=\columnwidth]{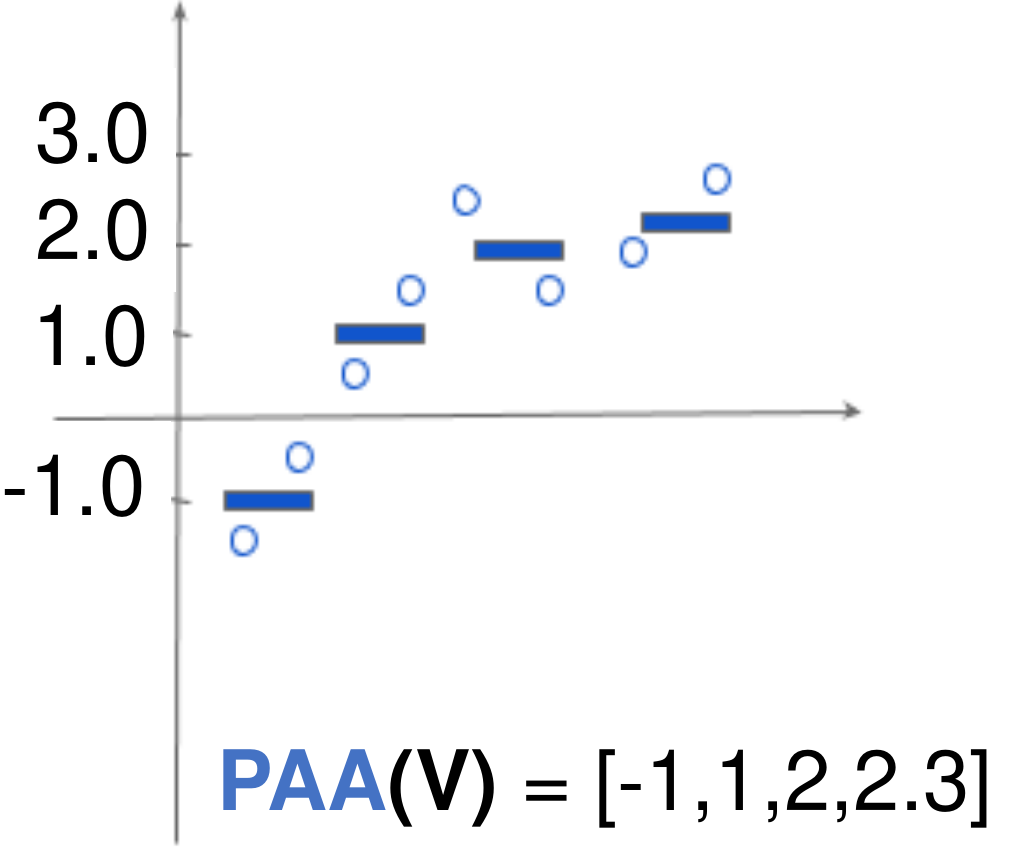}
		\caption{PAA}
		\label{fig:summarizations-paa}
	\end{subfigure}
	\begin{subfigure}{0.20\columnwidth}
		\centering
		\includegraphics[width=\columnwidth]{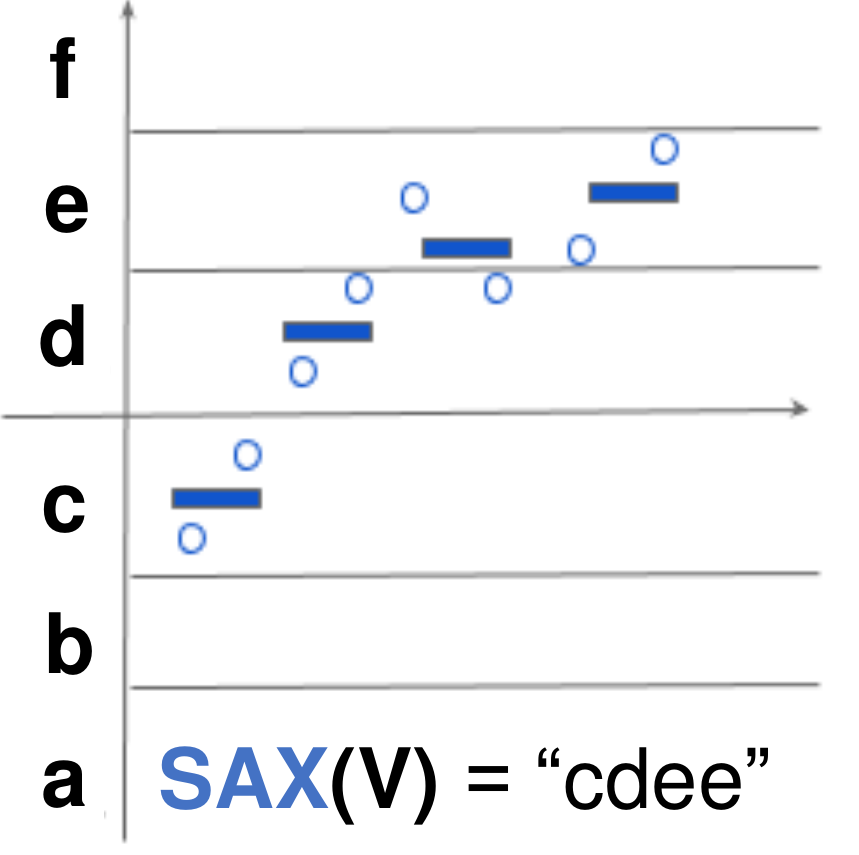}
		\caption{iSAX}
		\label{fig:summarizations-sax}
	\end{subfigure}
	\begin{subfigure}{0.24\columnwidth}
		\centering
		\includegraphics[width=\columnwidth]{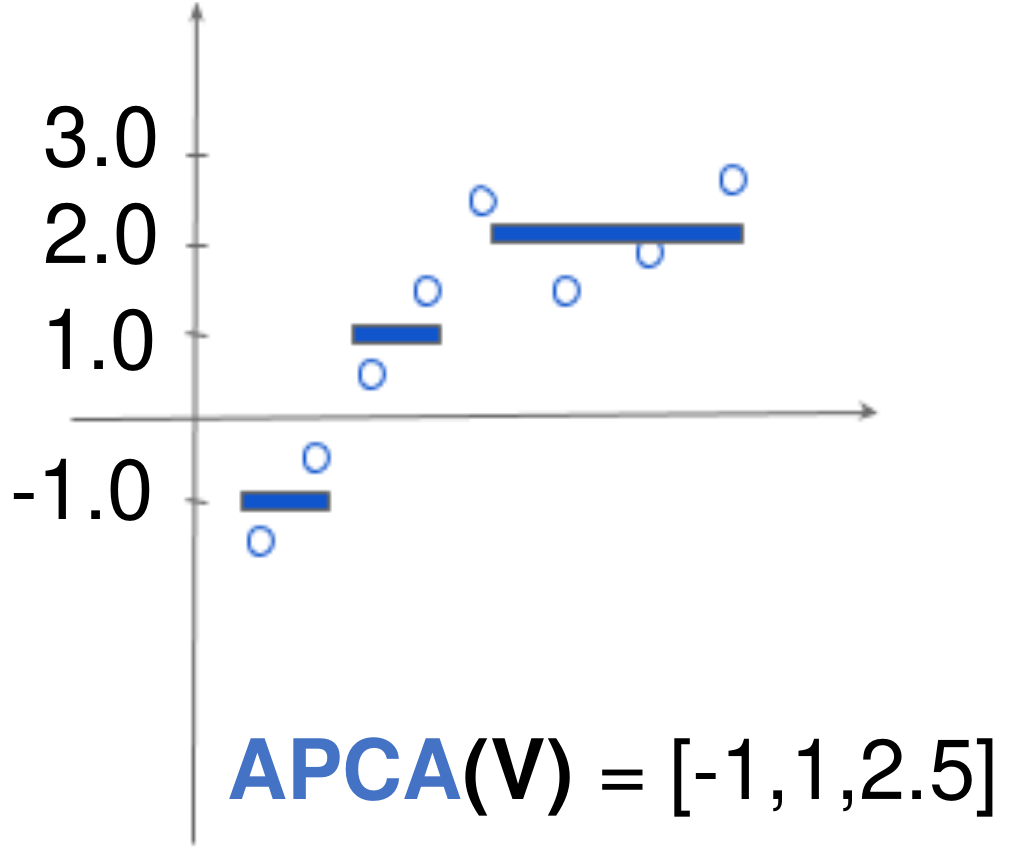}
		\caption{APCA}
		\label{fig:summarizations-apca}
	\end{subfigure}
	\begin{subfigure}{0.24\columnwidth}
		\centering
		\includegraphics[width=\columnwidth]{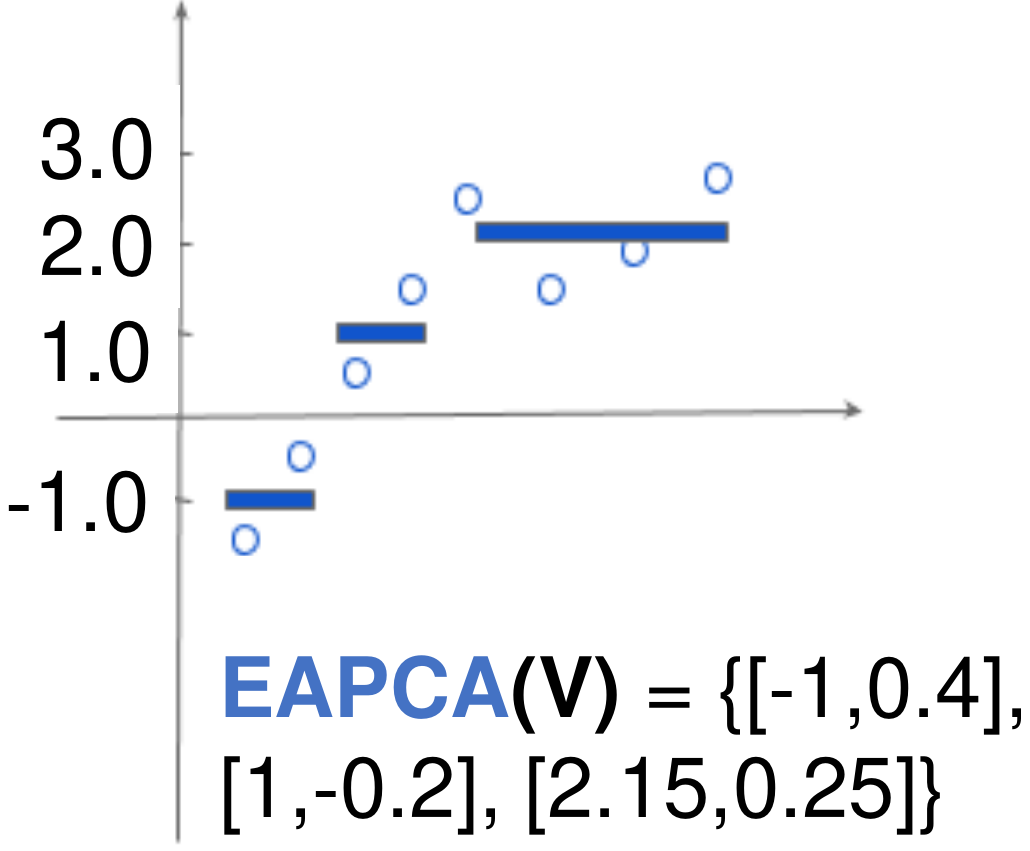}
		\caption{EAPCA}
		\label{fig:summarizations-eapca}
	\end{subfigure}
	\vspace*{-0.2cm}
	\caption{Summarization Techniques}
	\vspace*{-0.5cm}
	\label{fig:summarizations}
\end{figure}

\noindent {\bf Summarization Techniques.}
\hidevldb{Many dimensionality reduction techniques have been proposed in the literature. We briefly describe the techniques relevant to our work.
The Discrete Fourier Transform (DFT)~\cite{conf/fodo/Agrawal1993,conf/sigmod/Faloutsos1994,conf/sigmod/Rafiei1997,journal/corr/Rafiei1998} transforms a data series into the frequency domain and uses a subset of its coefficients as a summarization. The Fast Fourier Transform (FFT) algorithm is optimal for whole matching scenarios while the MFT algorithm~\cite{conf/icdsp/Albrecht1997} is more suited for subsequence matching queries.}
The Symbolic Aggregate Approximation (SAX)~\cite{conf/dmkd/LinKLC03} \hidevldb{is a representation technique that }first transforms a data series using Piecewise Aggregate Approximation (PAA)~\cite{journal/kais/Keogh2001}. It divides the series into equi-length segments and represents each segment with one floating-point value corresponding to the mean of all the points belonging to the segment (Fig.~\ref{fig:summarizations-paa}). The iSAX summarization reduces the footprint of the PAA representation by applying a discretization technique that maps PAA values to a discrete set of symbols (alphabet) that can be succinctly represented in binary form (Fig.~\ref{fig:summarizations-sax}). \ke{Following previous work, we use 16 segments~\cite{journal/pvldb/echihabi2018} and an alphabet size of 256~\cite{conf/kdd/shieh1998}}.

The Adaptive Piecewise Constant Approximation (APCA)~\cite{journal/acds/Chakrabarti2002} is a technique that approximates a series using variable-length segments. 
The approximation represents each segment with the mean value of its points (Fig.~\ref{fig:summarizations-apca}). 
The Extended APCA (EAPCA)~\cite{conf/vldb/Wang2013} 
represents each segment with both the mean ($\mu$) and standard deviation ($\sigma$) of the points belonging to it (Fig.~\ref{fig:summarizations-eapca}). 

\noindent {\bf Similarity Search Methods.} 
Exact similarity search methods are either sequential or index-based. 
The former compare a query series to each candidate series in the dataset, 
whereas the latter limit the number of such comparisons using efficient data structures and algorithms. 
An index is built using a summarization technique that supports lower-bounding~\cite{DBLP:conf/sigmod/FaloutsosRM94}. 
During search, a filtering step exploits the index to eliminate candidates in the summarized space with no false dismissals, and a refinement step verifies the remaining candidates against the query in the original high-dimensional space~\cite{conf/vldb/Ciaccia1997,conf/cikm/hakan2000, conf/icdm/Camerra2010,journal/vldb/Zoumpatianos2016,journal/edbt/Schafer2012,conf/vldb/Wang2013,conf/icmd/Beckmann1990,dpisaxjournal,ulisse,edbt21tutorial,vldb21tutorial,bigdata20tutorial}. We describe below the state-of-the-art exact similarity search techniques~\cite{journal/pvldb/echihabi2018}. \ke{Note that the scope of this paper is single-node systems, thus, we do not cover distributed approaches such as TARDIS~\cite{conf/icde/zhang2019} and DPiSAX~\cite{dpisaxjournal}.} 	

The UCR Suite~\cite{conf/kdd/Mueen2012} is an optimized sequential scan algorithm supporting the DTW and Euclidean distances for exact \emph{subsequence} matching. 
We adapted the original algorithm to support exact whole matching, used the optimizations relevant to the Euclidean distance (i.e., squared distances and early abandoning)\ke{, and developed a parallel version, called PSCAN, 
that exploits SIMD, multithreading and double buffering.}

The VA+file~\cite{conf/cikm/hakan2000} is a skip-sequential index that exploits a filter file containing quantization-based approximations of the high dimensional data. 
We refer to the VA+file variant proposed in~\cite{journal/pvldb/echihabi2018}, which exploits DFT 
(instead of the Karhunen\textendash Lo\`{e}ve Transform) 
for efficiency purposes~\cite{journal/acta/maccone2007}. 

The DSTree~\cite{conf/vldb/Wang2013} is a tree-based approach which uses the EAPCA segmentation. 
The DSTree intertwines segmentation and indexing, building an unbalanced binary tree. 
Internal nodes contain statistics about the data series belonging to the subtree rooted at it, and each leaf node is associated with a file on disk that stores the raw data series. 
The data series in a given node are all segmented using the same policy, but each node has its own segmentation policy, which may result in nodes having a different number of segments or segments of different lengths.\hidevldb{DSTree supports both lower and upper bounding distance.  
It uses both distances to determine the optimal splitting policy for each node.} 
During search, it exploits the lower-bounding distance $\mathrm{LB_{EAPCA}}$ 
to prune the search space. 
This distance is calculated using the EAPCA summarization of the query and the synopsis of the node, which refers to the EAPCA summaries of all series that belong to the node.
ParIS+~\cite{parisplus} is the first data series index for data series similarity search to exploit multi-core and multi-socket architectures. It is a tree-based index belonging to the iSAX family~\cite{evolutionofanindex}. Its index tree is comprised of several root subtrees, where each subtree is built by a single thread and a single thread can build multiple subtrees. During search, ParIS+ implements a parallel version of the ADS+ SIMS algorithm~\cite{journal/vldb/Zoumpatianos2016}, which prunes the search space using the iSAX summarizations and the \hidevldb{corresponding }lower-bounding distance $\mathrm{LB_{SAX}}$. 

\noindent{\bf Hercules vs. Related Work.} 
\ke{Hercules exploits the data-adaptive EAPCA segmentation proposed by DSTree~\cite{conf/vldb/Wang2013} 
	to cluster similar data series in the same leaf.} 
This results in a binary tree which cannot be efficiently constructed and processed by utilizing the simple parallelization techniques of ParIS+~\cite{parisplus}.
Specifically, the parallelization strategies of the index construction and query answering algorithms of ParIS+ 
exploit the large fanout of the root node, such that threads do not need to synchronize access to the different root subtrees. 
However, in Hercules, threads need to synchronize
from the root level. 
Therefore, Hercules uses a novel parallel index construction algorithm to build the tree.

In addition, unlike previous work, Hercules stores the raw data series belonging to the leaves using a two-level buffer management architecture. This scheme plays a major role in achieving the good performance exhibited by Hercules, 
but results in the additional complication of having to cope with the periodic flushing of HBuffer to the disk. This flushing is performed by a single thread (the flush coordinator) to maintain I/O cost low. 
Additional synchronization is required to properly manage all threads 
that populate the tree, 
until flushing is complete.
Note also that ParIS+ builds the index tree based on iSAX, thus, it only accesses the raw data once 
to calculate the iSAX summaries and insert the latter into the index tree. 
Node splits are also based on the summaries, which fit in-memory. 
However,
Hercules builds the index tree using the disk-based raw data, 
making the index construction parallelization more challenging. 
Finally, ParIS+ treats all queries in the same way, whereas Hercules adapts its query answering strategy 
to each query. 

Our experiments (Section~\ref{sec:experiments}) demonstrate that 
Hercules outperforms on query answering both DSTree and ParIS+ by up to 1-2 orders of magnitude across all query workloads.

\vspace{-0.1cm}

\vspace{-0.4cm}
\section{The Hercules Approach}
\label{sec:hercules_overview}
\subsection{\ke{Overview}}

\hidevldb{
Hercules is based on four key ideas: 1) leveraging two different summarization techniques, 
EAPCA to build a well-clustered tree structure and iSAX to accurately encode the raw series using low footprint;
2) optimizing memory management; 
3) carefully scheduling of external storage requests and other costly operations; 
and 4) taking advantage of multi-threading and SIMD capabilities of modern 
CPUs to accellerate the computation of CPU-intensive calculations.
\here{The above have already been mentioned in the intro, thus, they can
be removed to save space.}


} 


\ke{The Hercules pipeline  consists of two main}
stages: an \ke{\em index construction} stage, 
where the index tree is built 
and materialized to disk, and \ke{\em a query answering} stage where the index is used to answer  similarity search queries. 

The index construction stage involves {\em index building} and 
{\em index writing}. During {\em index building}, the raw series are read from disk and inserted into the appropriate leaf of the Hercules EAPCA-based index tree. 
As their synopses are required only for query answering,
to avoid contention at the internal nodes, we update only the synopses of the \emph{leaf} nodes during this
phase, while the synopses of the \emph{internal} nodes are updated during the index writing phase. 
\hidevldb{If inserting a data series to
a leaf node will cause it to exceed the {\em leaf threshold}, $\tau$, the node is split into two children, based on the raw series that have been stored in it.} The raw series belonging to all the leaf nodes are 
stored in a pre-allocated memory buffer called HBuffer.
Once all the data series are processed, index construction proceeds to 
index writing: the leaf nodes are post-processed 
to update their ancestors' synopses, and the iSAX summaries are calculated. 
Moreover, the index is materialized to disk into three files: 
(i) \ke{HTree}, containing the index tree; 
(ii) LRDFile, containing the raw data series; and 
(iii) LSDFile containing their iSAX summaries (in the same order as LRDFile's raw data).

Hercules's query answering stage consists of four main \ke{phases}:
(1) The index tree is searched heuristically to find initial approximate 
answers. (2) The index tree is pruned based on these initial answers and the EAPCA lower-bounding distances in order to build the list of candidate leaves (LCList). 
\hidevldb{This process guarantees to contain the exact answers (i.e., there are no false dismissals).} 
If the size of LCList is large (\hidevldb{meaning that the }EAPCA-based pruning was not effective), 
a single-thread skip-sequential search on LRDFile is preferred
(so phases 3 and 4 are skipped).
(3) Additional pruning is performed by multiple threads based on the iSAX summarizations of the data series in the candidate leaves of LCList.
The remaining data series are stored in SCList. 
If the size of SCList is large (\hidevldb{meaning that the }SAX-based pruning was low), 
a skip-sequential search using one thread is performed directly on LRDFile and phase 4 is skipped. (4) Multiple threads calculate real distances between the query and 
the series in SCList, and the series with the lowest distances 
are returned as final answers.
Thus, \hidevldb{during query answering, }Hercules prunes the search space using both the lower-bounding distances $\mathrm{LB_{EAPCA}}$~\cite{conf/vldb/Wang2013} and $\mathrm{LB_{SAX}}$~\cite{conf/dmkd/LinKLC03}.

\vspace{-0.3cm}
\subsection{\ke{The Hercules Tree}}

\begin{figure}[tb]
\captionsetup{justification=centering}	
\begin{subfigure}{\columnwidth}
\centering
\captionsetup{justification=centering}	
\includegraphics[width=0.9\columnwidth] {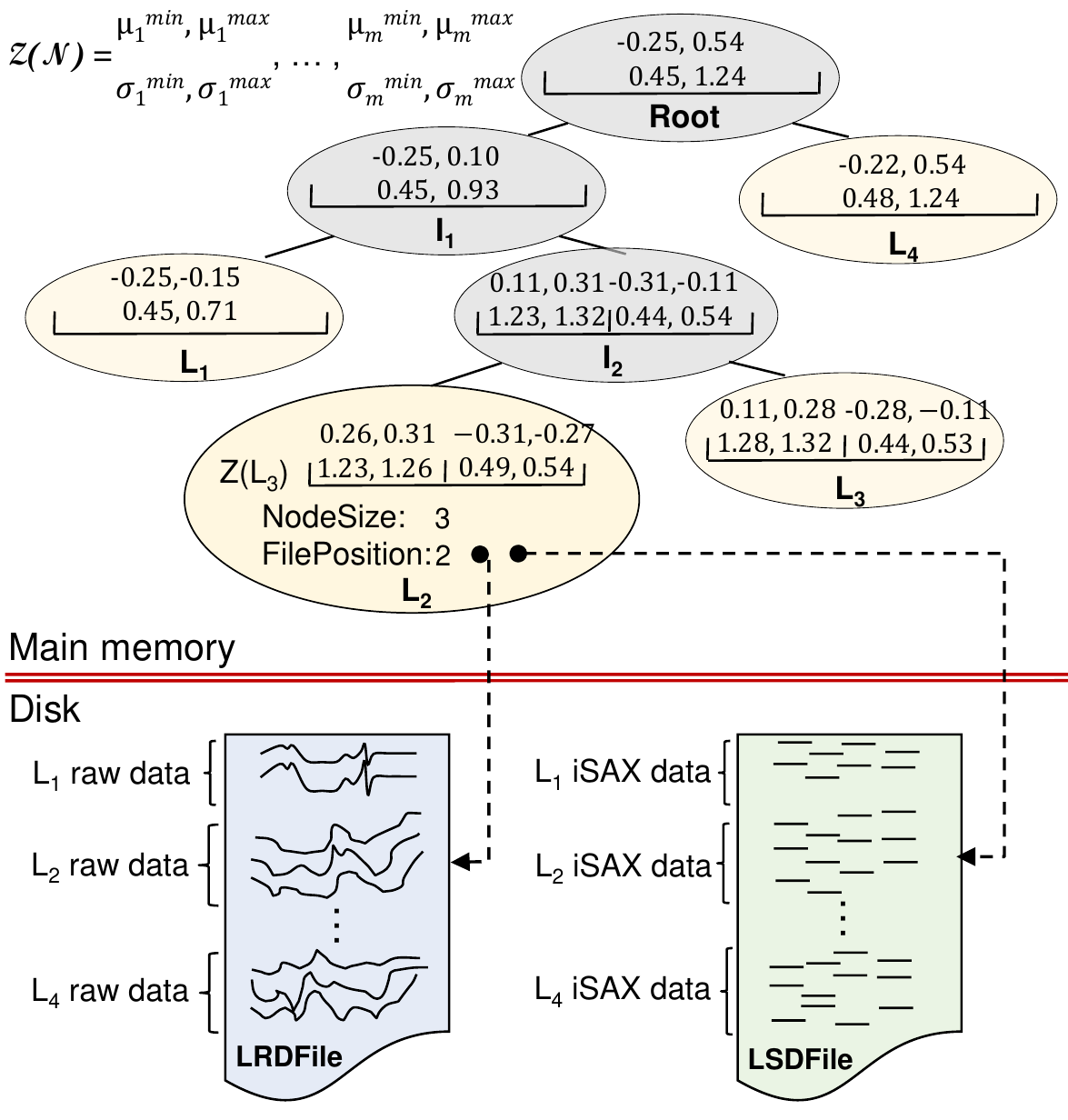}
\end{subfigure}
\vspace*{-0.3cm}
\caption{The Hercules Index Tree}  
\vspace*{-0.4cm}
\label{fig:hercules-tree}
\end{figure}

Figure~\ref{fig:hercules-tree} depicts the Hercules tree. 
Each node $\mathcal{N}$ contains: (i) the size, $\rho$, of the set $\mathbb{S}_\mathcal{N} = \{S_1,\dots,S_{\rho}\}$ of all series stored in the leaf descendants of $\mathcal{N}$ (or in $\mathcal{N}$ if it is a leaf); 
(ii) the segmentation $SG = \{r_1,...,r_m\}$ of $\mathcal{N}$, 
where $r_i$ is the right endpoint of $SG_i$, the $i$th segment of $SG$, with {\small $1 \le i \le m$, $ 1 \le r_1 <...< r_m = n$}, {\small $r_0 = 0$}
and $n$ is the length of the series; 
and (iii) a synopsis {\small $Z = (z_1, z_2, ..., z_m)$}, where {\small $z_i = \{\mu_i^{min}, \mu_i^{max}, \sigma_i^{min}, \sigma_i^{max}\}$} 
is the synopsis of the segment $SG_i$, 
{\small $\mu_i^{min} = min(\mu_i^{S_1},\dots,\mu_i^{S_\rho})$}, {\small $\mu_i^{max} = max(\mu_i^{S_1},\dots,\mu_i^{S_\rho})$}, {\small $\sigma_i^{min} = min(\sigma_i^{S_1},\dots,\sigma_i^{S_\rho})$}, 
and {\small $\sigma_i^{max} = max(\sigma_i^{S_1},\dots,\sigma_i^{S_\rho})$}. 

A leaf node is associated with a FilePosition which indicates the position of the leaf's raw data in LRDFile 
and that of the leaf's iSAX summaries in LSDFile. 
An internal node contains pointers to 
its children nodes and a splitting policy. 
When a leaf node $\mathcal{N}$ exceeds its capacity $\tau$ (\emph{leaf threshold}), it is split into two children nodes $N_l$ and $N_r$, 
which are leaf nodes, and $\mathcal{N}$ becomes an internal node.

Hercules exploits a splitting policy that (similarly to DSTree) allows the resolution of a node's summarization 
to increase along two dimensions: horizontally ({\it H-split}) and vertically ({\it V-split}), 
unlike the other data series indexes~\cite{isax2plus,journal/vldb/Zoumpatianos2016,journal/edbt/Schafer2012}, which allow either one or the other (for example, 
iSAX-based indexes only allow horizontal splitting). A node $\mathcal{N}$ is split by picking $SG_i$, 
one of the $m$ segments of $\mathcal{N}$, and using the mean (or the standard deviation) of the points belonging 
to $SG_i$ to redistribute the series in $\mathbb{S}_N$ among $N_l$ and $N_r$. 
In an H-split, both $N_l$ and $N_r$ have the same segmentation as $\mathcal{N}$ whereas in a V-split, 
the children nodes have one additional segment. In an H-split using the mean, 
the data series in $\mathbb{S}_N$ which have a mean value for segment $SG_i$ 
in the range [$\mu_i^{min},(\mu_i^{min} + \mu_i^{max})/2$) will be stored in $N_l$ 
and those whose range is in [$(\mu_i^{min} + \mu_i^{max})/2,\mu_i^{max}$] will be stored in $N_r$. 
An H-split using the standard deviation proceeds in the same fashion we just described, 
except it uses the values of standard deviations instead of mean values. 
A V-split first splits $SG_i$ into two new segments then applies an H-split on one of them. 
For example, in Figure~\ref{fig:hercules-tree}, node 
$\mathrm{I_{2}}$  has an additional segment compared to its parent 
$\mathrm{I_{1}}$, because it was split vertically, while the latter has the same segmentation as Root, because it was split horizontally. 
If we suppose that $\mathrm{I_{1}}$ was split using mean values, the data series that were originally stored in 
$\mathrm{\mathbb{S}_{I_{1}}}$ are now stored in
$\mathrm{\mathbb{S}_{L_{1}}}$ if their mean value $\mu_1 < -0.075$  
($\frac{-0.25+0.10}{2} = -0.075$ ), and to its right child otherwise.

\subsection{Index Construction}
\subsubsection{Overview}

\begin{figure*}[tb]
	\captionsetup{justification=centering}	
	\begin{subfigure}{\textwidth}
		\centering
		\captionsetup{justification=centering}	
		\includegraphics[width=0.90\textwidth] {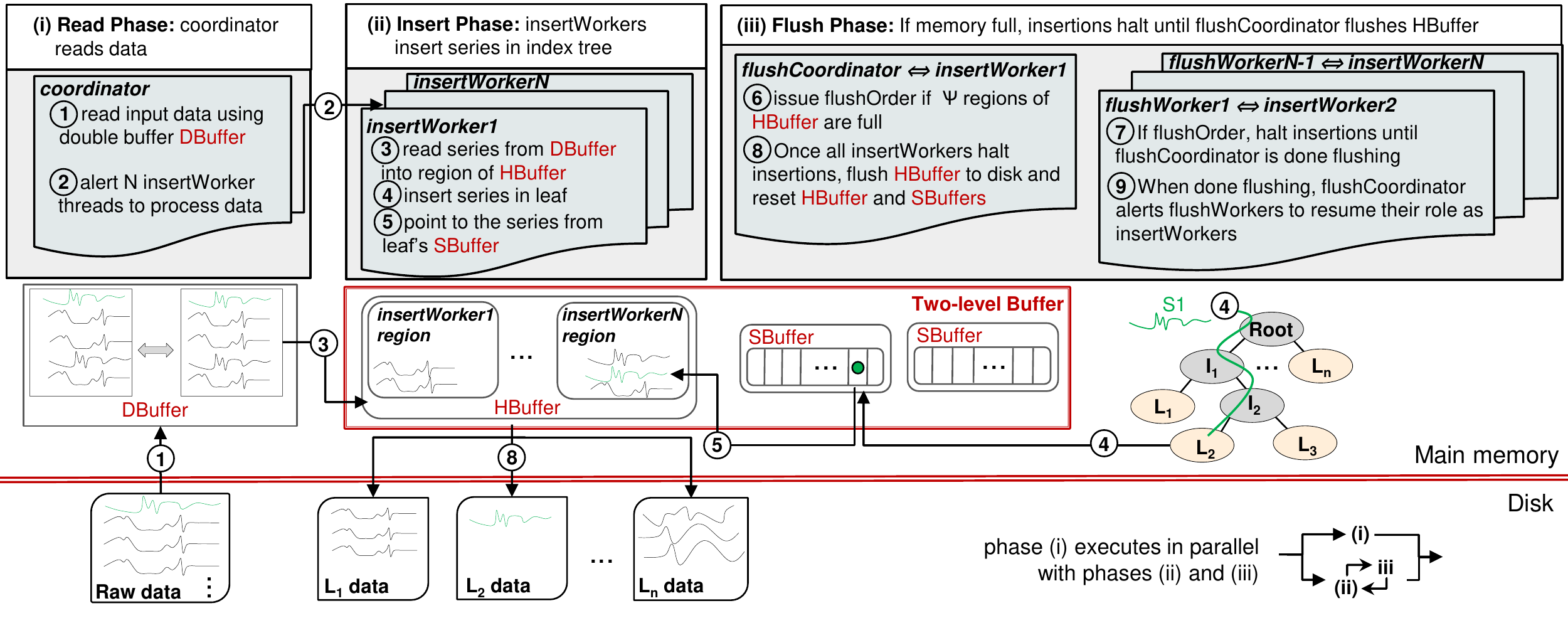}
	\end{subfigure}
	\caption{\ke{Hercules Index Building Workflow}}
	\label{fig:hercules-index-building}
\end{figure*}

\begin{figure}[!htb]
	\captionsetup{justification=centering}	
	\begin{subfigure}{\columnwidth}
		\centering
		\captionsetup{justification=centering}	
		\includegraphics[width=0.95\columnwidth] {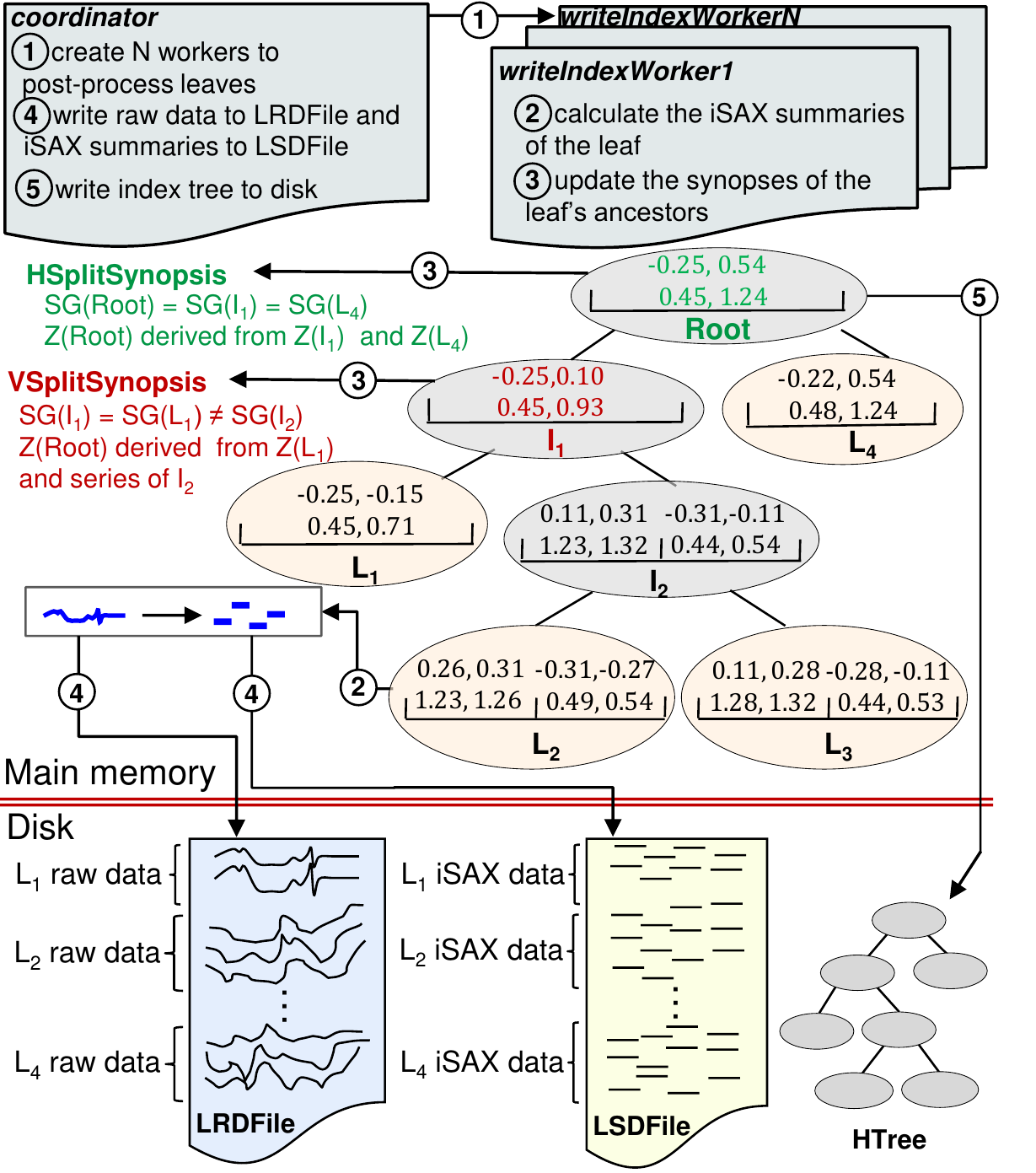}
	\end{subfigure}
\vspace*{-0.2cm}
	\caption{\ke{Hercules Index Writing Workflow}}  
\vspace*{-0.5cm}
	\label{fig:hercules-index-writing}
\end{figure}



\hidevldb{\subsubsection{Index Construction Workflows}}

%
Figure~\ref{fig:hercules-index-building} describes the index building workflow. \ke { {Index building} consists of three phases: (i) a {\em read} phase, (ii) an {\em insert} phase, and (iii) a {\em flush} phase}. A thread acting as the coordinator and a number of InsertWorker threads  operate on the raw data using a double buffer scheme, called DBuffer. 
Hercules leverages DBuffer 
to interleave the I/O cost of reading raw series from disk with 
CPU-intensive insertions into the index tree. \ke{The numbered circles in Figure~\ref{fig:hercules-index-building} denote the tasks performed by the threads.}

\ke{During the {\em read} phase}, the coordinator loads the dataset into memory in batches, reading a batch of series from the original file into the first part of DBuffer (1). It gives up the first part of DBuffer when it is full and continues storing data into the second part (2).


\ke{During the {\em insert} phase,} 
The InsertWorker threads\hidevldb{work in parallel to} read series from the part of the buffer that the coordinator 
does not work on (3) and insert them into the tree.  
Each InsertWorker traverses the index tree to find the appropriate leaf, 
routing left or right depending on the split policy of the visited node (4). 
Each tree leaf points to an SBuffer, an array of pointers to the raw data\hidevldb{series
which are} stored in HBuffer (5). We chose this architecture for performance reasons. 
Experiments showed that allocating a large memory 
buffer (HBuffer) at the start of the index creation 
and releasing it once all series have been inserted\hidevldb{,
as we do with HBuffer,} is more efficient than having each leaf pre-allocate its own memory buffer and release it when it is split, especially during the beginning of index construction where splits occur frequently.
Our design results in a small number of system calls and reduces
the occurrences of out-of-memory management issues
(when a program issues a large number of memory cleanup operations, 
the memory can be retained by the process for later reuse\hidevldb{, so it is not purged, i.e., it is not returned to the operating system that considers the memory to still be in use,  which may lead to an out-of-memory error~\cite{url/jemalloc}}).

Each InsertWorker has its own region in HBuffer and records the data series that it inserts there.  
When the number of full regions reaches a flush threshold, 
the {\em \ke{flush}} phase starts (6), where the data in HBuffer is flushed to disk. 
This is achieved by identifying one of the InsertWorkers as the FlushCoordinator 
to undertake the task of flushing. The rest of the InsertWorkers wait until the FlushCoordinator
is done (7). Careful synchronization 
is necessary between the FlushCoordinator and the rest of the InsertWorkers, which 
now play the role of the FlushWorkers. The FlushWorkers have to inform the FlushCoordinator when their region becomes full and synchronize 
with it to temporarily halt their execution (8) and resume it again after the flush phase completes (9). 
\hidevldb{
\here{It seems that either the HBuffer is necessary or the double buffering scheme,
	but not both. \ke{I guess we could use two DBSize chunks in HBuffer to serve as DBuffer. This will help us gain a small amount of RAM space but complicate the synchronization of the threads accessing HBuffer. I also do not think this will help performance. I personally think it is not worth the effort. I propose we discuss again for the journal version.}}
\here{PF: I agree that we should leave this for the journal. However (just for having
it in mind for the journal): I thought that 
using just one buffer would save from copying all the data set from the one 
buffer to the other. Isn't this expensive in terms of time? }
} 

\ke{Figure~\ref{fig:hercules-index-writing} describes the workflow of the index writing phase, which calculates in parallel the iSAX summaries and synopses of the internal nodes (1-3), and materializes HTree, LRDFile and LSDFile (4-5).}
Hercules stores the raw data series in LRDFile, 
as they are traversed in the tree leaves
during an inorder traversal. To improve the pruning degree, it also 
calculates iSAX summaries for the data series and stores them in an array in memory. 
At the end of every execution, it flushes this array into LSDFile, to avoid recalculating them during query answering. 
LSDFile follows the same order as LRDFile. 
If HBuffer can hold the full dataset, 
these operations do not require any disk access. 


\subsubsection{Index Building Algorithms}

\hidevldb{We now describe the different algorithms used during the index building phase.}

We start by providing the details of Algorithm~\ref{alg:BuildHerculesIndex}
which is executed by the coordinator. 
After the coordinator fills in the first part of DBuffer (line~\ref{line:bhi_read}),
it spawns the InsertWorkers (line~\ref{line:bhi_workers-end}). 
While the InsertWorkers process the data in one part of DBuffer, 
the coordinator fills in the other part of the buffer, by reading data series from disk in chunks of size $DBSize$
(lines~\ref{line:bhi_for2-start}-\ref{line:bhi_for2-end}). 
A $Toggle$ bit is used by the coordinator to alternate working on the two parts of DBuffer 
(line~\ref{line:bhi_Toggle}). 
Before starting to refill a
part of DBuffer, the coordinator has to wait for all InsertWorkers to finish 
processing the data series in it. This is achieved using the DBarrier (line~\ref{line:bhi_for2-end}). 
\ke{DBarrier is a Barrier object (line~\ref{line:DBarrier_def}, Algorithm~\ref{alg:BuildHerculesIndex}), i.e., 
it is an instance of an algorithm that forces asynchronous threads 
to reach a point of synchronization: when a thread reaches a Barrier object, 
it is blocked until all threads that are supposed to reach the Barrier object have done so.} Once all the data has been processed, the coordinator sets the 
Finished[Toggle] flag and waits for all InsertWorkers to finish their 
work (lines~\ref{line:bhi_finished-start}-\ref{line:bhi_finished-end}).
As the coordinator always works in the part of the buffer that the workers
will process next, it has to enter the barrier one more time (line~\ref{line:bhi_barrier2}) after it sets the Finished[Toggle] flag.



\begin{algorithm}[tb]
	{	\scriptsize		
		\SetAlgoLined
		\caption{{BuildHerculesIndex}}	
		\label{alg:BuildHerculesIndex}
		\KwIn{\textbf{File*} {\it File}, \textbf{Index*} {\it Idx}, \textbf{Integer} {\it NumThreads}, \textbf{Integer} {\it DataSize}}
		\textbf{Integer} {\it NumInsertWorkers} $\gets$ {\it NumThreads}-1\; 
		\textbf{Barrier} {\it DBarrier} for {\it NumThreads} threads\;			\label{line:DBarrier_def}
		\textbf{Barrier} {\it ContinueBarrier} for {\it NumInsertWorkers} threads\;
		\textbf{Barrier} {\it FlushBarrier} for {\it NumInsertWorkers} threads\;
        \ke{\textbf{Shared Integer} {\it InitialDBSize}}\;
		\textbf{Shared Integer} {\it DBSize[2]} $\gets$ $\{0,0\}$\; 		
		\textbf{Shared Integer} {\it DBCounter[2]} $\gets$ $\{0,0\}$\; 		
		\textbf{Shared Boolean} {\it FlushCounter} $\gets$ FALSE\; 
		\textbf{Shared Boolean} {\it FlushOrder} $\gets$ FALSE\; 
		\textbf{Shared Boolean} {\it Finished[2]} $\gets$ $\{$FALSE, FALSE$\}$\; 	
		\textbf{Shared Thread} {\it Workers[NumInsertWorkers]}\; 	
		initialize {\it Workers}\;					
		\vspace*{.1cm}
		\textbf{Static Bit} {\it Toggle} $\gets 0$\;
		\vspace*{.1cm}
		{\it DBSize[Toggle]} $\gets$ \texttt{min}\{{\it InitialDBSize, DataSize}\}\;	\label{line:DBSize 1}	
		read {\it DBSize[Toggle]} data series from {\it file} into {\it DBuffer[Toggle]} of {\it Idx}\;	\label{line:bhi_read}
		{\it Toggle} $\gets$  1-{\it Toggle}\;	    \label{line:bhi_Toggle 1}
		\For {j $\gets$ 0 \emph{\KwTo} NumInsertWorkers}{				
			each {\it Worker} in {\it Workers} runs an instance of				
			\texttt{InsertWorker}({\it Idx})\;						\label{line:bhi_workers-end}
		}		
		\For {i $\gets$ DBSize[1-Toggle]; i $<$ DataSize; i+=DBSize[Toggle]}{ \label{line:bhi_for2-start}
			{\it DBSize[Toggle]} $\gets$ \texttt{min}\{{\it InitialDBSize, DataSize-i}\}\;       \label{line:bhi_workers-start}
			read {\it DBSize[Toggle]} data series from {\it file} into {\it DBuffer[Toggle]} of {\it Idx}\;	\label{line:bhi_read 2}
			{\it DBCounter[Toggle]} $\gets$ $0$\; \label{line:bhi_dbcounter}
			{\it Toggle} $\gets$  1-{\it Toggle}\;	\label{line:bhi_Toggle}
			{\it Coordinator} reaches {\it DBarrier}\;   \label{line:bhi_DBarrier2}
			\label{line:bhi_for2-end}    			
		}
		{\it Finished[Toggle]} $\gets$ TRUE\; \label{line:bhi_finished-start}
		{\it Coordinator} reaches {\it DBarrier}\; \label{line:bhi_barrier2}
		
		wait for all the InsertWorkers to finish\;\label{line:bhi_finished-end}
	} 
	
\end{algorithm}


\begin{algorithm}[tb]
	{   \scriptsize
		\SetAlgoLined
		\caption{\texttt{InsertWorker}}
		\label{alg:InsertWorker}				
		\KwIn{\textbf{Index*} \textit{Idx}}
		\textbf{Node} {\it Root} $\gets$ Root of {\it Idx}\;
		\textbf{Bit} {\it Toggle} $\gets$ $0$\; 
		\textbf{Integer} {\it Pos} $\gets$ $0$\; 
		\textbf{Float*} {\it S\textsubscript{raw}}\;

		\vspace*{.1cm}
		\While{!Finished[Toggle]}{ \label{alg:iw_finished}
			\If{\textnormal{InsertWorker's buffer has at least {\it DBSize[Toggle]} empty slots}}{ \label{line:iw_ifs-start}
				{\it Pos} $\gets$ \texttt{FetchAdd}({\it DBCounter[Toggle]}, 1)\; \label{line:iw_fetchset}
				\While {({\it Pos} $<$ {\it DBSize[Toggle]})} 
				{						\label{alg:iw_while}
					{\it S\textsubscript{raw}} $\gets$ {\it DBuffer[Toggle][Pos]}\; \label{line:iw_dbuffer}
					\texttt{InsertSeriesToNode}({\it Idx, Root, S\textsubscript{raw}})\;    \label{line:InsertSeries}
					{\it Pos} $\gets$ \texttt{FetchAdd}({\it DBCounter[Toggle]}, 1)\;	\label{alg:fetchadd} 
				}
			}	\label{line:iw_ifs-end}					
			InsertWorker blocks on {\it DBarrier}\;	\label{alg:iw_dbarrier}	 
			\If {\textnormal {InsertWorker is FlushCoordinator}}     
			{						\label{alg:iw_flushing_starts}
				\texttt{FlushCoordinator}({\it Idx})\;
			}
			\Else 
			{
				\texttt{FlushWorker}()\;  \label{alg:iw_flushing_ends}
			}		       		
			{\it Toggle} $\gets$ 1-{\it Toggle}\; 			\label{alg:iw_togggle}
		}
	} 
\end{algorithm}

\begin{algorithm}[tb]
	{   \scriptsize
		\SetAlgoLined
		\caption{\texttt{FlushCoordinator}}
		\label{alg:FlushCoordinator}		
		\KwIn{\textbf{Index*} {\it Idx}}
		\textbf{Integer Volatile} {\it Cnt}\;
		\textbf{Integer} {\it Tmp}\;
		\vspace*{.1cm}		
		{\it Workers[WorkerID].ContinueHandShake} $\gets$ TRUE\; \label{line:fc_handshare-true}
		\For {\textnormal{each {\it Worker} in {\it Workers}}}{	\label{line:fc_handshare-begin}
			\While{!Worker.ContinueHandShake}{
				\For {Tmp $\gets$ 0; Tmp $<$ BUSYWAIT; Tmp++} {
					{\it Cnt++};					
				}	
			}	
		} \label{line:fc_handshare-end}
		\If{\textnormal{(FlushCoordinator's region in HBuffer is full) OR ({\it FlushCounter} $>=$ 	FLUSH\_THRESHOLD)}}{    \label{line:full}
			{\it FlushOrder} $\gets$ TRUE\; \label{line:fc_order}
		}
		{\it FlushCounter $\gets 0$}\;						
		FlushCoordinator blocks on {\it ContinueBarrier}\;				\label{alg:fc_cb}		
		{\it Workers[WorkerID].ContinueHandShake} $\gets$ FALSE\; 		
		\If {\it FlushOrder}
		{								\label{alg:fc_if}
			materialize to disk the data in {\it Idx}'s leaves and reset soft and hard buffers\;	\label{alg:fc_flush}				
			{\it FlushOrder} $\gets$ FALSE \;		
			FlushCoordinator blocks on {\it FlushBarrier}\;		\label{alg:fc_fb}			    			
		}
	} 
\end{algorithm}			

\begin{algorithm}[tb]
	{   \scriptsize
		\SetAlgoLined
		\caption{\texttt{FlushWorker}}
		\label{alg:FlushWorker}		
		\ke{\Comment{a worker's buffer is full if it cannot hold at least {\it DBSize[InitialDBSize]} series}}\;	\label{line:comment}		
		\If{\textnormal{FlushWorker's region in HBuffer is full}}{	\label{line:fw_if1}
			\texttt{FetchAdd}({\it FlushCounter}, 1)\;	 \label{line:fw_fetchadd}
		} 	
		{\it Workers[WorkerID].ContinueHandShake} $\gets$ TRUE\; \label{line:fw_handshake-t}
		FlushWorker blocks on {\it ContinueBarrier}\; \label{line:fw_cbarrier}				
		{\it Workers[WorkerID].ContinueHandShake} $\gets$ FALSE\; 	\label{line:fw_handshake-f}
		\If {FlushOrder} {			\label{line:fw_if}
			FlushWorker blocks on {\it FlushBarrier}\;\label{line:fw_fbarrier}			    		
		}
	} 
\end{algorithm}			

Each InsertWorker executes 
Algorithm~\ref{alg:InsertWorker}. 
The execution alternates between insertion phases, where all InsertWorkers
insert data series in the tree, and flushing phases where the first of the InsertWorkers (that plays the role of FlushCoordinator) flushes all buffers to disk, while the other FlushWorkers block waiting for FlushCoordinator to complete the flushing phase.
InsertWorkers keep track of the part of DBuffer they work on in the variable $Toggle$. 

In Algorithm~\ref{alg:InsertWorker}, each InsertWorker
first checks if it has enough space in HBuffer 
to store at least $DBSize$ series (lines~\ref{line:iw_ifs-start}-\ref{line:iw_ifs-end}).
Next, the InsertWorker \hidevldb{simply }reads a data series from DBuffer[Toggle] (line~\ref{line:iw_dbuffer})
and calls \texttt{InsertSeriesToNode} to insert this data series in the index tree. 
To synchronize reads from DBuffer[Toggle], InsertWorkers use a FetchAdd variable
DBCounter[Toggle]. DBCounter[Toggle] is incremented on line~\ref{line:iw_fetchset}
and it is reset to $0$ by the coordinator (line~\ref{line:bhi_dbcounter} in Algorithm~\ref{alg:BuildHerculesIndex}).
To synchronize with the coordinator, InsertWorkers reach the DBarrier 
once they have finished processing the data in the current part of DBuffer, 
or have exhausted their capacity in HBuffer (line~\ref{alg:iw_dbarrier}). 

To execute the flushing protocol, 
the FlushCoordinator executes an instance of Algorithm~\ref{alg:FlushCoordinator}, 
whereas FlushWorkers execute an instance of Algorithm~\ref{alg:FlushWorker}. 
The FlushCoordinator and FlushWorkers synchronize using the ContinueHandShake bits, the ContinueBarrier and the FlushBarrier. 
The FlushCoordinator reads the FlushCounter to determine the number of FlushWorkers that have exceeded the capacity 
of their regions in HBuffer (for loop at line~\ref{line:fc_handshare-begin}).
Each InsertWorker uses the ContinueHandShake bit to notify the coordinator if its part in HBuffer is full, in which case it increments FlushCounter. The FlushCoordinator waits until it sees that all these bits are set (lines~\ref{line:fc_handshare-begin}-\ref{line:fc_handshare-end}).
If the FlushCounter reaches a threshold, or the FlushCoordinator itself is full, FlushOrder is set to TRUE (line~\ref{line:fc_order}). The FlushCoordinator waits at the ContinueBarrier to let FlushWorkers know that it has made its decision 
and resets its own ContinueHandShake bit to FALSE. If FlushOrder is TRUE, 
it flushes the raw data in HBuffer to disk, resets the SBuffer pointers in the leaves, and resets FlushCounter to zero and FlushOrder to FALSE. The FlushCoordinator informs FlushWorkers that it has finished the flushing by reaching the FlushBarrier.


In Algorithm~\ref{alg:FlushWorker}, FlushWorkers increase the FetchAdd variable FlushCounter if their region in HBuffer 
is full (line~\ref{line:fw_fetchadd}) and flip the ContinueHandShake bit to TRUE (line~\ref{line:fw_handshake-t}). FlushWorkers need to wait until the FlushCoordinator has received all handshakes and issued the order to flush or continue insertions, before they reach the ContinueBarrier (lines~\ref{line:fw_cbarrier}). The ContinueHandShake is then set back to FALSE (line~\ref{line:fw_handshake-f}). If the FlushCoordinator has set FlushOrder to TRUE, FlushWorkers wait at the FlushBarrier (line~\ref{line:fw_fbarrier}).

\begin{algorithm}[tb]
	{   \scriptsize
		\SetAlgoLined		
		\caption{$\texttt{InsertSeriesToNode}$}
		\label{alg:InsertSeriesToNode}
		\KwIn{\textbf{Index*} {\it Idx}, \textbf{Node*} $\scriptstyle \mathcal{N}$, \textbf{Float*} {\it S\textsubscript{raw}} }
		\vspace*{.1cm}		
		$\scriptstyle \mathcal{N}$ $\gets$ \texttt{RouteToLeaf}($\mathcal{N}$, {\it S\textsubscript{raw}})\;  \label{line:is2n_route}
		acquire lock on $\scriptstyle \mathcal{N}$\; \label{line:is2n_lock-acquire}
		\While {\textnormal{not} $\scriptstyle \mathcal{N}$.IsLeaf}{   \label{line:is2n_handoverhand-start}
			release lock on $\scriptstyle \mathcal{N}$\; \label{release 1}
			$\scriptstyle \mathcal{N}$ $\gets$ \texttt{RouteToLeaf}($\scriptstyle \mathcal{N}$, {\it S\textsubscript{raw}})\;
			acquire lock on $\scriptstyle \mathcal{N}$\;		\label{acquire lock}	
		}\label{line:is2n_handoverhand-end}
		update the synopsis of $\scriptstyle \mathcal{N}$ using {\it S\textsubscript{raw}}\; \label{line:is2n_update}
		add {\it S\textsubscript{raw}} in this thread's region of HBuffer and a pointer to it in  $\scriptstyle \mathcal{N}$'s SBuffer\;	\label{line:is2n_append}	
		\If{$\scriptstyle \mathcal{N}$ is full}{ \label{line:is2n_full-leaf}
			{\it Policy} $\gets$ \texttt{getBestSplitPolicy}($\scriptstyle \mathcal{N}$)\;  	 \label{line:is2n_policy}
			create two children nodes for $\scriptstyle \mathcal{N}$ according to {\it Policy}\;  	\label{line:is2n_split}		
			get all data series in $\scriptstyle \mathcal{N}$ from memory and disk (if flushed)\;  		
			distribute data series among the two children nodes and update the node synopses accordingly\;
			$\scriptstyle \mathcal{N}$.{\it IsLeaf} $\gets$ FALSE\; \label{line:is2n_internal}
		}	
		release lock on $\scriptstyle \mathcal{N}$\;	\label{line:is2n_lock-release}
	} 
\end{algorithm}
Algorithm~\ref{alg:InsertSeriesToNode} describes the steps taken by each thread
to insert one data series into the index. First, the thread traverses the current index tree 
to find the correct leaf to insert the new data series (line~\ref{line:is2n_route}), then
it attempts to acquire a lock on the leaf node (line~\ref{line:is2n_lock-acquire}). 
Another thread that has 
already acquired a lock on this same leaf might have split the node; therefore, 
it is important to verify that the node is still a leaf before appending data to it. 
 Once a leaf is reached, its synopsis is updated with
the data series statistics (line~\ref{line:is2n_update}) and the latter is appended to it (line~\ref{line:is2n_append}). 
The lock on the leaf is then released (line~\ref{line:is2n_lock-release}).  
In the special case when a leaf node reaches its maximum capacity (line~\ref{line:is2n_full-leaf}), the best
splitting policy is determined following the same heuristics as in the DSTree~\cite{conf/vldb/Wang2013} (line~\ref{line:is2n_policy}) and the node is split by creating
two new children nodes \hidevldb{and pointing to them }(line~\ref{line:is2n_split}). 
The series in the split 
node are fetched from memory and disk (if they have been flushed to disk), then
redistributed  among the left and right children nodes according to the node's 
splitting policy. 
Once all \hidevldb{the data }series have been stored in the correct leaf,
the split node becomes an internal node (line~\ref{line:is2n_internal}).

\subsubsection{Index Writing Algorithms}

We now describe the algorithms used in the index writing phase 
which materializes the index tree and the raw data. 
This phase also updates the synopsis of the internal nodes and calculates the iSAX summaries 
of the raw data series (which are also materialized). 

\begin{algorithm}[tb]
	{   \scriptsize
		\SetAlgoLined
		\caption{\texttt{WriteHerculesIndex}}
		\label{alg:WriteHerculesIndex}
		\KwIn{\textbf{Index*} {\it Idx}, \textbf{File*} {\it File}, \textbf{Integer} {\it NumThreads}}	
		\textbf{Shared Integer} {\it LeafCounter} $\gets$ 0\;
		%
		\vspace*{.1cm}
		write {\it Idx.Settings} to {\it File}\; \label{line:whi_settings}		
		\vspace*{.1cm}
		\For {j $\gets$ 0 \emph{\KwTo} {\it NumThreads-1}}{		\label{line:whi_for}
			create a thread to execute an instance of
			\texttt{WriteIndexWorker}({\it Idx, LeafCounter})\; \label{line:whi_wiw}
		}							
		\vspace*{.1cm}
		\texttt{WriteLeafData}({\it Idx})\; 
		\vspace*{.1cm}
		\For {j $\gets$ 0 \emph{\KwTo} {\it NumThreads-1}}{			
			wait until all WriteIndexWorker threads are done;
		}
		\vspace*{.1cm}
		\texttt{WriteIndexTree}({\it Idx, Idx.Root, File})\;		\label{alg:WriteIndexTree}
	} 
\end{algorithm}

Algorithm~\ref{alg:WriteHerculesIndex} 
which implements this phase is invoked by a thread called the WriteIndexCoordinator.
The coordinator first saves the index settings used to build the index which
include the data series length, the dataset size, and the leaf threshold (line~\ref{line:whi_settings}). 
Recall that for performance reasons,  
the synopses of the internal nodes were not updated during index building (Algorithm~\ref{alg:InsertSeriesToNode}).
The WriteIndexCoordinator creates a number of workers (WriteIndexWorkers) to concurrently update these synopses and to calculate the iSAX summaries of the raw data series. 
At the same time, the WriteIndexCoordinator calls \texttt{WriteLeafData} to materialize the data contained in the leaves that have been fully processed by the WriteIndexWorkers. 
After all WriteIndexWorkers are done, the coordinator calls 
\texttt{WriteIndexTree} to materialize the data of the internal nodes.

Each WriteIndexWorker (Algorithm~\ref{alg:WriteIndexWorker}) processes one leaf at a time, 
calculating the \ke{iSAX} summaries of the leaf's raw data  and updating the synopses of its ancestor nodes (line~\ref{line:wiw_process}). 
Then, it  notifies the WriteIndexCoordinator that it completed
processing this leaf (line~\ref{line:wiw_true}). 
Once all leaves have been processed, the worker terminates. 

\begin{algorithm}[tb]
	{   \scriptsize
		\SetAlgoLined
		\caption{\texttt{WriteIndexWorker}}
		\label{alg:WriteIndexWorker}
		\KwIn{\textbf{Index*} {\it Idx}, \textbf{Shared Integer} {\it LeafCounter}}	
		\textbf{Integer} $j$ $\gets$ 0\; 
		\textbf{Node} $\scriptstyle \mathcal{L}$\; 
		\vspace{.1cm}
		$j \gets$ \texttt{FetchAdd}({\it LeafCounter}, 1)\;			\label{line:wiw_fetchadd1}
		$\scriptstyle \mathcal{L} \gets$ {\it j}th leaf of the index tree (based on inorder traversal)\;
		\While{j < Idx.NumLeaves}{			\label{line:wiw_while} 
			\texttt{ProcessLeaf}({\it Idx},$\scriptstyle \mathcal{L}$)\; \label{line:wiw_process}
			$\scriptstyle \mathcal{L}$.{\it processed} $\gets$ TRUE\; \label{line:wiw_true}
			wait until $\scriptstyle \mathcal{L}$.{\it written} is equal to TRUE\; \label{line:wiw_written}
			{\it j} $\gets$ \texttt{FetchAdd}({\it LeafCounter}, 1)\;    \label{line:wiw_fetchadd2}
		}	    
	} 
\end{algorithm}

For each processed leaf of the index tree, \texttt{WriteLeafData} \hidevldb{(Algorithn~\ref{alg:WriteLeafData}) }simply stores the leaf's raw data 
and their iSAX summaries into two disk files, called LRDFile and LSDFile, respectively. 
%
%

We next discuss the post-processing of each individual leaf $\mathcal{L}$\hidevldb{ (Algorithm~\ref{alg:flushLeafUpdateAncestors})}, 
which consists mainly of the calculation of the \ke{iSAX} summaries of the data series belonging in $\mathcal{L}$\hidevldb{ (line~\ref{line:flua_sax})} 
and the update of the synopses of the ancestors of $\mathcal{L}$\hidevldb{ (lines~\ref{line:flua_vs}-\ref{line:flua_nvs})}. 
\hidevldb{
\begin{algorithm}[tb]
	{   \scriptsize
		\SetAlgoLined
		\caption{\texttt{ProcessLeaf}}
		\label{alg:flushLeafUpdateAncestors}
		\KwIn{\textbf{Index} $Idx$, \textbf{Node} $\mathcal{L}$}		
		\For {each data series $S_{raw}$ in $\mathcal{L}$}{
			$S_{sax}$ $\gets$ the iSAX representation of $S_{raw}$\;  \label{line:flua_sax}
			{\texttt{VSplitSynopsis}($\mathcal{L}$, $S_{raw}$)}\; \label{line:flua_vs}
		}		
		{\texttt{HSplitSynopsis}($\mathcal{L}$)}\;	\label{line:flua_nvs}	 	
	} 
\end{algorithm}
} 
Recall that the synopsis of a node $\mathcal{N}$ consists of the synopses of all segments in $\mathcal{N}$'s segmentation, where each segment is represented using the minimum/maximum mean and standard deviation of all points within this segment among all series that traversed $\mathcal{N}$.
Therefore, updating the synopsis of an internal node $\mathcal{N}$ requires updating the synopses of all its segments. 
The operations required to do so depend on the type 
of the split that the node $\mathcal{N}$ has undergone. If a node has been split vertically, the synopsis of the segment that was vertically split is calculated gradually by repeatedly calling \texttt{VSplitSynopsis} (Algorithm~\ref{alg:VSplitSynopsis})
on each raw data series, whereas the synopsis of its other segments can be derived from those of its children nodes by calling \texttt{HSplitSynopsis} (Algorithm~\ref{alg:HSplitSynopsis}). So \texttt{HSplitSynopsis} is invoked only once for each leaf node. 

\begin{algorithm}[tb]
	{   \scriptsize
		\SetAlgoLined
		\caption{\texttt{VSplitSynopsis} }
		\label{alg:VSplitSynopsis}
		\KwIn{\textbf{Node} $\scriptstyle \mathcal{N}$, \textbf{Float *} {\it S\textsubscript{raw} }}		
		\textbf{Segment} {\it SG}\;
		\vspace*{.1cm}		 
		\While {$\scriptstyle \mathcal{N}$ \textnormal{is not null}}{		\label{alg:vss_while}
			\If {$\scriptstyle \mathcal{N}$ \textnormal{has a split vertical segment} } {								
				{\it SG} $\gets$ the split segment of $\scriptstyle \mathcal{N}$\;
				{\it S\textsubscript{sketch}} $\gets$ 			\texttt{CalcMeanSD}({\it S\textsubscript{raw}, SG\textsubscript{start}, SG\textsubscript{end}})\;						
				\textnormal{acquire lock on} $\scriptstyle \mathcal{N}$\;				
				\textnormal{update the min/max mean and sd of} SG with S\textsubscript{sketch}.Mean and S\textsubscript{sketch}.SD\;		\label{alg:vss_update}				
				release lock on $\scriptstyle \mathcal{N}$\;				
		}
			{$\scriptstyle \mathcal{N}$} $\gets$ $\scriptstyle \mathcal{N}$.{\it Parent}\;										
	}
}
\end{algorithm}

\begin{algorithm}[tb]
	{   \scriptsize
		\SetAlgoLined
		\caption{\texttt{HSplitSynopsis}}
		\label{alg:HSplitSynopsis}
		\KwIn{\textbf{Node} $\scriptstyle \mathcal{N}$}		
		{\textbf{Segment} {\it SG\textsubscript{p}}, \textbf{Node} $\scriptstyle \mathcal{P}$ }\;
		\vspace*{.1cm}
		$\scriptstyle \mathcal{P}$ $\gets$ $\scriptstyle \mathcal{N}$.Parent\;
		
		\While {$\scriptstyle \mathcal{P}$ \textnormal{is not null}}{		\label{alg:hss_while}
			\For {\textnormal{each non vertically split segment} SG\textsubscript{p} in $\scriptstyle \mathcal{P}$}{								
				{\it SG\textsubscript{n}} $\gets$ the corresponding segment of {\it SG} in $\scriptstyle \mathcal{N}$\;
				acquire lock on $\scriptstyle \mathcal{N}$\;				
				update the min/max mean and sd of 				{\it SG\textsubscript{p}} with the min/max mean and sd of 				{\it SG\textsubscript{n}} \;			\label{alg:hss_update}				
				release lock on $\scriptstyle \mathcal{N}$\;												                                       
			}
			$\scriptstyle \mathcal{N}$ $\gets$$\scriptstyle \mathcal{P}$\;									
			$\scriptstyle \mathcal{P}$ $\gets$ $\scriptstyle \mathcal{N}$.{\it Parent}\;									
		}		
	}
\end{algorithm}
In the case of a horizontal split, the synopsis of $\mathcal{N}$ can be derived entirely from those of its children (Algorithm~\ref{alg:HSplitSynopsis}).
\hidevldb{
\here{The IndexTreeWorkers could traverse the tree without using the array
	implemented by the coordinator. \ke{The array needs to be sorted in order of FilePosition in LRDFile while the traversal needs the pq to be sorted in increasing lb distance. I think this is a minor space optimization that will not help performance.}}
\here{PF: Why isn't creating and sorting the array expensive in terms of time?} 
} 

Routine \texttt{WriteIndexTree} \hidevldb{(Algorithm~\ref{alg:WriteIndexTree}) }materializes the tree nodes. It applies a recursive (Postorder) tree traversal, updating the size of each node and writing the node's data into $File$.

\hidevldb{
\begin{algorithm}[tb]
	{   \scriptsize
		\SetAlgoLined
		\caption{\texttt{WriteIndexTree}}
		\label{alg:WriteIndexTree}
		\KwIn{\textbf{Index*} $Idx$, \textbf{Node*} $\mathcal{N}$, \textbf{File*} $File$}		
		\If {$\mathcal{N}$ is an internal node}{
			\texttt{WriteIndexTree}($Idx$,$\mathcal{N}.leftChild$,$File$)\;
			\texttt{WriteIndexTree}($Idx$,$\mathcal{N}.rightChild$,$File$)\;
			$\mathcal{N}.size$ $\gets$ $\mathcal{N}.leftChild.size$ + $\mathcal{N}.rightChild.size$\;
		}
		write $\mathcal{N}$ info to $File$\;	
	} 
\end{algorithm}
} 

\subsection{Query Answering}

\subsubsection{\ke{Overview}}

\begin{figure}[tb]
	\captionsetup{justification=centering}	
	\begin{subfigure}{\columnwidth}
		\centering
		\captionsetup{justification=centering}	
		\includegraphics[width=0.9\columnwidth] {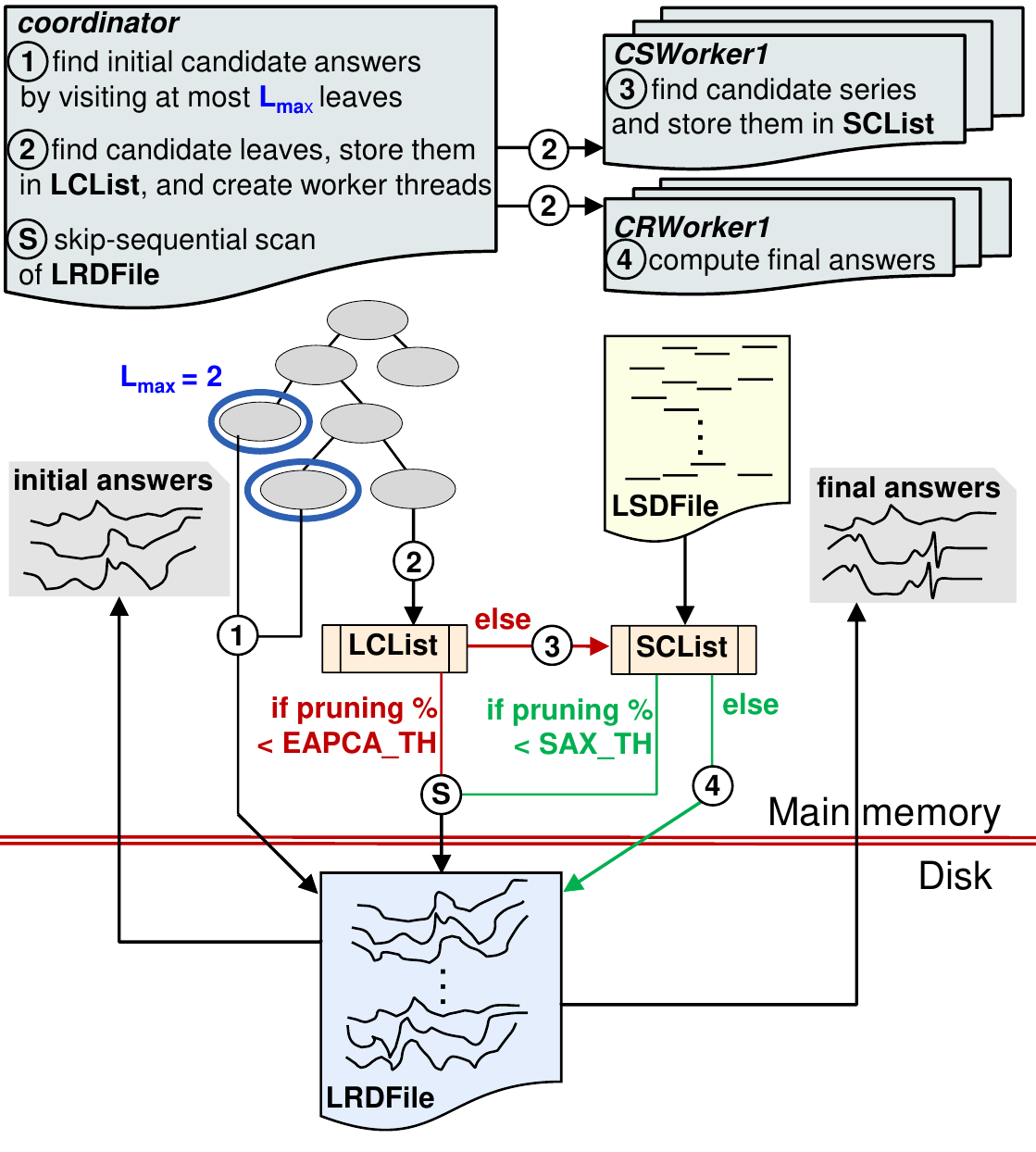}
	\end{subfigure}
	\vspace*{-0.2cm}
	\caption{Query Answering Workflow}  
	\vspace*{-0.2cm}
	\label{fig:query-answering-workflow}
\end{figure}

Figure~\ref{fig:query-answering-workflow} describes query answering 
in Hercules, which consists of four main steps. The first and second 
steps are run sequentially, whereas the others are run in 
parallel. The distance calculations in all steps are performed using SIMD.

In the first step, when a kNN query  
$S_Q$ arrives, the index tree is traversed to visit at most  $\mathrm{{L_{max}}}$
leaves. When a leaf $\mathcal{L}$ is visited, the series that belong to it are 
read from LRDFile and the Euclidean distance (ED) between $S_Q$
and every series in $\mathcal{L}$ is calculated. 
The $k$ best-so-far neighbors, i.e., the $k$ series with the smallest Euclidean distances to $S_Q$,
are stored in array $Results$. 
The leaves are chosen based on a 
priority queue $PQ$, which  initially contains just the index tree Root node. 
The algorithm  iteratively removes a node $\mathcal{N}$ from the priority queue, and if the $\mathrm{LB_{EAPCA}}$ 
distance of $S_Q$ to $\mathcal{N}$ is larger than the current
value of $\mathrm{BSF_{k}}$, the $k^{th}$ smallest 
ED distance in $Results$, the node is pruned. 
Otherwise, if $\mathcal{N}$ is 
a leaf, it is added to $PQ$, and if it is an internal node, its children are added
to $PQ$ if their $\mathrm{LB_{EAPCA}}$ distance to $S_Q$ is smaller than $\mathrm{BSF_{k}}$. Once  $\mathrm{{L_{max}}}$
leaves are visited, the first step terminates returning the array $Results$ with 
$k$ initial answers. 
This step is called approximate search.

The second step takes place entirely in-memory with no ED distance calculations, 
thus, the current $\mathrm{BSF_{k}}$ is not updated. 
This step resumes processing $PQ$ as in step one, except when a leaf is visited: it is either pruned, based on $\mathrm{BSF_{k}}$, 
or added to a list of candidate leaves, called LCList, in increasing order of the leaves file positions in LRDFile. 
If the pruning is smaller than a threshold, EAPCA\_TH, a skip-sequential scan is performed on LRDFile and the final
results are \ke{stored} in $Results$. The skip-sequential scan seeks the file position
of the first leaf in LCList, calculates the ED distances between the data in this leaf
and $S_Q$ and updates $Results$ if applicable. Similarly, subsequent leaves are processed in
the order they appear in LCList, provided they cannot be pruned using $\mathrm{BSF_{k}}$.

If the size of LCList is relatively small, Hercules proceeds to the third step,\hidevldb{ which} also 
\hidevldb{takes place }entirely in-memory. 
Threads process LCList 
in parallel. Each thread picks a leaf $\mathcal{L}$ from LCList, loads the iSAX summaries  of $\mathcal{L}$'s  data series from LSDFile, which is pre-loaded 
with the index tree, 
and calculates the $\mathrm{LB_{SAX}}$ distance 
between $S_Q$ and each iSAX summary. 
If the iSAX summary cannot be pruned, 
the file position of the series and its $\mathrm{LB_{SAX}}$ distance are 
added to SCList. 
Note that the file position of a series in LRDFile is
the same as that of its iSAX summary in LSDFile.
This step terminates once all the leaves in LCList 
are processed. 
If the pruning is smaller than a threshold, SAX\_TH, one thread performs a skip-sequential scan 
on LRDFile and the final results are stored in $Results$\hidevldb{ following the same skip-sequential approach described earlier}. 

Note that a skip-sequential scan on LRDFile is favored when pruning is low for efficiency. 
It incurs as many random I/O operations as the number of non pruned leaves, whereas applying the second filtering step using $\mathrm{LB_{SAX}}$ on a large number of series incurs as many random I/O operations as the number of non-pruned series. 
The EAPCA\_TH and SAX\_TH thresholds are tuned experimentally, and exhibit a stable behavior (cf. Section~\ref{sec:experiments}).

If the size of SCList is relatively small, Hercules proceeds to 
the fourth step, which 
computes the final results based on SCList and LRDFile. 
Multiple threads process SCList in parallel. Each thread picks a series from
SCList and compares its $\mathrm{LB_{SAX}}$ against the current $\mathrm{BSF_{k}}$. If the series cannot 
be pruned, the thread loads the series corresponding to this summary from LRDFile, 
and calculates the Euclidean distance between this series and $S_Q$, updating 
$Results$ when appropriate. Once SCList has been processed entirely, the
final answers are returned in $Results$.
 
\subsubsection{Query Answering Algorithms}

Algorithm~\ref{alg:exactKNN} outlines the $k$NN exact nearest neighbor search with Hercules. 
It takes as arguments, the query series $S_Q$, the index $Idx$, 
a threshold which determines the maximum number of leaves, $L_{max}$, that can be visited during the approximate search, the number $k$ of neighbors to return, and two thresholds, EAPCA\_TH and SAX\_TH, which determine whether multi-threading will be used  (and will be discussed later).

\begin{algorithm}[tb]
	{	\scriptsize
		\SetAlgoLined
		\caption{\texttt{Exact-kNN}}
		\label{alg:exactKNN}
		\KwIn{\textbf{Query} {\it S\textsubscript{Q}}, \textbf{Index*} {\it Idx}, \textbf{Integer} 
			{\it L\textsubscript{max}}, \textbf{Integer} {\it k}, 
			\textbf{Float} EAPCA\_TH, \textbf{Float} SAX\_TH } 
		\textbf{Priority queue} {\it PQ};	\Comment{initially empty}\\\label{line:eks_inits}		
		\textbf{Array} {\it Results} containing {\it k} structs of type {\bf Result} with fields {\it Dist} and {\it Pos} each;	\Comment{initially, each array element stores $\langle \infty$, NULL $\rangle$}\\ 
		\textbf{Float} {\it BSF\textsubscript{k}}\;
		\textbf{Float} {\it eapca\_pr, sax\_pr}\;
		\textbf{List*} {\it LCList}, {\it SCList}\; \label{line:eks_inite}
		\vspace*{.1cm}
		\texttt{Approx-kNN}({\it S\textsubscript{Q}, Idx, PQ, L\textsubscript{max},  k, Results})\;				
  	    {\it BSF\textsubscript{k}} $\gets$ {\it Results[k-1].Dist}\; 
		\texttt{FindCandidateLeaves}({\it S\textsubscript{Q}, Idx, PQ, BSF\textsubscript{k},  LCList})\;	\label{line:eks_blcl}	
		{\it eapca\_pr} $\gets$ 1-\texttt{size}({\it LCList})/(total leaves of index tree)\;
		
		\If {{\it eapca\_pr} < EAPCA\_TH} {
			perform a skip-sequential scan on {\it Idx.LRDFile} and update {\it Results}\; \label{line:eks_eapca}
		}		
		\Else{
			\texttt{FindCandidateSeries}({\it S\textsubscript{Q}, Idx, BSF\textsubscript{k}, LCList, SCList})\;	\label{line:eks_bscl}	 					
			{\it sax\_pr} $\gets$ 1-\texttt{size}({\it SCList})/(total series of {\it Idx})\;
			\If {sax\_pr < SAX\_TH} {
				perform a skip-sequential scan on {\it Idx.LRDFile}\; \label{line:eks_sax}
			}	
			\Else{
				\texttt{ComputeResults}({\it S\textsubscript{Q}, Idx, k, Results, SCList})\; \label{line:eks_rscl}									
			}
		}
		return {\it Results}\;		
	} 
\end{algorithm}

The algorithm uses an array $Results$ to store the current $k$ best-so-far answers
at each point in time, two array-lists, namely LCList and SCList, which 
hold the candidate leaves and series, respectively, and a priority queue $PQ$ in which the priority is based on the $\mathrm{LB_{EAPCA}}$ distance of the query to a given node (lines~\ref{line:eks_inits}-\ref{line:eks_inite}). 

First, an approximate $k$NN search is performed by calling the function \texttt{Approx-kNN}, which \ke{stores in the array $Results$}, the $k$ approximate neighbors for query $S_Q$ by visiting at most $\mathrm{{L_{max}}}$ leaves. 
The variable $\mathrm{BSF_{k}}$ is initialized with the real distance of the $k^{th}$ neighbor to $S_Q$ 
and is used by the \texttt{FindCandidateLeaves} function to prune the search space. 
The non-pruned leaves are stored in LCList (line~\ref{line:eks_blcl}). 
If the size of LCList is large enough (so that $eapca\_pr$ is smaller than the EAPCA\_TH threshold), 
then a skip-sequential scan on LRDFile is performed using 
one thread \ke{and 
$Results$ is updated with the best answers} (line~\ref{line:eks_eapca}). 


Otherwise, \texttt{FindCandidateSeries} applies a second filter using the \ke{iSAX} representation 
of the series belonging to the leaves in LCList using the $\mathrm{LB_{SAX}}$ distance, 
and stores the non-pruned candidate series in SCList (line~\ref{line:eks_bscl}). 
If the size of SCList is large, i.e., that the pruning based on $\mathrm{LB_{SAX}}$ is low, 
then a skip-sequential scan on LRDFile is performed using 
one thread (line~\ref{line:eks_sax}). 
Otherwise, \texttt{ComputeResults} (line~\ref{line:eks_rscl}) loads the series in SCList from disk 
and calculates their real ED distance to the query returning the $k$ series with the minimum ED distance 
to $S_Q$ as the final answers. 
\ke{All real and lower-bounding distance calculations used exploit SIMD for efficiency (following~\cite{parisplus,messijournal}).}

\hidevldb{In what follows,} The building blocks of Algorithm~\ref{alg:exactKNN} are as follows. 


\hidevldb{\subsubsection{Step 1: Finding Initial Approximate Answers}}
\noindent{\bf Step 1: Finding Initial Approximate Answers.}
Algorithm~\ref{alg:approxKNN} finds $k$ approximate first baseline answers and stores them in the array $Results$ in increasing order of the real Euclidean distance. It visits a maximum of $L_{max}$ leaves, where $L_{max}$ is a parameter provided by the user and is 1 by default. In line \ref{line:aks_dq}, it pops the element in $PQ$ with the highest priority, i.e., the node with the lowest $\mathrm{LB_{EAPCA}}$ to the query. To understand the reason behind allocating a higher priority to a smaller $\mathrm{LB_{EAPCA}}$, recall that $\mathrm{LB_{EAPCA}}$ is a lower-bounding distance. Therefore, if a popped node has an $\mathrm{LB_{EAPCA}}$ value greater than the current $\mathrm{BSF_{k}}$ answer, then search can terminate since any series in the subtree rooted at this node has a real distance that is greater than or equal to $\mathrm{LB_{EAPCA}}$, and thus cannot improve $\mathrm{BSF_{k}}$ (line~\ref{line:aks_break}). Besides, since the priority in $PQ$ is based on the minimum $\mathrm{LB_{EAPCA}}$, all remaining nodes in $PQ$ can be pruned because their lower-bounding distances will also be greater than $\mathrm{BSF_{k}}$. Note that we use SIMD operations to speed up the calculations of the $\mathrm{LB_{EAPCA}}$ distances.

\begin{algorithm}[tb]
	{   \scriptsize
		\SetAlgoLined
		\caption{\texttt{Approx-kNN}}
		\label{alg:approxKNN}		
		\KwIn{\textbf{Query} {\it S\textsubscript{Q}}, \textbf{Index*} {\it Idx}, \textbf{Priority Queue} {\it PQ}, \textbf{Integer} {\it L\textsubscript{max}}, \textbf{Integer} {\it k}, \textbf{Result*} {\it Results}}
		\textbf{PQElement} $q$ \Comment{PQElement has two fields, $\scriptstyle \mathcal{N}$ and {\it Dist}}\;
		\vspace*{.1cm}
		\textbf{Float} {\it RootLB\textsubscript{EAPCA}} $\gets$ $\mathtt{CalculateLB_{EAPCA}}$({\it S\textsubscript{Q},  Idx.Root})\;			
		\textbf{Integer} {\it VisitedLeaves} $\gets$ $0$\;
		\vspace*{.1cm}
		add {\it Idx.Root} to {\it PQ} with priority {\it RootLB\textsubscript{EAPCA}}\; \label{line:aks_inite} 
		\While{VisitedLeaves < L\textsubscript{max} \textnormal{AND NOT} \texttt{IsEmpty}({\it PQ})} { \label{line:aks_dq}	
			$\langle \scriptstyle \mathcal{N}$, {\it LB\textsubscript{EAPCA}} $\rangle$ $\gets$ \texttt{DeleteMinPQ}({\it PQ})\;     \label{line:eks_examine}
			\If {LB\textsubscript{EAPCA} $>$ BSF\textsubscript{k}}{			\label{line:pruning}
				break; \label{line:aks_break}
			}
			\If {$\scriptstyle \mathcal{N}$ \textnormal{is a leaf}}{  			\label{line:aks_leaf}
				\For {\textnormal{each {\it S} in} $\scriptstyle \mathcal{N}$}{          \label{line:aks_lrds}
					{\it RealDist} $\gets$ \texttt{CalculateRealDist}({\it S\textsubscript{Q}, S})\;
					\If {{\it RealDist} $<$ {\it BSF\textsubscript{k}}} {       \label{line:pruning1}
						create new struct {\it result} of type {\bf Result}\;			
						{\it result.Dist} $\gets$ {\it RealDist}\;
						{\it result.Pos} $\gets$ Position of {\it S} in {\it LRDFile}\;	
						add {\it result} to {\it Results}\;	 \label{line:aks_lrde}			
						{\it BSF\textsubscript{k}} $\gets$ {\it Results[k-1].Dist}\;    \label{line:aks_BSF}
					}	                 
				} 	   
				{\it VisitedLeaves} $\gets$ {\it VisitedLeaves}+1\;
			}                  			
			\Else{
				\For {each {\it Child} in $\scriptstyle \mathcal{N}$}{ \label{line:aks_internal}
					{\it LB\textsubscript{EAPCA}} $\gets$ \texttt{CalculateLB\textsubscript{EAPCA}}({\it S\textsubscript{Q}, Child})\;
					\If {LB\textsubscript{EAPCA} $<$  BSF\textsubscript{k}}{			\label{line:pruning2}
						add {\it Child} to {\it PQ} with priority {\it LB\textsubscript{EAPCA}};    \label{line:aks_child}
					}                  
				}        
			} 
		}
	} 
\end{algorithm}
If the non-pruned node is a leaf (line~\ref{line:aks_leaf}), then the series in this leaf are read from LRDFile, their real Euclidean distances to the query are calculated and the $Results$ array is updated if applicable (lines~\ref{line:aks_lrds}-\ref{line:aks_lrde}). The algorithm stops improving the $k$ answers once the leaves threshold is reached.

Instead, if the non-pruned node is an internal node (line~\ref{line:aks_internal}), its children are added to $PQ$ if their $\mathrm{LB_{EAPCA}}$ is smaller than the current value of $\mathrm{BSF_{k}}$. Then the algorithm resumes processing the $PQ$ unless it is empty and \ke{stores} the approximate $k$ answers in \ke{$Results$ to help the exact algorithm prune the search space}.

\hidevldb{\subsubsection{Step 2: Finding Candidate Leaves}}
\noindent{\bf Step 2: Finding Candidate Leaves.}
\begin{algorithm}[tb]
	{   \scriptsize
		\SetAlgoLined
		\caption{\texttt{FindCandidateLeaves}}
		\label{alg:buildLCList}
		\KwIn{\textbf{Query} {\it S\textsubscript{Q}}, \textbf{Index*} {\it Idx}, \textbf{Priority queue}  {\it PQ}, \textbf{Float} {\it BSF\textsubscript{k}}, \textbf{List*} {\it LCList}}
		\vspace*{.1cm}
		
		\While{\textnormal{NOT} \texttt{IsEmpty}({\it PQ})}{  
			$\langle \scriptstyle \mathcal{N}$, {\it LB\textsubscript{EAPCA}} $\rangle$ $\gets$ \texttt{DeleteMin}({\it PQ})\; \label{line:bll_pq}
			\If {LB\textsubscript{EAPCA} $>$ BSF\textsubscript{k}}{			\label{line:bll_prune}
				break;  \label{line:bll_break}
			}                 
			\If {$\scriptstyle \mathcal{N}$ \textnormal{is a leaf}}{                       
				add $\langle \scriptstyle \mathcal{N},{\it LB\textsubscript{EAPCA}} \rangle$ to {\it LCList}\; \label{line:bll_leaf}
			}
			\Else {
				\For {\textnormal{each} {\it Child} in $\scriptstyle \mathcal{N}$}{   \label{line:bll_ins}      
					{\it LB\textsubscript{EAPCA}} $\gets$ \texttt{CalculateLB\textsubscript{EAPCA}}({\it S\textsubscript{Q},Child})\;
					\If { LB\textsubscript{EAPCA} $<$ BSF\textsubscript{k}}{
						add {\it Child} to {\it PQ} with priority {\it LB\textsubscript{EAPCA}}\; \label{line:bll_ine} 
					}                 
				}
			}        
		}
		sort the leaves in {\it LCList} in increasing order of their position in the {\it Idx.LRDFile}\;  \label{line:bll_sort} 				
	} 
\end{algorithm}
Once Algorithm~\ref{alg:approxKNN} finds $k$ initial approximate answers, the ED distance between $S_Q$ and the $k^{th}$ answer, called $\mathrm{BSF_{k}}$, is used by Algorithm~\ref{alg:buildLCList} to prune the search space and populate LCList with candidate leaves. Search in the index tree resumes with the remaining nodes in $PQ$ (line~\ref{line:bll_pq}), i.e., nodes that were visited by algorithm~\ref{alg:approxKNN} are not accessed again. If the current node's $\mathrm{LB_{EAPCA}}$ is larger than the $\mathrm{BSF_{k}}$ distance, the algorithm terminates (line~\ref{line:bll_break}). Otherwise, it adds non-pruned leaves into LCList (line~\ref{line:bll_leaf}) and non-pruned internal nodes into the priority queue $PQ$ (lines~\ref{line:bll_ins}-\ref{line:bll_ine}). 
Note that leaves are treated differently in Algorithms~\ref{alg:approxKNN} and~\ref{alg:buildLCList}: in the former, the series of the leaves are loaded from disk and the real distances are calculated between the series and the query, updating $Results$ as necessary, whereas in the latter, $Results$ is not updated and pointers to the leaves are stored for further processing, so the disk is not accessed in this case.

Once all nodes in $PQ$ have been processed, the candidate leaves in LCList are sorted in increasing order of their Position in LRDFile (line~\ref{line:bll_sort}). This is to reduce the overhead of disk random I/O by ensuring that data pages are visited in the order that they are laid out on disk.

\hidevldb{\subsubsection{Step 3: Finding Candidate Series}}
\noindent{\bf Step 3: Finding Candidate Series.}
\hidevldb{
\begin{algorithm}[tb]
	{   \scriptsize
		\SetAlgoLined
		\caption{\texttt{FindCandidateSeries}}
		\label{alg:buildSCList}
		\KwIn{\textbf{Query} $S_Q$, \textbf{Index*} $Idx$, \textbf{Float} $BSF_{k}$, \textbf{List*} $LCList$, \textbf{List*} $SCList$} 
		\vspace*{.1cm}
								
		\For {$i$ $\gets$ 1 to $num\_threads$}{
			create thread $t_i$ to execute \texttt{CSWorker}($S_Q$, $Idx.LSDFile$, $BSF_{k}$, $i$, $LCList$, $SCList$); \label{line:bsl_threads}	
			\label{line:bsl_inie}
		}		
		\For {$i$ $\gets$ 1 to $num\_threads$}{ 
			wait until thread $t_i$ terminates\;
		}
		Return $SCList$\; \label{line:bsl_return}	
	} 
\end{algorithm}
} 
While the previous building blocks of \hidevldb{the }exact search \hidevldb{algorithm }use SIMD to efficiently calculate the real and lower-bounding distances, they run using a single thread. 
\hidevldb{We now describe }\texttt{FindCandidateSeries} is a \hidevldb{{\bf ??? previous " is a" ???}Algorithm~\ref{alg:buildSCList}, the }multi-threaded algorithm that processes the candidate leaves in LCList to populate SCList with candidate data series. 
Each thread executes an instance of \texttt{CSWorker}\hidevldb{ (line~\ref{line:bsl_threads})}. 
Once all threads have finished execution, the algorithm \ke{stores all candidate data series in SCList}\hidevldb{ (line~\ref{line:bsl_return})}.

\begin{algorithm}[tb]
	{   \scriptsize
		\SetAlgoLined
		\caption{\texttt{CSWorker}}
		\label{alg:buildSCListWorker}
		\KwIn{\textbf{Query} {\it S\textsubscript{Q}}, \textbf{Char **} {\it LSDFile}, \textbf{Float} {\it BSF\textsubscript{k}}, {\bf Integer} {\it id}, \textbf{List*} {\it LCList}, \textbf{List*} {\it SCList}}
		\textbf{Shared Integer} {\it LCL\textsubscript{Idx}} $\gets$ $0$\;						
		\textbf{List*} {\it SCL\textsubscript{local}} \Comment{thread's local list, initially empty}\;						\label{line:bsw_lcl}
		\textbf{Node} $\scriptstyle \mathcal{L}$\; 
		
		\vspace*{.1cm}
		j $\gets$ \texttt{FetchAdd}({\it LCL\textsubscript{Idx}})\; 		\label{line:bsw_Idx}			
		\While {j $<$ LCList.size}{ 
			$\scriptstyle \mathcal{L}$ $\gets$ {\it LCList[j]}\; 		\label{line:bsw_while}			
			\For {\textnormal{each iSAX summary} S\textsubscript{sax} \textnormal{in} $\scriptstyle \mathcal{L}$} {
				{\it LB\textsubscript{SAX}} $\gets$ \texttt{calculateLB\textsubscript{SAX}}(\texttt{PAA}({\it S\textsubscript{Q}}),{\it S\textsubscript{sax}})\; \label{line:bsw_sax}
				\If { LB\textsubscript{SAX} $<$ BSF\textsubscript{k}}{			\label{line:bsw_pruning}
					add $({\it S\textsubscript{sax}.Pos,LB\textsubscript{SAX}})$ to {\it SCL\textsubscript{local}}\; \label{line:bsw_add}
				}                   
			}
			{\it j}  $\gets$ \texttt{FetchAdd} ({\it LCL\textsubscript	{Idx}})\; 							        
		}
		store in {\it SCList[id]} a pointer to {\it SCL\textsubscript{local}}\;  \label{line:bsw_store}
	} 
\end{algorithm}

In Algorithm~\ref{alg:buildSCListWorker}, each CSWorker maintains its candidate series in a local list called $SCL_{local}$ (line~\ref{line:bsw_lcl}). It processes one candidate leaf at a time using a FetchAdd operation on the variable $LCL_{Idx}$ for concurrency control (line~\ref{line:bsw_Idx}). 
For each leaf, it reads each of the \ke{iSAX} summaries of this leaf from LSDFile, which is stored in-memory, and calculates  the $\mathrm{LB_{SAX}}$ of each summary to the query's PAA (line~\ref{line:bsw_sax}). If a summary cannot be pruned, its Position in LSDFile and its $\mathrm{LB_{SAX}}$ are added to the thread's local list $SCL_{local}$. The local list is accessible from the global list SCList via the thread's identifier $id$ (line~\ref{line:bsw_store}). Recall that the summaries in LSDFile and raw series in LRDFile are stored in the same order, so the Position of a series in LRDFile is equal to the Position of its \ke{iSAX} summary in LSDFile.


\hidevldb{\subsubsection{Step 4: Computing Final Results}}
\noindent{\bf Step 4: Computing Final Results.}
Routine \texttt{ComputeResults} \hidevldb{Algorithm~\ref{alg:refineSCList} }performs the last step in exact search, spawning multiple threads 
to refine SCList and \ke{store} the exact $k$ neighbors of the query series $S_Q$ in \ke{$Results$}. 
Each thread executes an instance of \texttt{CRWorker} (Algorithm~\ref{alg:refineSCListWorker}) to process one by one candidate series from its local list $SCList[id]$.
As long as there are unprocessed elements in SCList, each CRWorker reads the Position and $\mathrm{LB_{SAX}}$ of a candidate series from its local SCList (line~\ref{line:rsw_scl}). If the series cannot be pruned based on the current $k^{th}$ best answer (line~\ref{line:rsw_min}), it is loaded from LRDFile on disk (line~\ref{line:rsw_disk}), and its real Euclidean distance to the query is calculated using efficient SIMD calculations (line~\ref{line:rsw_dist}). 
$Results$
is atomically updated if needed (lines~\ref{line:rsw_ifs}-\ref{line:rsw_ife}). 

\begin{algorithm}[tb]
	{   \scriptsize
		\SetAlgoLined
		\caption{\texttt{CRWorker}}
		\label{alg:refineSCListWorker}
		\KwIn{\textbf{Query} {\it S\textsubscript{Q}}, \textbf{Float**} {\it LRDFile}, \textbf{Integer} {\it k}, \textbf{\ke{Result}*} {\it Results}, \textbf{List*} {\it SCList}}
		\vspace*{.1cm}
		\For {\textnormal{each} j \textnormal{in 1 to} SCList[id].size}{		\label{line:rsw_for}
			{\it BSF\textsubscript{k}} $\gets$ {\it Results[k-1].Dist}\;				
			({\it Pos, LB\textsubscript{SAX}}) $\gets$ {\it SCList[id][j]}\; 			\label{line:rsw_scl}
			\If {LB\textsubscript{SAX} $<$ BSF\textsubscript{k} }{ 	 	\label{line:rsw_min}			
				{\it DS} $\gets$ data series with location {\it Pos} in {\it LRDFile}\;		\label{line:rsw_disk}
				{\it RealDist} $\gets$ \texttt{calculateRealDist}({\it S\textsubscript{Q}, DS})\; \label{line:rsw_dist}
				\If { RealDist $<$ BSF\textsubscript{k}}{ 	 
					create new {\it result} of type {\bf Result}\; \label{line:rsw_ifs}			
					{\it result.Dist} $\gets$ {\it RealDist}\;
					{\it result.Pos} $\gets$ {\it Pos}\;	
					atomically add {\it result} to {\it Results} \Comment{a readers-writers lock is used}\; \label{line:rsw_ife}							               		
				}		
			}
		}
	} 
\end{algorithm}

When all threads spawned by \texttt{ComputeResults} \hidevldb{(Algorithm~\ref{alg:refineSCList}) }have finished execution, the $Results$ array will contain the final answers to the kNN query $S_Q$, i.e., the $k$ series with the minimum real Euclidean distance to $S_Q$.

\section{Experimental Evaluation}
\label{sec:experiments}
\subsection{\ke{Framework}}

\noindent{\bf Setup.}
We compiled all methods with GCC 6.2.0 under Ubuntu Linux 16.04.2 with their default compilation flags; optimization level was set to 2. 
Experiments were run on a server with 2 Intel Xeon E5-2650 v4 2.2GHz CPUs,
(30MB cache, 12 cores, 24 hyper-threads),
75GB 
of RAM (forcing all methods to use the disk, since our datasets are larger), 
and 10.8TB (6 x 1.8TB) 10K RPM SAS hard drives 
in RAID0 with a throughput of 1290 MB/sec.


\noindent{\textbf{Algorithms.}}
We evaluate Hercules against the recent state-of-the-art similarity search approaches for data series: DSTree*~\cite{url/DSSeval} as the best performing single-core single-socket method (note that this is an optimized C implementation that is 4x faster than the original DSTree~\cite{conf/vldb/Wang2013}), ParIS+~\cite{parisplus} as the best performing multi-core multi-socket technique, VA+file~\cite{url/DSSeval} as the best skip-sequential algorithm, and PSCAN (our parallel implementation of UCR-Suite~\cite{conf/kdd/Mueen2012}) as the best optimized sequential scan. 
PSCAN exploits SIMD, multithreading and a double buffer in addition to all of UCR-Suite's ED optimizations.
Data series points are 
represented using single precision values and methods based on fixed summarizations use 16 dimensions. 

\noindent{\textbf{Datasets.}}
We use synthetic and real datasets. Synthetic datasets, called $Synth$, were generated as random-walks using a summing process with steps following a Gaussian distribution (0,1). 
Such data model financial time series~\cite{DBLP:conf/sigmod/FaloutsosRM94} and have been widely used in the literature~\cite{DBLP:conf/sigmod/FaloutsosRM94,isax2plus,conf/kdd/Zoumpatianos2015}. 
We also use the following three real datasets:
(i) \emph{SALD}~\cite{url/data/eeg} contains neuroscience MRI data and includes 200 million data series of size 128; 
(ii) \emph{Seismic}~\cite{url/data/seismic}, contains 100 million data series of size 256 representing earthquake recordings at seismic stations worldwide; and
(iii) \emph{Deep}~\cite{url/data/deep1b} comprises 1 billion vectors of size 96 representing deep network embeddings of images, extracted from the last layers of a convolutional neural network. 
The Deep dataset is the largest publicly available real dataset of deep network embeddings. 
We use a 100GB subset which contains 267 million embeddings. 

\noindent{\textbf{Queries.}}
All our query workloads consist of 100 query series run asynchronously, 
representing an interactive analysis scenario, where the queries are not known in advance~\cite{Palpanas2019,DBLP:conf/edbt/GogolouTPB19,conf/sigmod/gogolou20}.
Synthetic queries were generated using the same random-walk generator as the $Synth$ dataset (with a different seed, reported in~\cite{url/DSSeval2}). 
For each dataset, we use five different query workloads of varying difficulty: $1\%$, $2\%$, $5\%$, $10\%$ and $ood$. The first four query workloads are generated by randomly selecting series from the dataset and perturbing them using  different levels of Gaussian noise ($\mu$ = 0, $\sigma^2$ = 0.01-0.1), in order to produce queries having different levels of difficulty, following the ideas in~\cite{johannesjoural2018}. The queries are labeled with the value of $\sigma^2$ expressed as a percentage 1\%-10\%. 
The $ood$ (out-of-dataset) queries are a 100 queries randomly selected from the raw dataset and excluded from indexing. 
Since $Deep$ includes a real workload, the $ood$ queries for this dataset are selected from this workload.

Our experiments cover k-NN queries, where k $\in [1,100]$. 
\hidevldb{Following common practice, all datasets and queries are z-normalized~\cite{journal/dmkd/Keogh2003,conf/kdd/Zoumpatianos2015,johannesjoural2018}.} 


\noindent{\textbf{Measures.}} We use two measures:
%
\noindent(1) \emph{Wall clock time} measures input, output and total execution times (CPU time is calculated by subtracting I/O time from the total time). 
\noindent(2) Pruning using the \emph{Percentage of accessed data}.

\noindent{\textbf{Procedure.}}
Experiments involve two steps: index building and query answering. Caches are fully cleared before each step, and stay warm between consecutive queries.
For large datasets that do not fit in memory, the effect of caching is minimized for all methods. 
All experiments use workloads of 100 queries. 
Results reported for workloads of 10K queries are extrapolated: we discard the $5$ best and $5$ worst queries of the original 100 (in terms of total execution time), and multiply the average of the 90 remaining queries by 10K. 

Our code and data are available online~\cite{url/Hercules}.

\subsection{Results}
\label{sec:new-results}


\hidevldb{\subsubsection{Parameterization}}
\noindent{\bf Parameterization.}
\label{ssec:parametrization}
We initially tuned all methods (graphs omitted for brevity). 
The optimal parameters for DSTree* and VA+file are set according to~\cite{journal/pvldb/echihabi2018} and those for ParIS+ are chosen per~\cite{conf/bigdata/peng18}. 
For indexing, DTree* uses a 60GB buffer and a 100K leaf size, VA+file uses a 20GB buffer and 16 DFT symbols, and  
ParIS+ uses a 20GB buffer, a 2K leaf size, and a 20K double buffer size. 

\ke{For Hercules, we use a leaf size of 100K, a buffer of 60GB, a DBSize of 120K, 24 threads during index building with a flush threshold of 12, and 12 threads during index writing. During query answering, we use 24 threads, and set $\mathrm{L_{max} = 80}$, EAPCA\_TH = 0.25 and SAX\_TH = 0.50. The above default settings were used across all experiments in this paper. The detailed tuning results can be found~\cite{url/Hercules}.}



\begin{figure}[tb]
	\captionsetup{justification=centering}	
	\centering
	\begin{subfigure}{0.49\columnwidth}
		\captionsetup{justification=centering}	
		\includegraphics[width=\columnwidth] {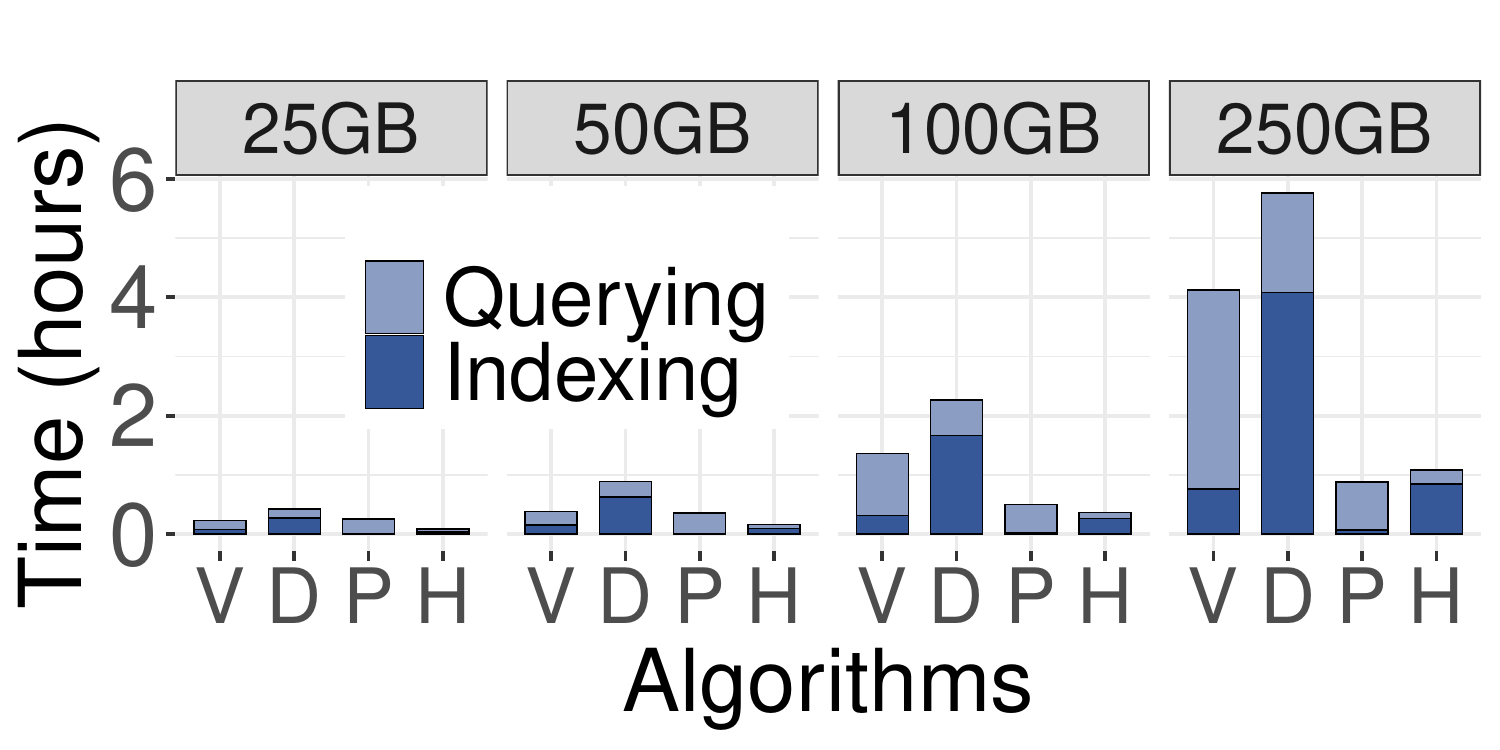}
		\caption{Idx + 100 Queries}  
		\label{exact:varysize:hdd:1}
	\end{subfigure}
	\begin{subfigure}{0.49\columnwidth}
		\captionsetup{justification=centering}	
		\includegraphics[width=\columnwidth] {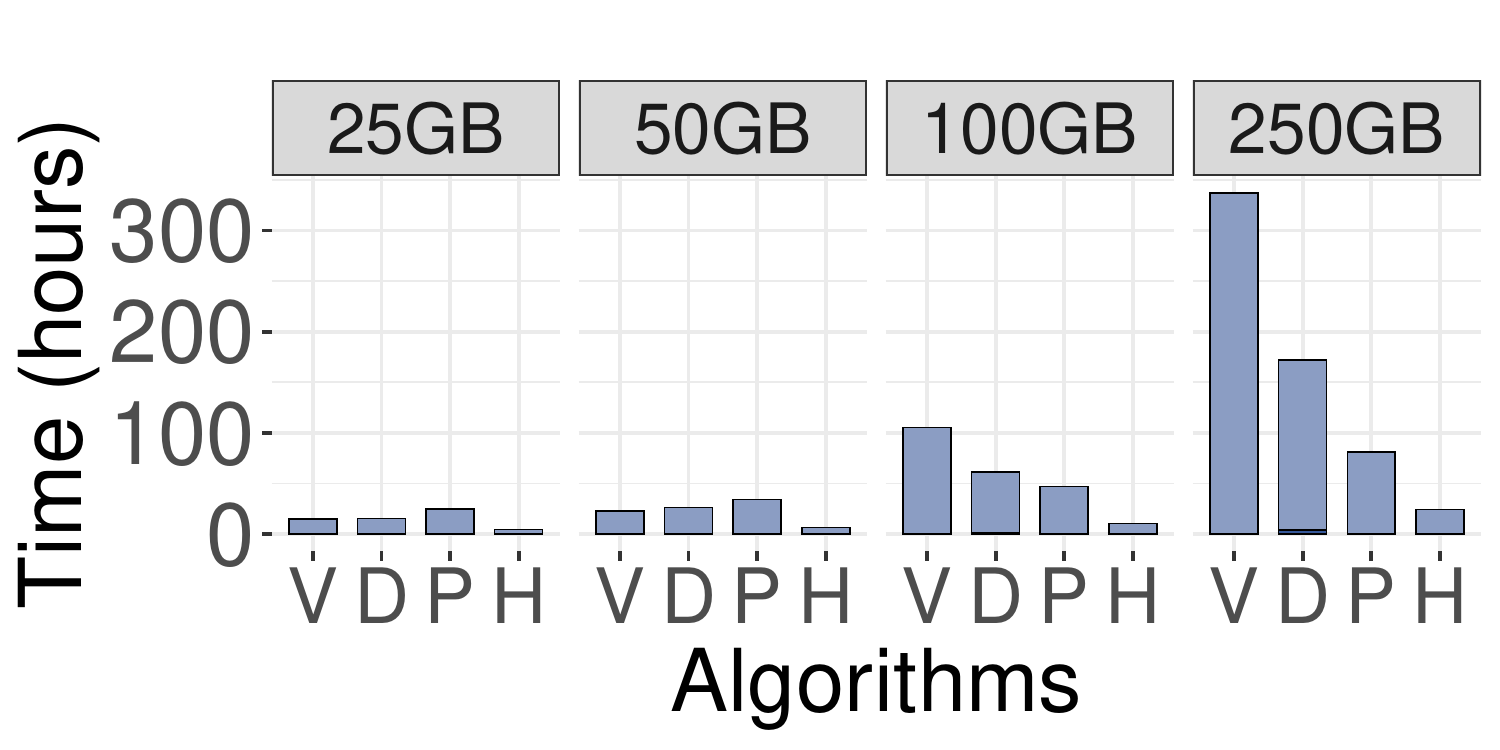}
		\caption{Idx + 10K Queries}  
		\label{exact:varysize:hdd:10K}
	\end{subfigure}\\
	\caption{Scalability with increasing dataset sizes (100 1NN exact queries)}
	\vspace*{-0.2cm}
	\label{exact:varysize:hdd}
\end{figure}

\begin{figure}[tb]
	\captionsetup{justification=centering}
	\centering	
	\begin{minipage}{\columnwidth}
		\begin{subfigure}{\columnwidth}
			\captionsetup{justification=centering}	
			\includegraphics[width=\columnwidth] {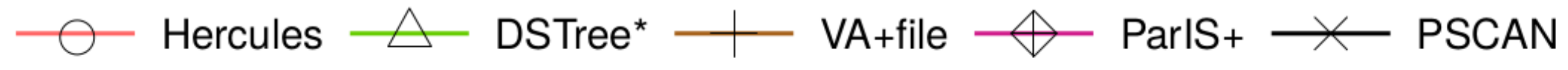}
			\label{fig:idx:time:comParIS+on:Hercules:ParIS+:andromache:coldcache}
		\end{subfigure}\\
		\vspace{-0.5cm}
		\end{minipage}
	\begin{minipage}{0.49\columnwidth}				
		\begin{subfigure}{\columnwidth}
		\captionsetup{justification=centering}	
		\includegraphics[width=\columnwidth] {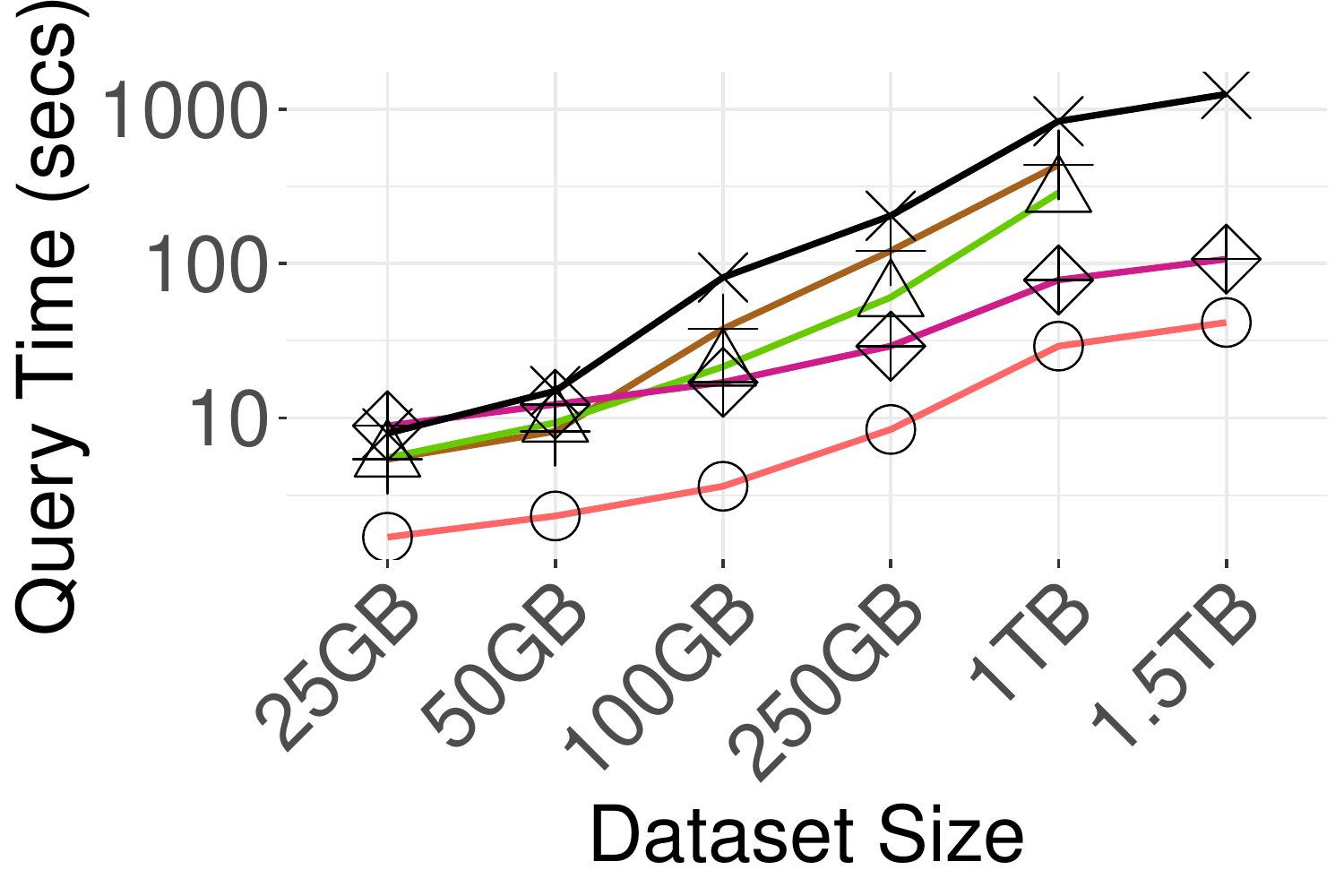}
		\end{subfigure}	
		\vspace{-0.2cm}
		\caption{\ke{Scalability with dataset size}}
		\label{exact:proc:synth:hdd}
	\end{minipage}	
	\begin{minipage}{0.49\columnwidth}
		\begin{subfigure}{\columnwidth}
			\includegraphics[width=\columnwidth] {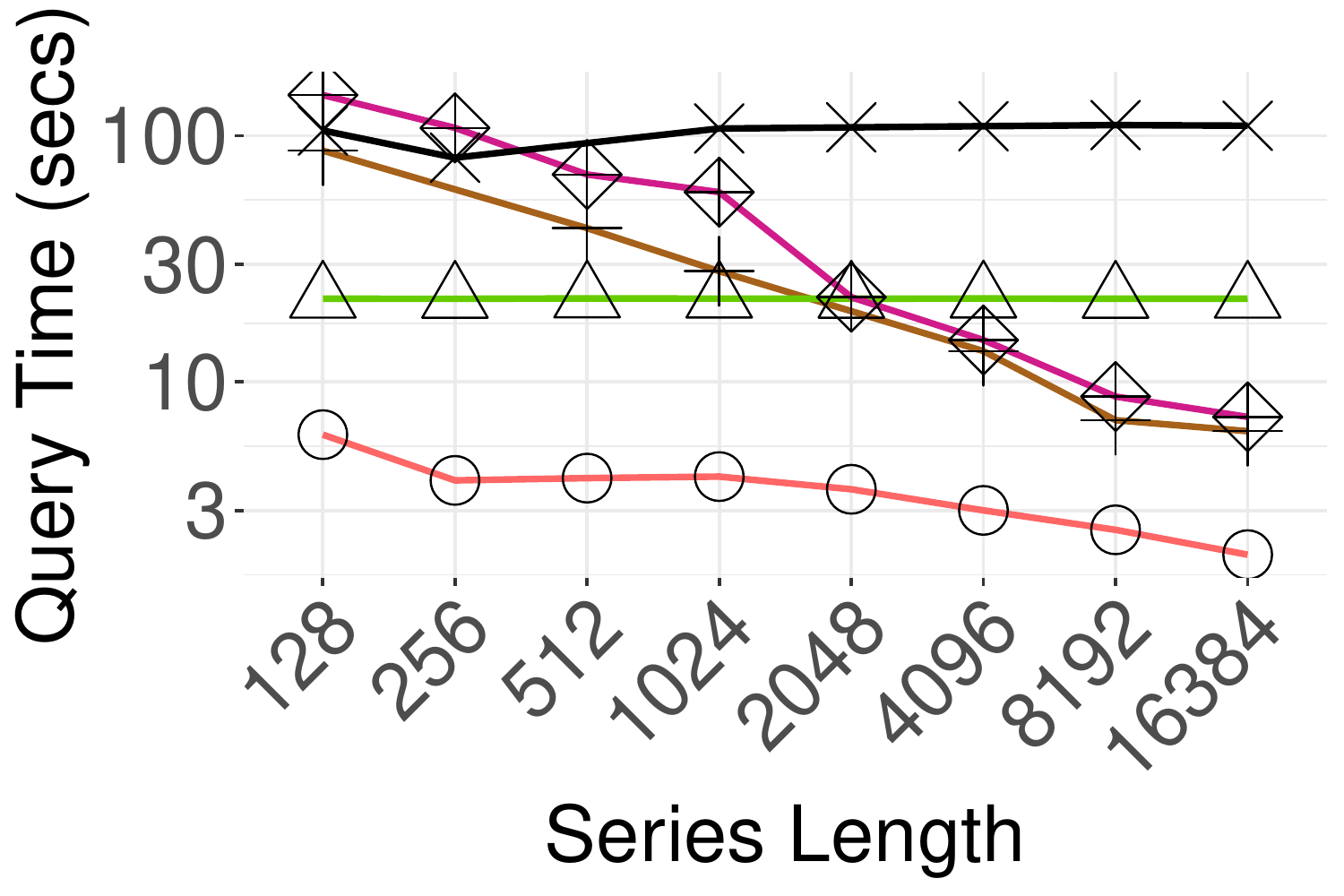}
		\end{subfigure} 
			\vspace{-0.2cm}

		\caption{\ke{Scalability with series length}}
		\label{exact:proc:varylength:hdd}
	\end{minipage}
	

\end{figure}

\hidevldb{\subsubsection{Scalability with Increasing Dataset Size}}
\noindent{\bf Scalability with Increasing Dataset Size.}
We now evaluate the performance of Hercules on in-memory and out-of-core datasets. We use four different dataset sizes such that two fit in-memory (25GB and 50GB) and two are out-of-core (100GB and 250GB). We report the combined index construction and query answering times for 100 and 10K exact 1NN queries. Figure~\ref{exact:varysize:hdd}  demonstrates the robustness of Hercules across all dataset sizes. 
Compared to DSTree*, Hercules is 3x-4x faster in index construction time, and 1.6x-10x faster in query answering.
The only scenario where Hercules does not win is against ParIS+ on the 250GB dataset and the small query workload (Figure~\ref{exact:varysize:hdd:1}). 
However, on the large query workload of 10K queries, Hercules outperforms ParIS+ by a factor of 3.
This is because Hercules invests more on index construction (one-time cost), in order to significantly speed-up query answering (recurring cost).  
\ke{We run an additional scalability experiment with two very large datasets (1TB and 1.5TB), and measure the average runtime for one 1-NN query (over a workload of 100 queries). Figure~\ref{exact:proc:synth:hdd} shows that Hercules outperforms all competitors including the optimized parallel scan PSCAN. The 1.5TB results for DSTree* and VA+file are missing because their index could not be built (out-of-memory issue for VA+file; indexing taking over 24 hours for DSTree*).}

\hidevldb{\subsubsection{Scalability with Increasing Series Length}}
\noindent{\bf Scalability with Increasing Series Length.}
In this experiment, we study the performance of the algorithms as we vary the data series length.
Figure~\ref{exact:proc:varylength:hdd} shows that Hercules (bottom curve) consistently performs \ke{5-10x}
faster than the best other competitor, i.e., DSTree* for series of length 128-1024, and VA+file (followed very closely by Paris+) for series of length 2048-16384. \ke{Hercules outperforms PSCAN by at least one order of magnitude on all lengths.}

\begin{figure}[tb]
	\captionsetup{justification=centering}	
	\centering
	\begin{subfigure}{0.49\columnwidth}
		\captionsetup{justification=centering}	
		\includegraphics[width=\columnwidth] {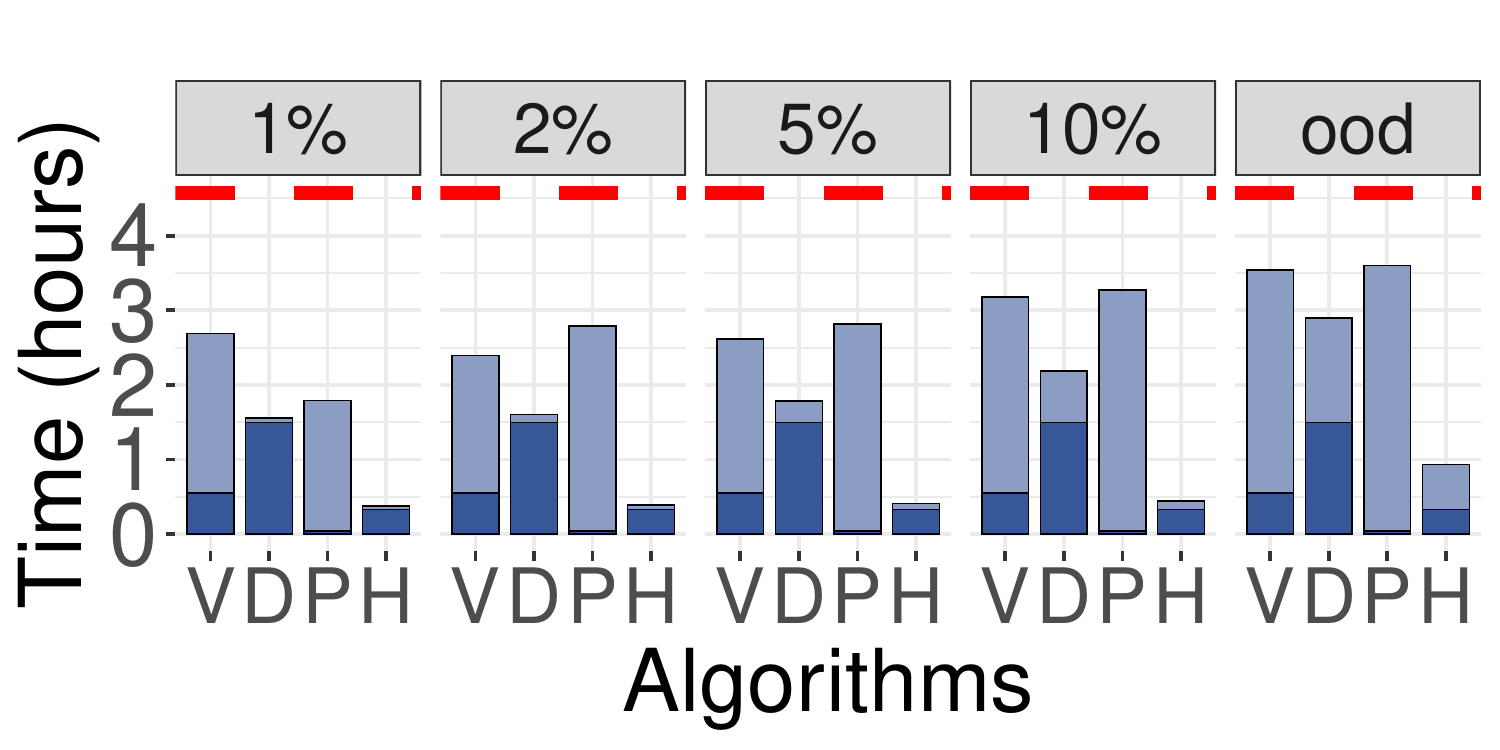}
		\caption{Sald (100 Queries)}  
		\label{exact:easyhard:index+queries:hdd:sald:100}
	\end{subfigure}
	\begin{subfigure}{0.49\columnwidth}
		\captionsetup{justification=centering}	
		\includegraphics[width=\columnwidth] {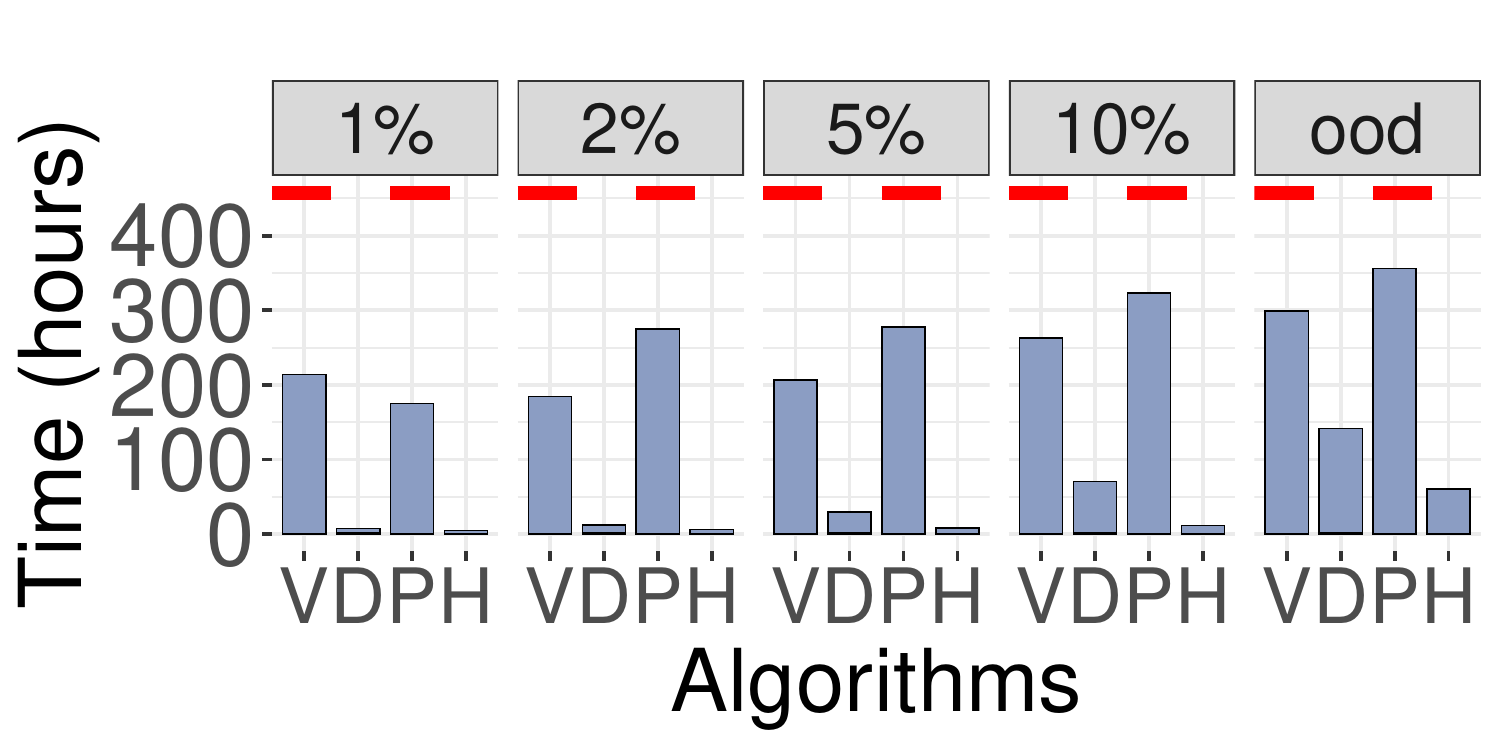}
		\caption{Sald (10K Queries)}  
		\label{exact:easyhard:index+queries:hdd:sald:10K}
	\end{subfigure}
	\begin{subfigure}{0.49\columnwidth}
		\captionsetup{justification=centering}	
		\includegraphics[width=\columnwidth] {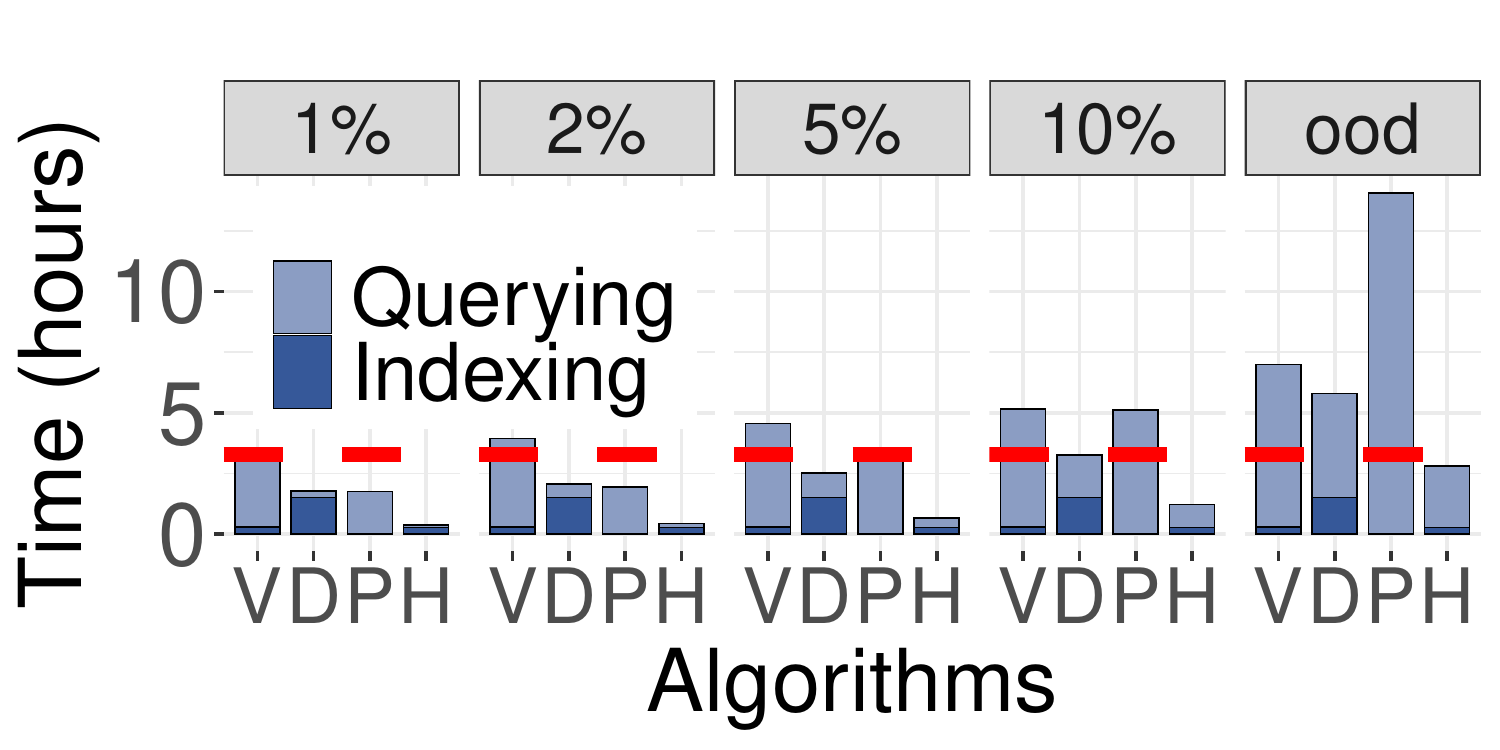}
		\caption{Seismic (100 Queries)}  
		\label{exact:easyhard:index+queries:hdd:seismic:100}
	\end{subfigure}
	\begin{subfigure}{0.49\columnwidth}
		\captionsetup{justification=centering}	
		\includegraphics[width=\columnwidth] {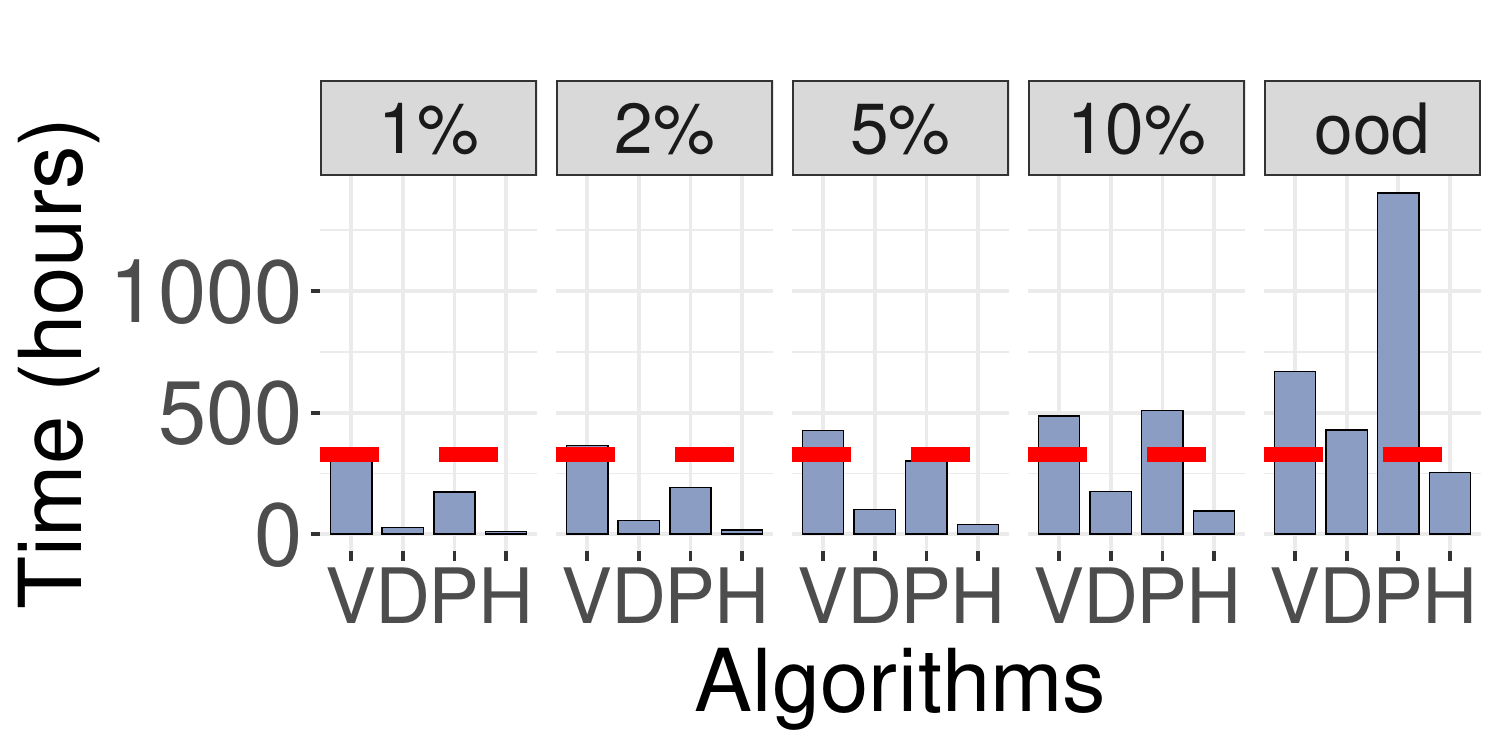}
		\caption{Seismic (10K Queries)}  
		\label{exact:easyhard:index+queries:hdd:seismic:10K}
	\end{subfigure}
	\begin{subfigure}{0.49\columnwidth}
		\captionsetup{justification=centering}	
		\includegraphics[width=\columnwidth] {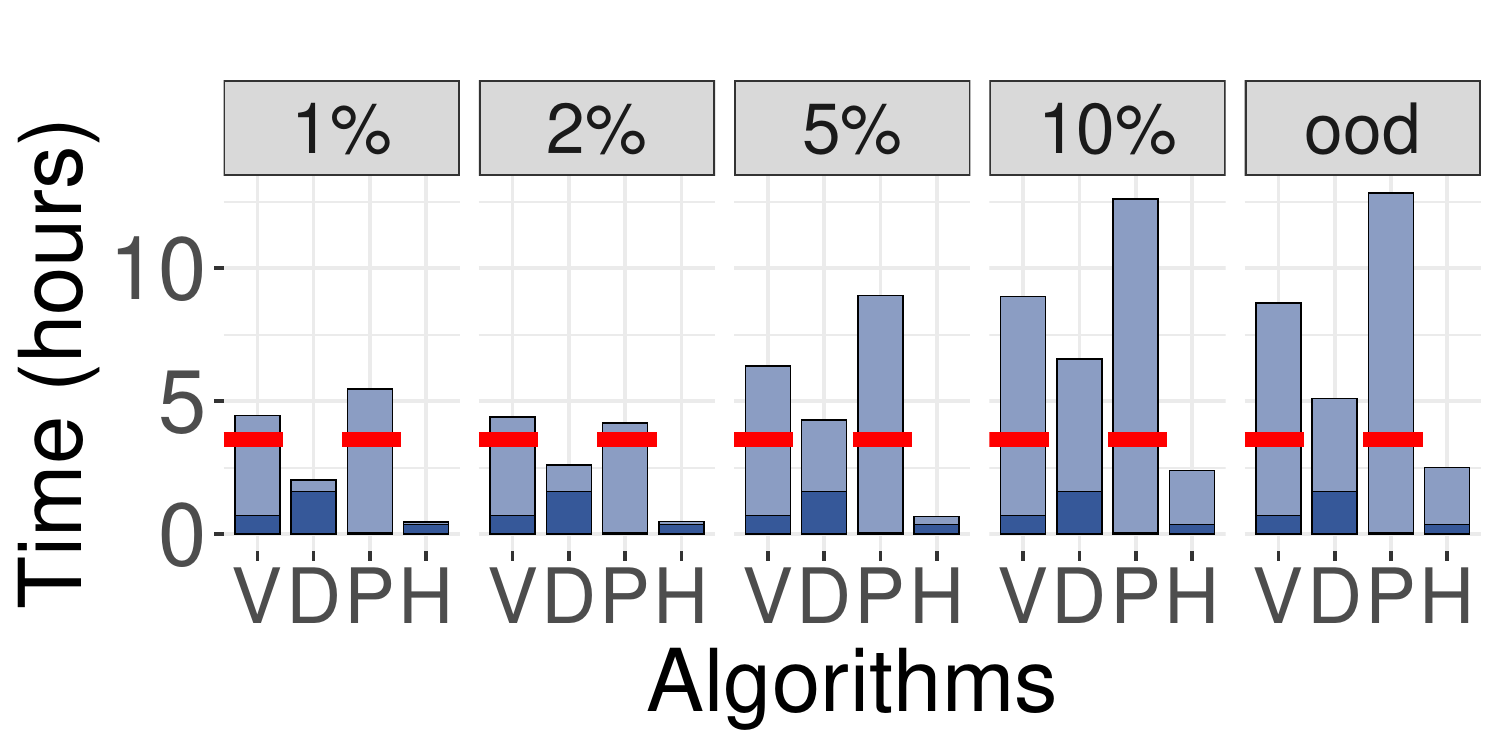}
		\caption{Deep (100 Queries)}  
		\label{exact:easyhard:index+queries:hdd:deep:100}
	\end{subfigure}
	\begin{subfigure}{0.49\columnwidth}
		\captionsetup{justification=centering}	
		\includegraphics[width=\columnwidth]
		{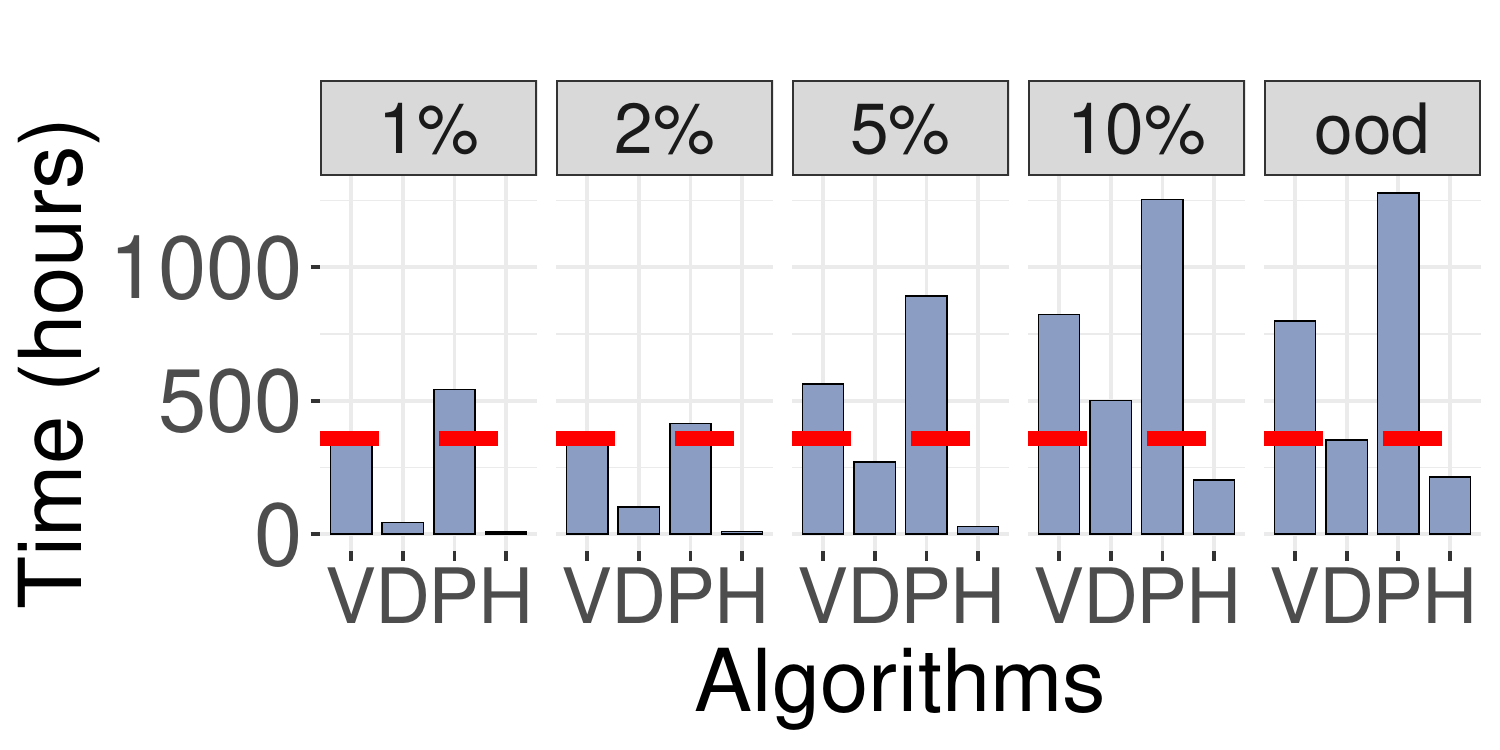}
		\caption{Deep (10K Queries)}  
		\label{exact:easyhard:index+queries:hdd:deep:10K}
	\end{subfigure}
\vspace*{-0.2cm}
	\caption{\ke{Scalability with query difficulty (combined indexing and query answering times)}  
	}  
\vspace*{-0.2cm}
	\label{exact:easyhard:index+queries:hdd}
\end{figure}

\hidevldb{\subsubsection{Scalability with Increasing Query Difficulty}}
\noindent{\bf Scalability with Increasing Query Difficulty.}
We now evaluate the performance of Hercules in terms of index construction time, and query answering on 100 and 10K exact 1NN queries of varying difficulty over the real datasets. 
Figure~\ref{exact:easyhard:index+queries:hdd} shows the superiority of Hercules against its competitors 
for all datasets and query workloads. 
Observe that Hercules is the only method that, across all experiments, builds an index and answers 100/10K exact queries before the sequential scan (red dotted line) finishes execution.

\begin{figure*}[th]
	\captionsetup{justification=centering}	
	\centering
	\begin{subfigure}{0.16\textwidth}
		\captionsetup{justification=centering}	
		\includegraphics[width=\textwidth] {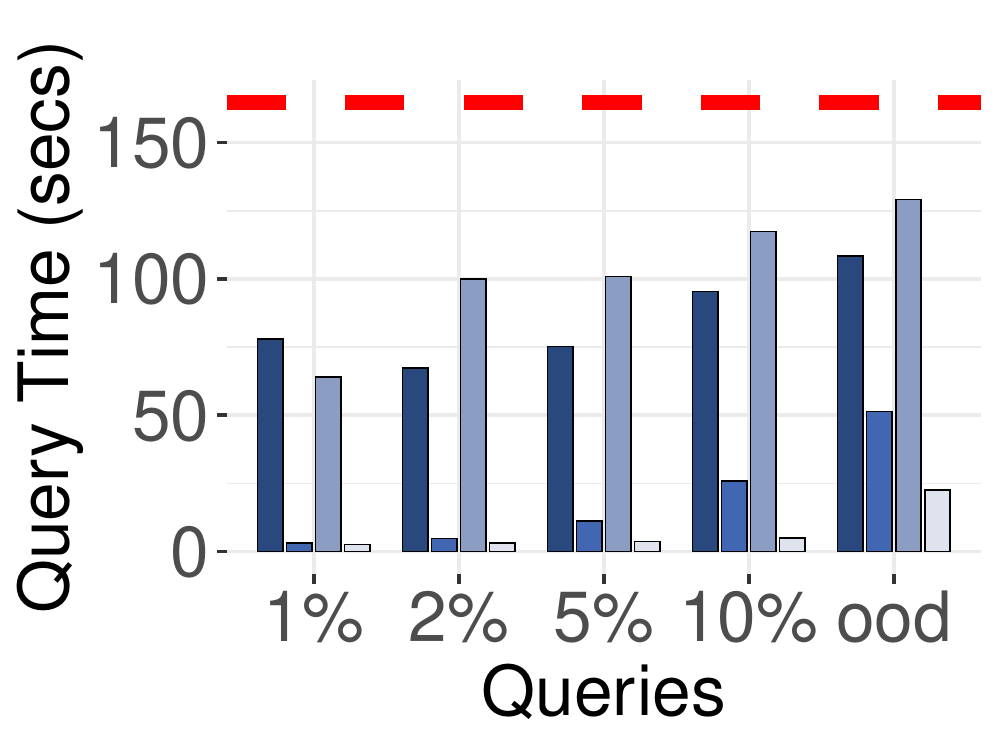}
		\caption{Time (SALD)}  
		\label{exact:easyhard:queries:hdd:sald:time}
	\end{subfigure}
	\begin{subfigure}{0.16\textwidth}
		\captionsetup{justification=centering}	
		\includegraphics[width=\textwidth]		 {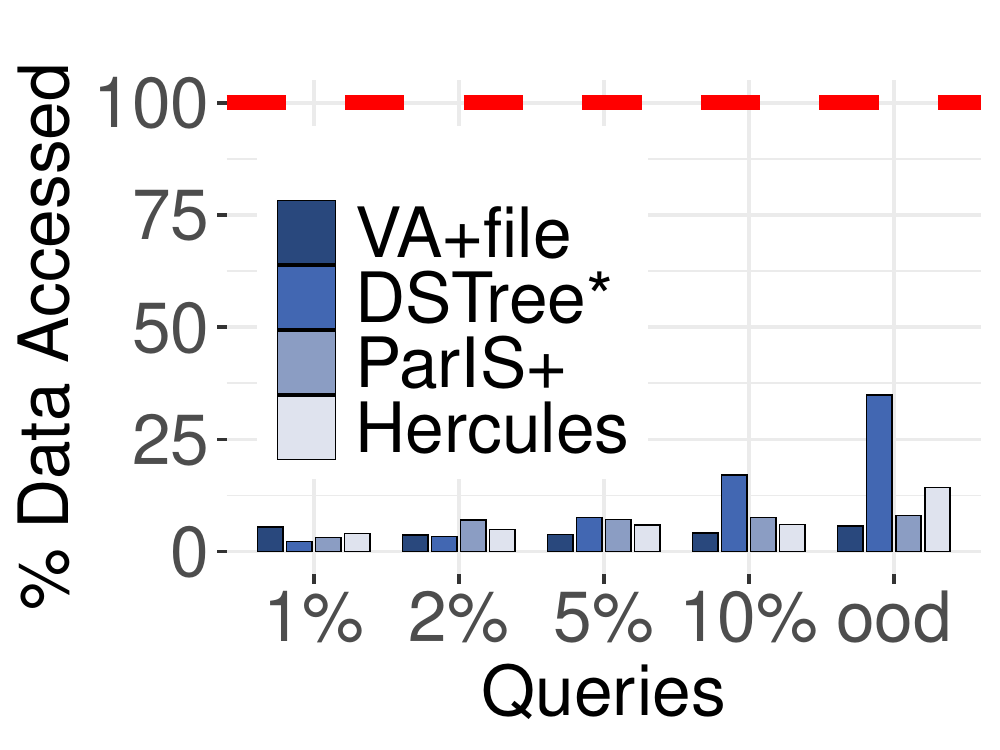}
		\caption{Data (SALD)}  
		\label{exact:easyhard:queries:hdd:sald:data}
	\end{subfigure}
	\begin{subfigure}{0.16\textwidth}
		\captionsetup{justification=centering}	
		\includegraphics[width=\textwidth] {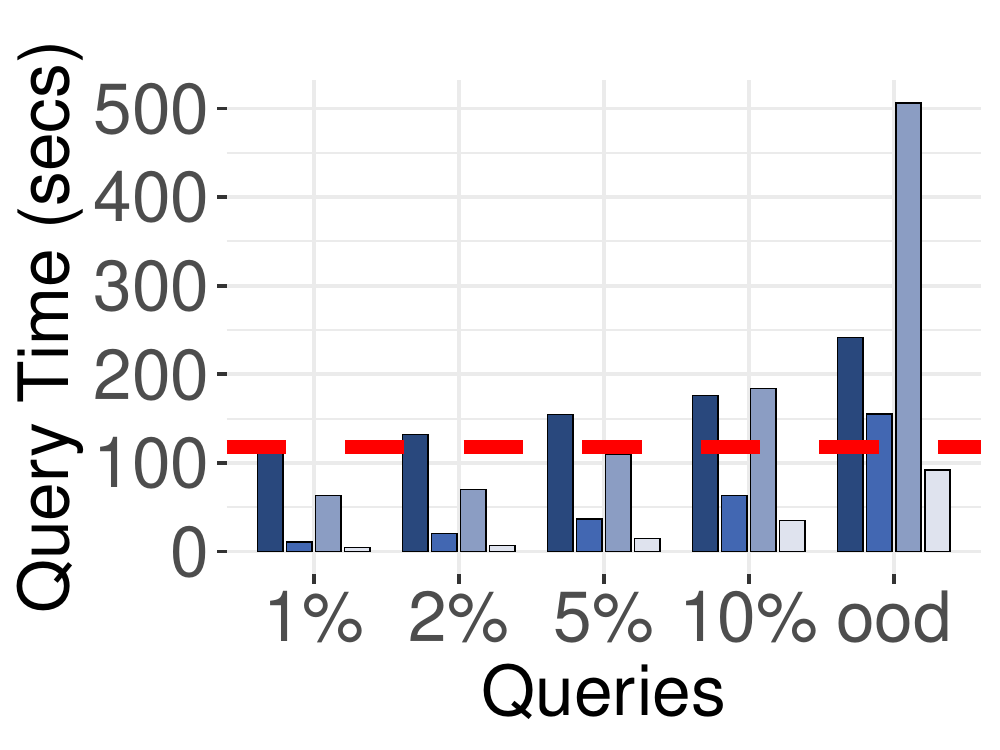}
		\caption{Time (Seismic)}  
		\label{exact:easyhard:queries:hdd:seismic:time}
	\end{subfigure}
	\begin{subfigure}{0.16\textwidth}
		\captionsetup{justification=centering}	
		\includegraphics[width=\textwidth] {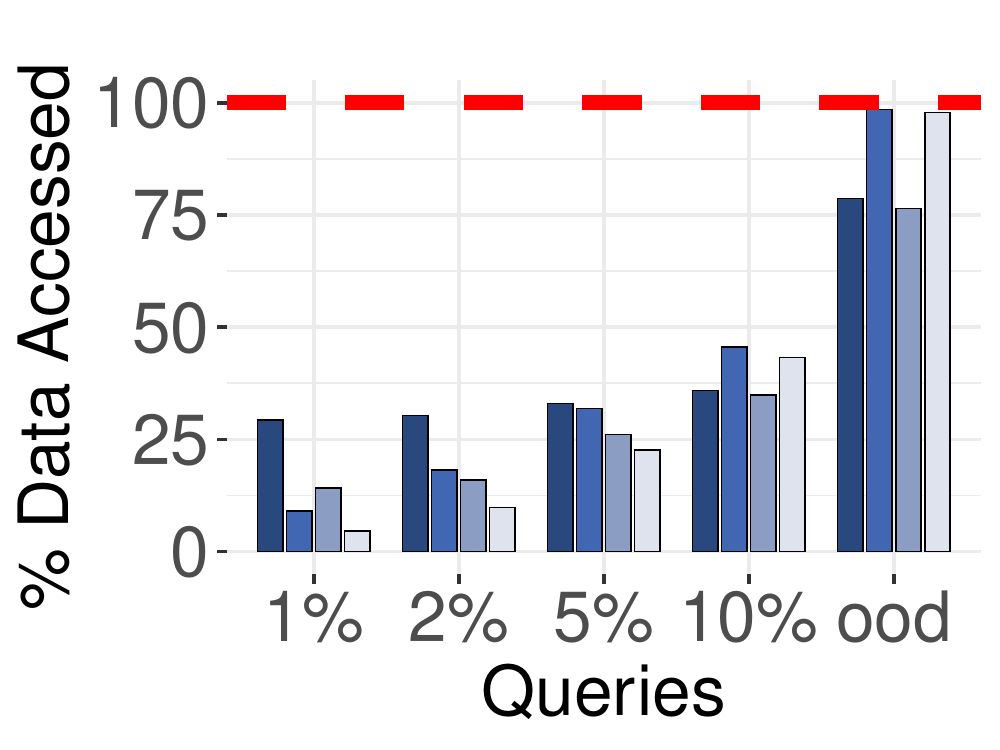}
		\caption{Data (Seismic)}  
		\label{exact:easyhard:queries:hdd:seismic:data}
	\end{subfigure}
	\begin{subfigure}{0.16\textwidth}
		\captionsetup{justification=centering}	
		\includegraphics[width=\textwidth] {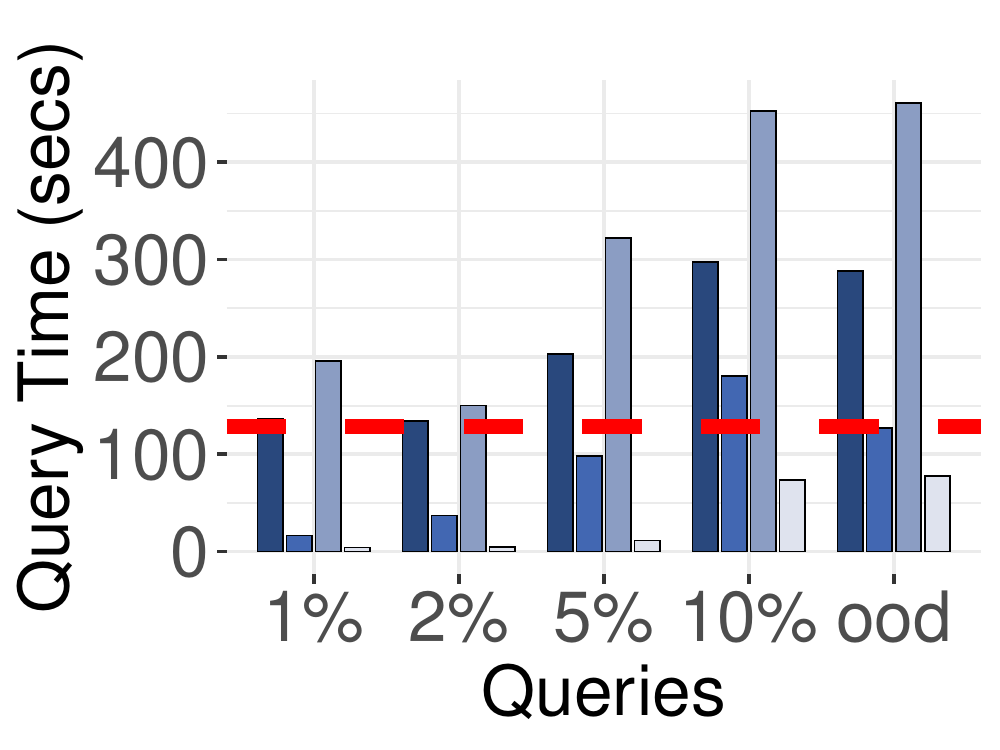}
		\caption{Time (Deep)}  
		\label{exact:easyhard:queries:hdd:deep:time}
	\end{subfigure}
	\begin{subfigure}{0.16\textwidth}
		\captionsetup{justification=centering}	
		\includegraphics[width=\textwidth] {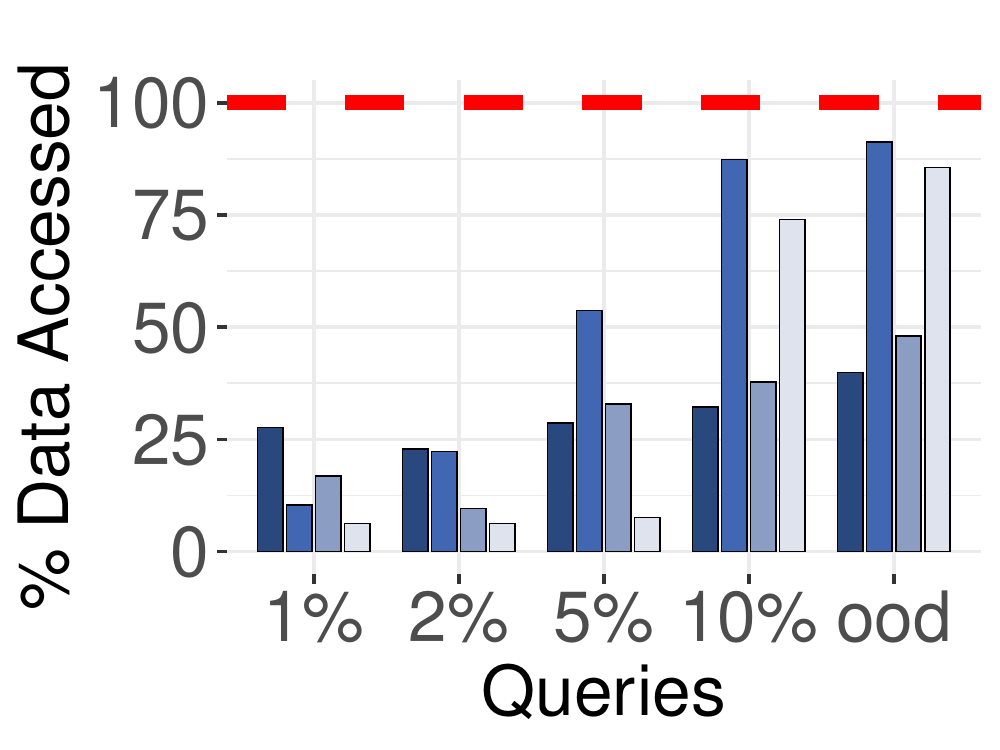}
		\caption{Data (Deep)}  
		\label{exact:easyhard:queries:hdd:deep:data}
	\end{subfigure}
\vspace*{-0.2cm}
	\caption{\ke{Scalability with query difficulty 
		(average query answering time and data accessed)}
	}  
\vspace*{-0.2cm}
	\label{exact:easyhard:queries:hdd}
\end{figure*}

In Figure~\ref{exact:easyhard:queries:hdd}, we focus on the query answering performance for 1NN exact queries of varying difficulty. 
We report the query time and the percentage of data accessed, averaged per query.
We observe that Hercules is superior to its competitors across the board. 
On the SALD dataset, 
Hercules is more than 2x faster than DSTree* on all queries, and 50x faster than ParIS+ on the easy (1\%, 2\%) to medium-hard queries (5\%). 
Hercules maintains its good performance on the hard dataset $ood$, where it is 6x faster than ParIS+ despite the fact that it accesses double the amount of data on disk. 
We observe the same trend on the Seismic dataset: Hercules is the only index performing better than a sequential scan on the hard $ood$ dataset although it accesses 96\% of the data. 
This is achieved thanks to two key design choices: the pruning thresholds based on which Hercules can choose to use a single thread to perform a skip-sequential scan (rather than using multiple threads to perform a large number of random I/O operations), and the LRDFile storage layout that leads to sequential disk I/O. 
The Deep dataset is notoriously hard~\cite{url/faiss,journal/tpami/ge2014,journal/pvldb/echihabi2018,journal/tkde/li19}, and we observe that indexes, except Hercules, start degenerating even on the easy queries (Figure~\ref{exact:easyhard:queries:hdd:deep:time}).

\begin{figure}[tb]
	\captionsetup{justification=centering}	
	\centering
	\begin{subfigure}{0.9\columnwidth}
		\captionsetup{justification=centering}	
		\includegraphics[width=\columnwidth] {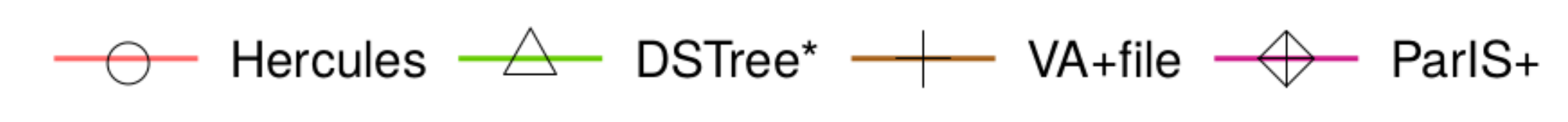}
		\label{fig:idx:time:comParIS+on:Hercules:ParIS+:andromache:coldcache}
	\end{subfigure}\\
     \vspace{-0.5cm}
	\begin{subfigure}{0.33\columnwidth}
		\centering
		\captionsetup{justification=centering}	
		\includegraphics[width=\columnwidth] {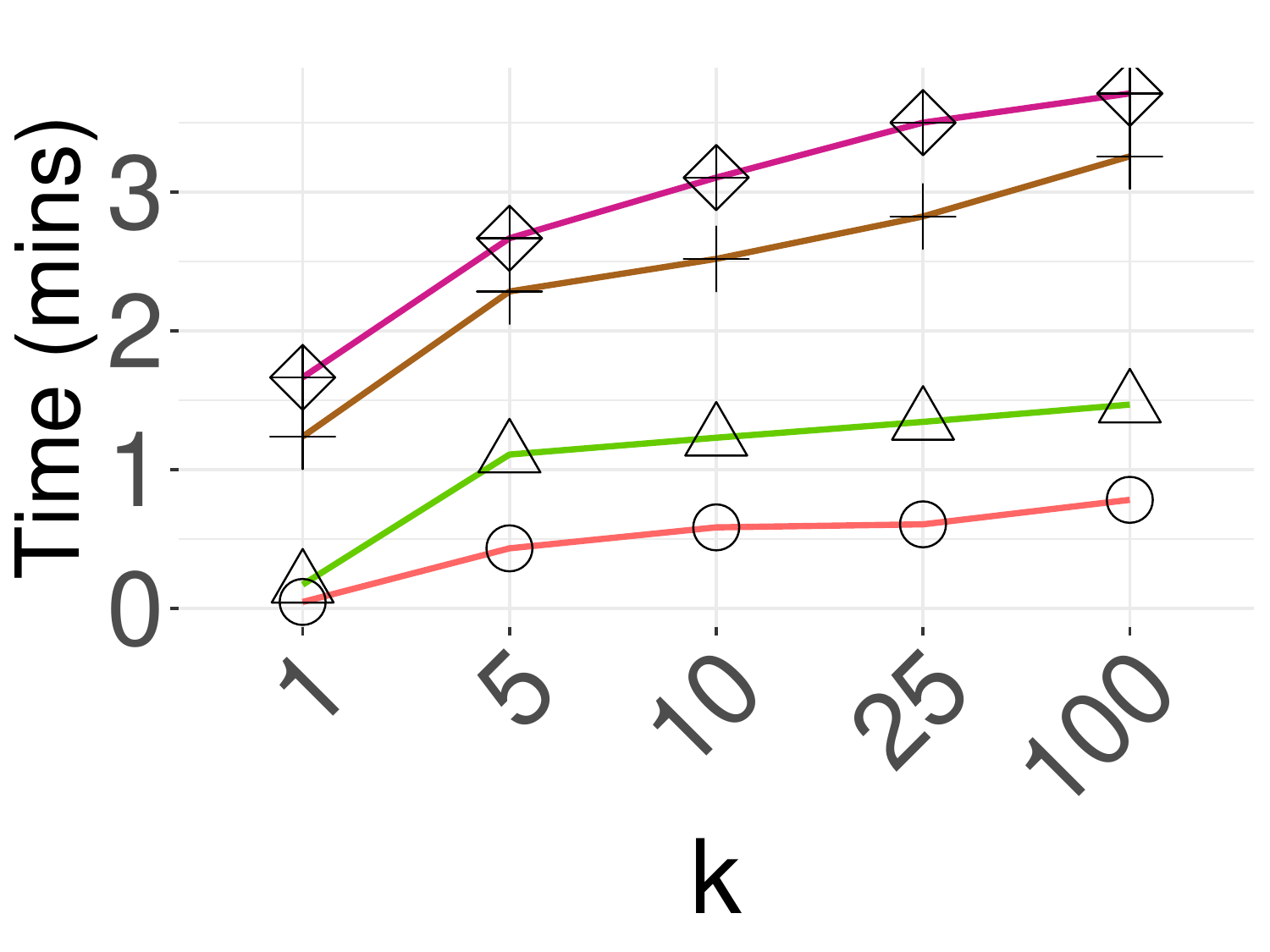}
		\caption{Time (SALD)}
		\label{exact:varyk:hdd:sald:time}
	\end{subfigure} 
	\begin{subfigure}{0.32\columnwidth}
		\centering
		\captionsetup{justification=centering}	
		\includegraphics[width=\columnwidth] {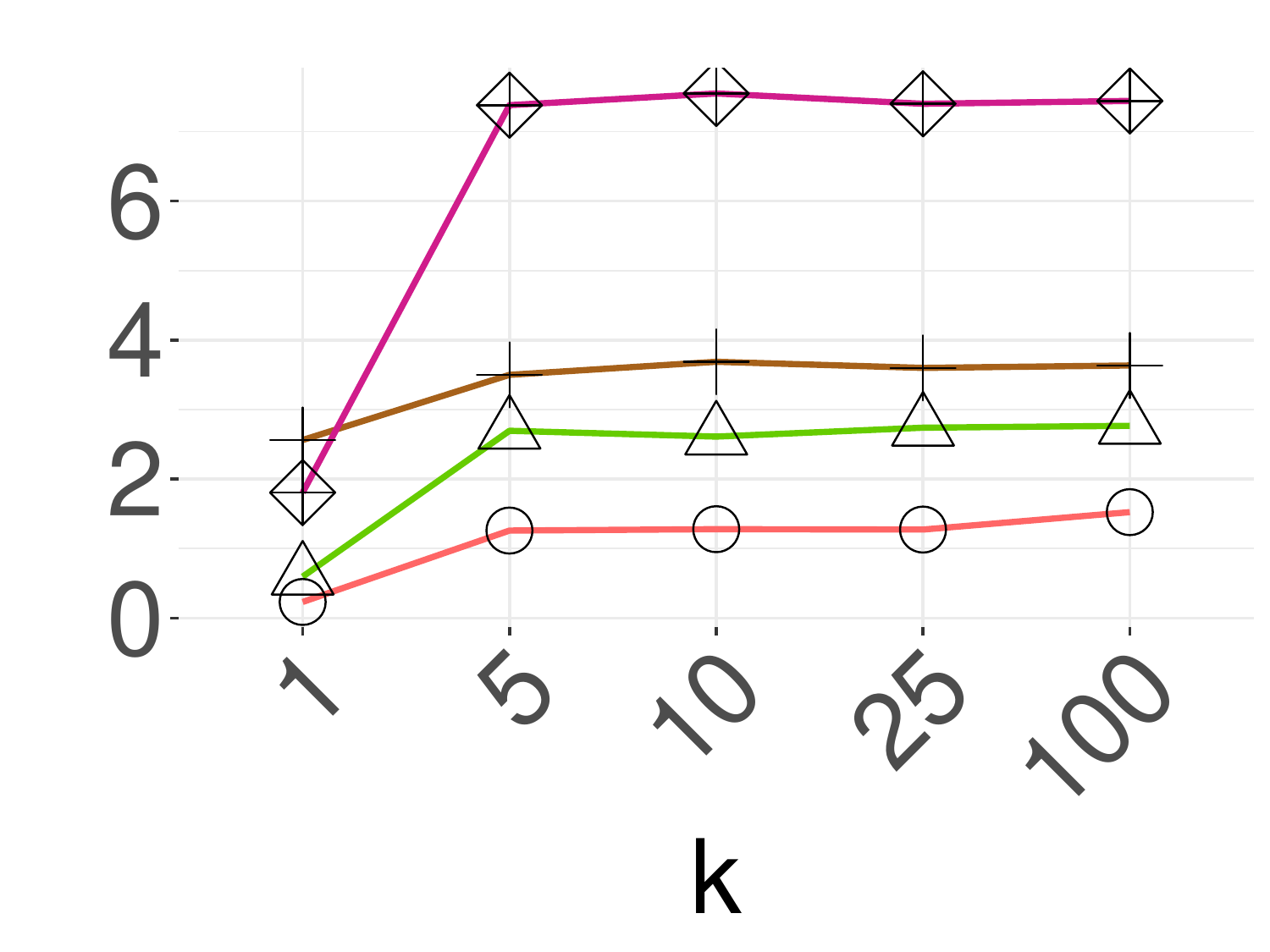}
		\caption{Time (Seis.)}  
		\label{exact:varyk:hdd:seismic:time}
	\end{subfigure} 
	\begin{subfigure}{0.32\columnwidth}
		\centering
		\captionsetup{justification=centering}	
		\includegraphics[width=\columnwidth] {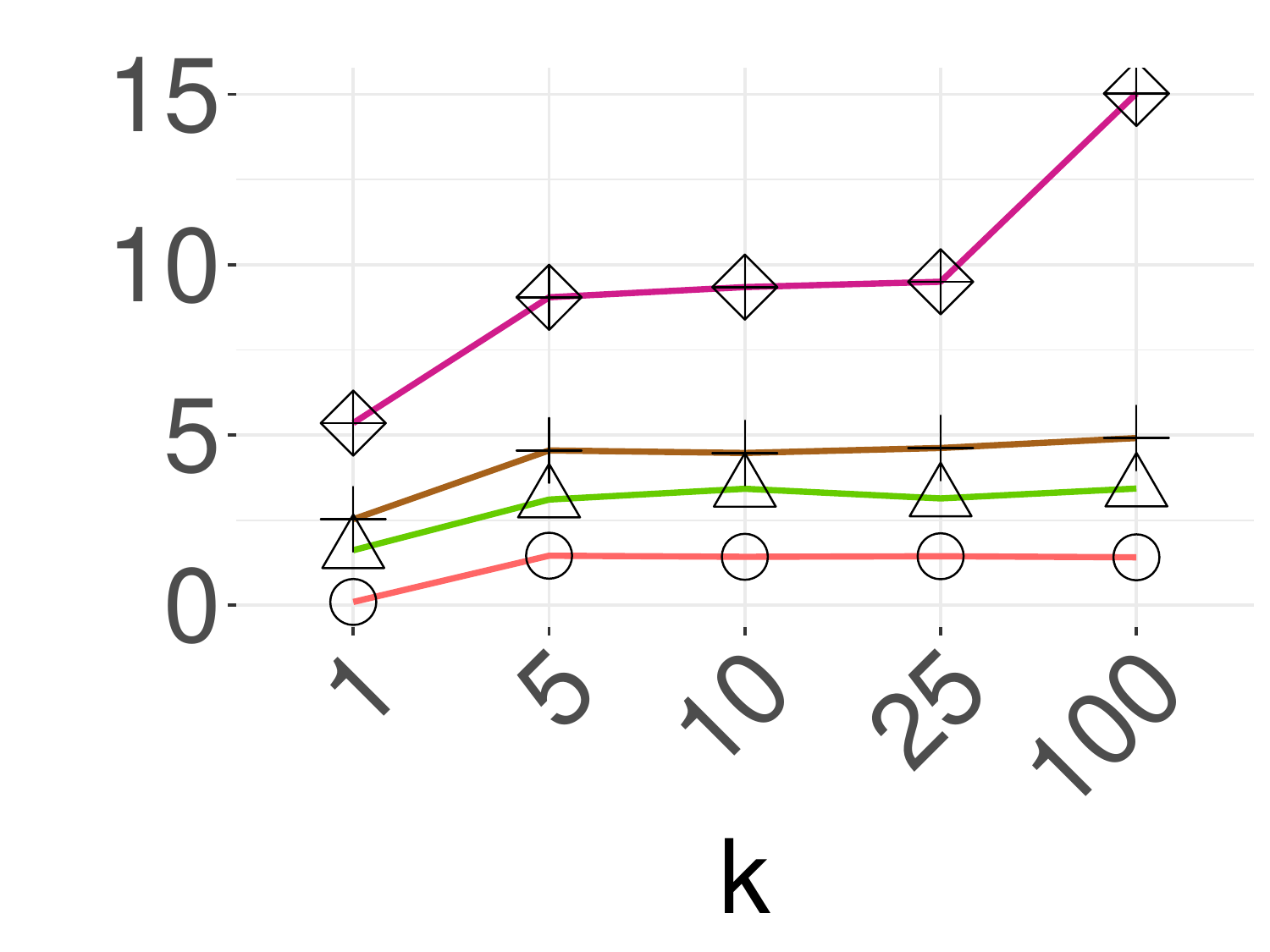}
		\caption{Time (Deep)}  
		\label{exact:varyk:hdd:deep:time}
	\end{subfigure} 
	\begin{subfigure}{0.33\columnwidth}
		\centering
		\captionsetup{justification=centering}	
		\includegraphics[width=\columnwidth] {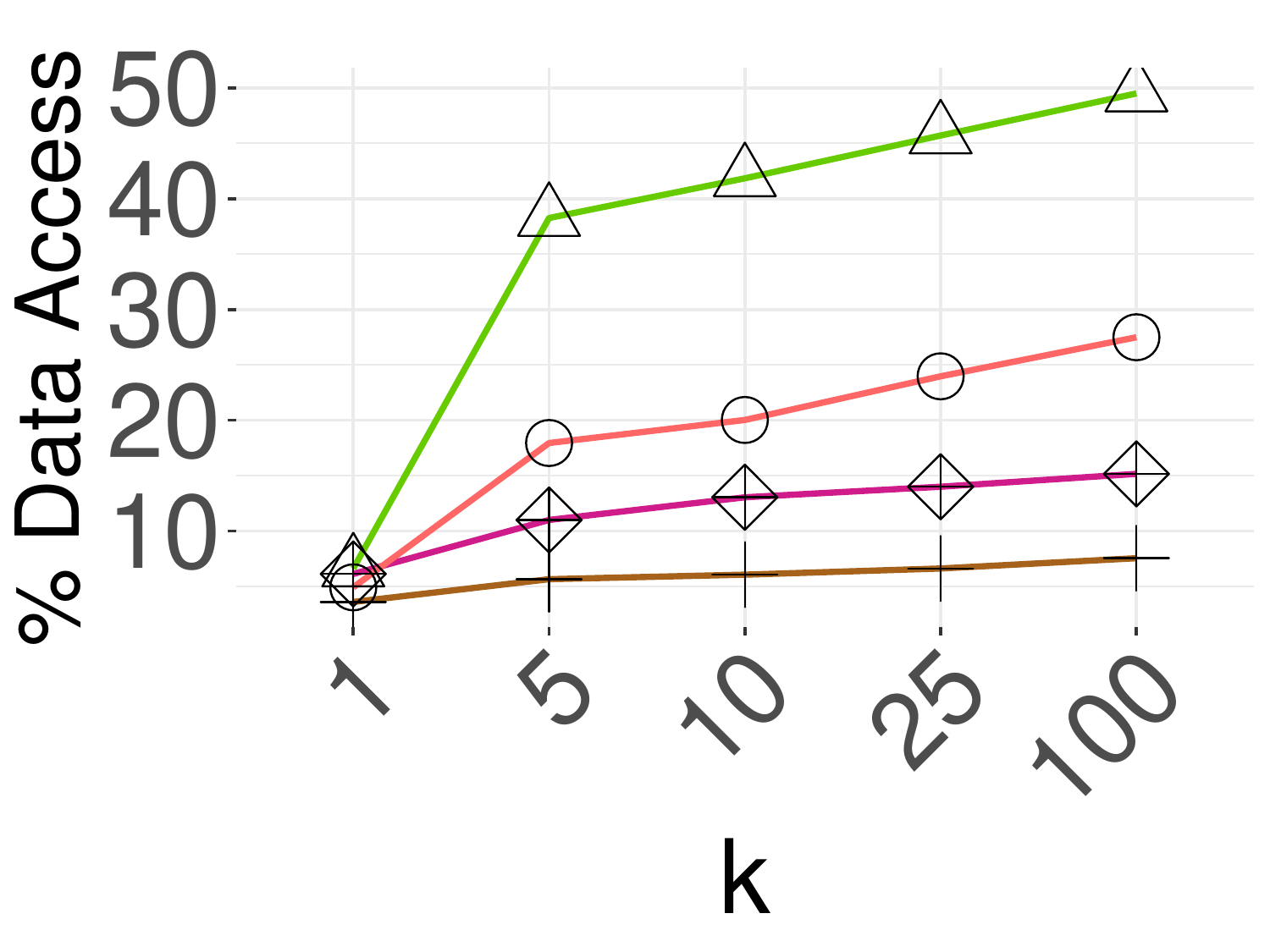}
		\caption{Data (SALD)}  
		\label{exact:varyk:hdd:sald:data}
	\end{subfigure} 
	\begin{subfigure}{0.32\columnwidth}
		\centering
		\captionsetup{justification=centering}	
		\includegraphics[width=\columnwidth] {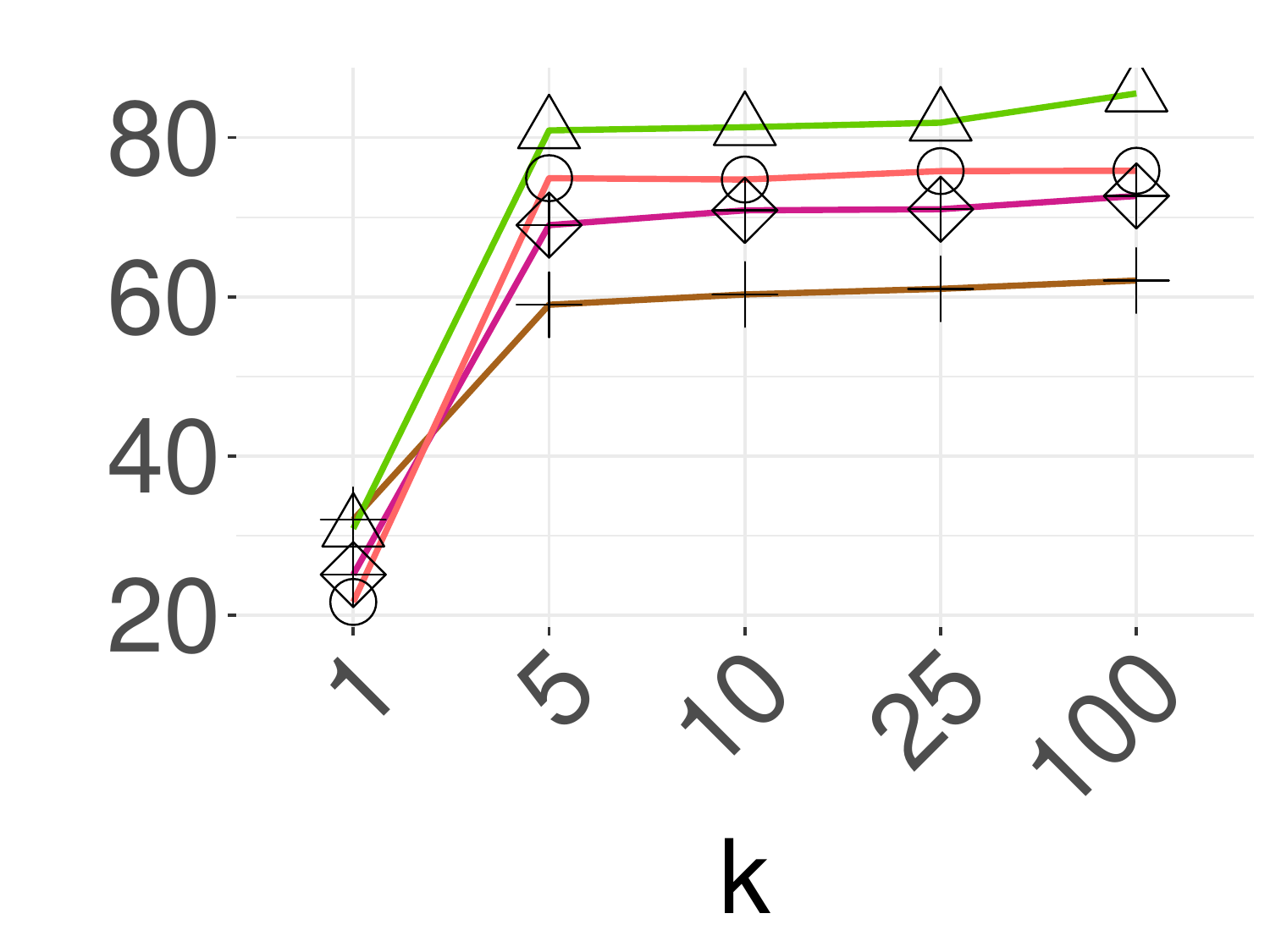}
		\caption{Data (Seis.)}  
		\label{exact:varyk:hdd:seismic:data}
	\end{subfigure} 
	\begin{subfigure}{0.32\columnwidth}
		\centering
		\captionsetup{justification=centering}	
		\includegraphics[width=\textwidth] {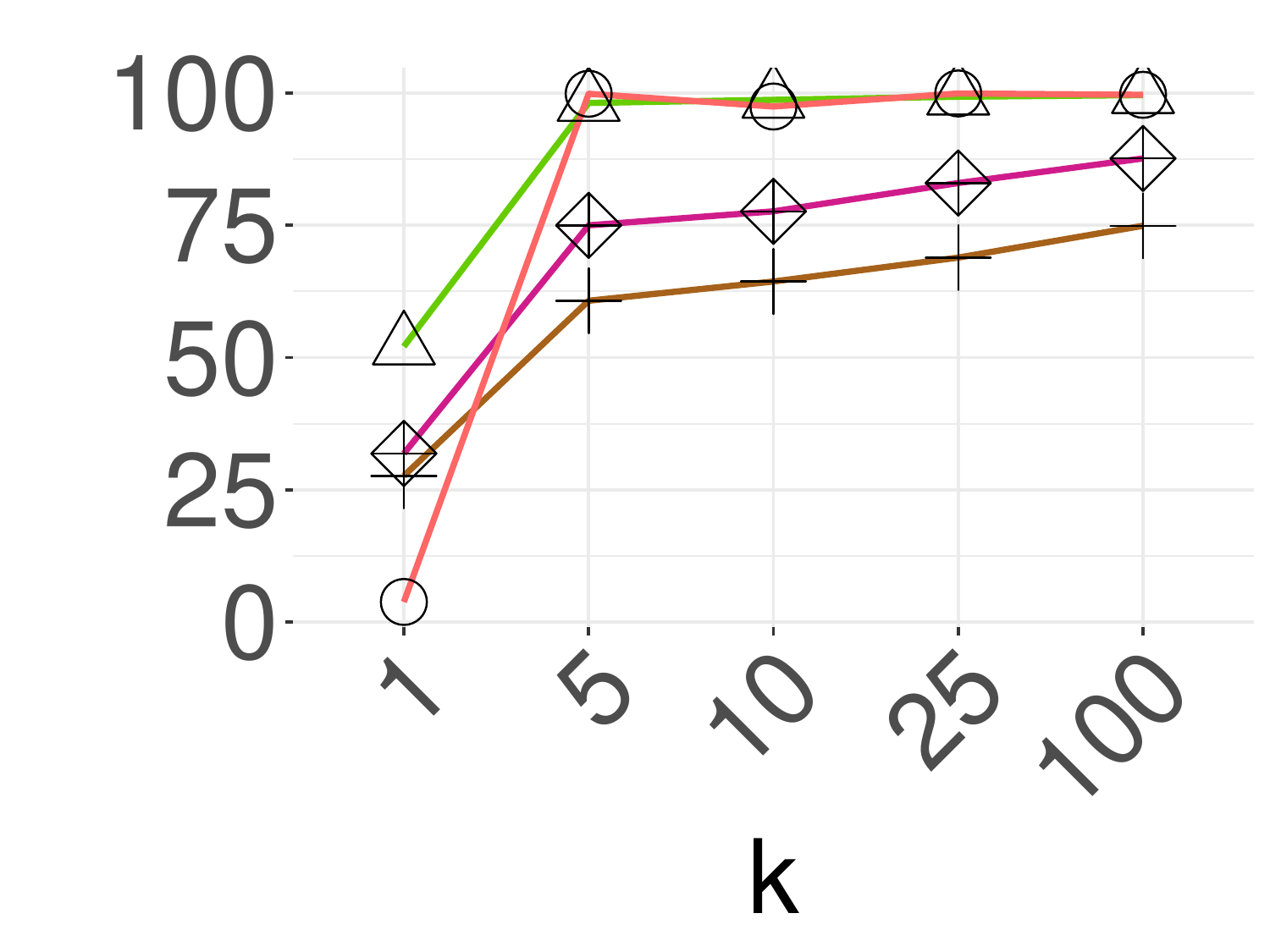}
		\caption{Data (Deep)}  
		\label{exact:varyk:hdd:deep:data}
	\end{subfigure} 
\vspace*{-0.2cm}
	\caption{\ke{Scalability with increasing k} 
	}  
\vspace*{-0.1cm}
	\label{exact:varyk:hdd}
\end{figure}

\begin{figure}[tb]
	\captionsetup{justification=centering}
	\centering	
	\begin{subfigure}{0.35\columnwidth}
	\centering
	\captionsetup{justification=centering}	
	\includegraphics[width=\textwidth] {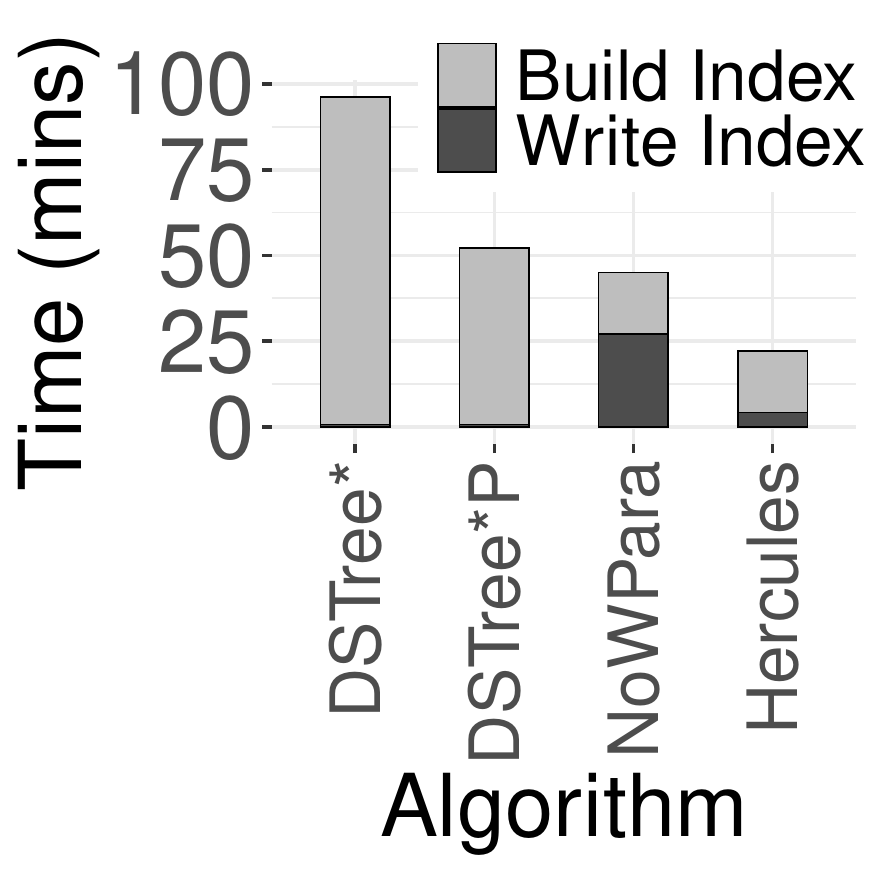}
	\caption{Indexing}  
	\label{exact:ablation:hdd:idx:deep}
	\end{subfigure} 
	\begin{subfigure}{0.63\columnwidth}
	\centering
	\captionsetup{justification=centering}	
	\includegraphics[width=\columnwidth] {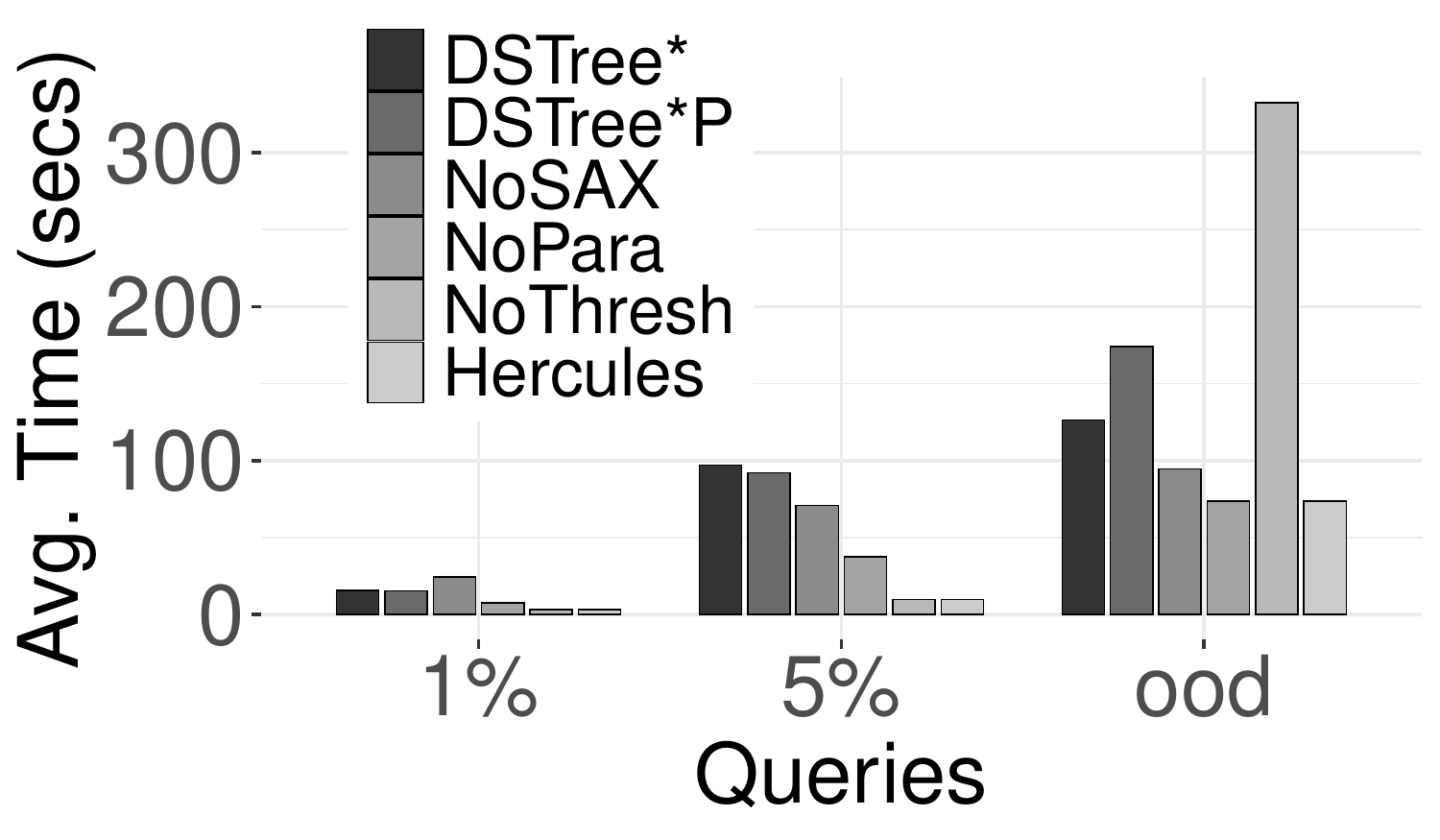}
	\caption{Query Answering}  
	\label{exact:ablation:hdd:proc:deep}
	\end{subfigure} 
	\caption{Ablation Study}  
	\label{exact:ablation:hdd}
	\vspace*{-0.4cm}
\end{figure}

Overall, Hercules performs query answering 5.5x-63x 
faster than ParIS+, and 1.5x-10x 
faster than DSTree*.
Once again, Hercules is the only index that wins across the board, performing 1.3x-9.4x faster than the second-best approach. 
\noindent{\bf Scalability with Increasing k.}
In this scenario, we choose the medium-hard query workload $5\%$ for synthetic and real datasets of size 100GB, and vary the number of neighbors k in the kNN exact query workloads. We then measure the query time and percent of data accessed, averaged over all queries. 
Figure~\ref{exact:varyk:hdd} shows that Hercules wins across the board for all values of $k$. 
Note that finding the first neighbor is the most costly operation for DSTree* and Hercules, while the performance of ParIS+ deteriorates as the number of neighbors increases. 
This is due to ParIS+ employing a skip-sequential query answering algorithm, with the raw data of the neighbors of a query being located anywhere in the dataset file, whereas in DSTree* and Hercules these data 
reside within the same subtree. 
(Note also that the skip-sequential algorithm used by Hercules operates on the LRDFile, which stores contiguously the data of each leaf.)

\hidevldb{\subsubsection{Ablation Study}}
\noindent{\bf Ablation Study.} 
\label{sec:ablation}
In this experiment, we study the individual (not cumulative) effect on performance when removing each of the main building blocks of Hercules.
Figure~\ref{exact:ablation:hdd} summarizes the results \hidevldb{of the ablation study }on the Deep dataset (the other datasets and query workloads show similar trends~\cite{url/Hercules}\hidevldb{, and are omitted for brevity}). 
Figure~\ref{exact:ablation:hdd:idx:deep} shows the total index building and \hidevldb{index }writing times for DSTree*, our parallelization of DSTree* (DSTree*P), Hercules without parallelization of index writing (NoWPara), and Hercules. Note that although DSTree*P exploits parallelism, it still incurs a very large index building cost. This is because insert workers need to lock entire paths (from the root to a leaf) for updating node statistics, causing a large synchronization overhead. 
In NoWPara, threads only lock leaf nodes since statistics of internal nodes are updated at the index writing phase, leading to a faster index building phase. 
However, the index writing phase is slower because of the additional calculations. 
By parallelizing index writing \emph{bottom-up}, Hercules avoids some of the synchronization overhead, and achieves a much better performance. 

We evaluate query answering in Figure~\ref{exact:ablation:hdd:proc:deep}, on three workloads of varying difficulty. 
We first remove the iSAX summarization and rely only on EAPCA for pruning (NoSAX), then we remove parallelization altogether from Hercules (NoPara), then we evaluate Hercules without the pruning thresholds (NoThresh), i.e., Hercules applies the same parallelization strategy to each query regardless of its difficulty. 
We observe that using only EAPCA pruning (NoSAX) always worsens performance, regardless of the query difficulty. 
Besides, the parallelization strategy used by Hercules is effective as it improves query efficiency (better than NoPara) for easy and medium-hard queries and has no negative effect on hard queries. 
Finally, we observe that the pruning thresholds contribute significantly to the efficiency of Hercules (better than NoThresh) on hard workloads (i.e., ood), and have no negative effect on the rest.
\section{Conclusions and Future Work}
\label{sec:conclusions}

\hidevldb{In this paper, we {\bd ??? remove following "We}}We proposed Hercules, a novel index for exact similarity search over large data series collections. 
We demonstrated the superiority \hidevldb{and robustness }of Hercules against the state-of-the-art \hidevldb{methods }using an extensive experimental evaluation: Hercules performs between 1.3x-9.4x faster than the best competitor (which is different for the various datasets and workloads\hidevldb{ in our experiments}), and is the first index that achieves better performance than the optimized serial-scan algorithm across \emph{all} workloads. 
In our future work, we will study in detail the behavior of Hercules in approximate query answering (with/without deterministic/probabilistic quality guarantees~\cite{journal/pvldb/echihabi2019})\hidevldb{, and also enhance Hercules with an auto-tuning mechanism that can determine automatically the optimal parameters for different hardware platforms and query workloads}.

\begin{acks}
\vspace*{0.3cm}
We thank Botao Peng for sharing his ParIS+ code. Work partially supported by program Investir l'Avenir and Univ. of Paris IdEx Emergence en Recherche ANR-18-IDEX-0001 and EU project NESTOR (MSCA {\#}748945).
\end{acks}

\balance

\bibliographystyle{ACM-Reference-Format}
\bibliography{ref,pargis,parisinmemory}  

\end{document}